\documentclass[fleqn,usenatbib]{mnras}

\usepackage[T1]{fontenc}
\usepackage{ae,aecompl}
\usepackage[authoryear]{natbib}
\usepackage{booktabs}
\usepackage[para,online,flushleft]{threeparttable}

\usepackage{graphicx}	
\usepackage{amsmath}	
\usepackage{amssymb}	
\usepackage{multirow}
\usepackage{mathtools}
\usepackage{IEEEtrantools}
\usepackage{float}
\usepackage{rotating}
\usepackage{bm}
\usepackage{footmisc}

\usepackage{booktabs}
\usepackage{threeparttable}

\usepackage{hyperref}
\usepackage{txfonts}
\usepackage{scalefnt}

\LetLtxMacro{\oldtextsc}{\textsc}
\renewcommand{\textsc}[1]{\oldtextsc{\scalefont{1.2}#1}}

\newcommand{\cloudy}{\textsc{cloudy}}
\newcommand{\Msun}{\hbox{M$_{\rm{\odot}}$}}
\newcommand{\cloudspec}{\textsc{cloudspec}}
\newcommand{\synspec}{\textsc{synspec}}

\newcommand{\idl}{\textsc{idl}}

\usepackage{relsize}

\newcommand{\mup}{\hbox{$m_{\rm{up}}$}}
\newcommand{\Mcl}{\hbox{$M_{\rm{cl}}$}}
\newcommand{\Mgal}{\hbox{$M_{\rm{gal}}$}}
\newcommand{\tprime}{\hbox{$t^\prime$}}
\newcommand{\hii}{\hbox{H{\sc ii}}}
\newcommand{\hi}{\hbox{H{\sc i}}}
\newcommand{\zsun}{\hbox{${Z}_{\odot}$}}
\newcommand{\zpsun}{\hbox{${Z}_{\odot}^0$}}
\newcommand{\subISM}{\textnormal{\tiny \textsc{ism}}}
\newcommand{\zism}{\hbox{$Z_\subISM$}}

\newcommand{\nh}{\hbox{$n_{\mathrm{H}}$}}
\newcommand{\nhicm}{\hbox{$n_{\mathrm{H}}^{\mathrm{ICM}}$}}
\newcommand{\Nhicm}{\hbox{$N_{\mathrm{HI}}^{\mathrm{ICM}}$}}
\newcommand{\Us}{\hbox{$U_{\rm{S}}$}}
\newcommand{\xid}{\hbox{$\xi_{\rm{d}}$}}
\newcommand{\xidsol}{\hbox{$\xi_{\rm{d\odot}}$}}
\newcommand{\CO}{\hbox{C/O}}
\newcommand{\COsol}{\hbox{(C/O)$_\odot$}}
\newcommand{\NO}{\hbox{N/O}}
\newcommand{\NOsol}{\hbox{(N/O)$_\odot$}}
\newcommand{\OHgassol}{\hbox{(O/H)$_{\odot,\mathrm{gas}}$}}
\newcommand{\NOgassol}{\hbox{(N/O)$_{\odot,\mathrm{gas}}$}}
\newcommand{\COgassol}{\hbox{(C/O)$_{\odot,\mathrm{gas}}$}}

\newcommand{\lyb}{\hbox{H-Ly$\beta$}}
\newcommand{\ovid}{\hbox{O\,\textsc{vi}\,$\lambda\lambda1032,1038$}}

\newcommand{\ciia}{\hbox{C\,\textsc{ii}\,$\lambda\lambda1036,1037$}}
\newcommand{\nv}{\hbox{N\,\textsc{v}\,$\lambda1240$}}
\newcommand{\nvd}{\hbox{N\,\textsc{v}\,$\lambda\lambda1239,1243$}}
\newcommand{\lya}{\hbox{H-Ly\,$\alpha$}}
\newcommand{\siliia}{\hbox{Si\,\textsc{ii}\,$\lambda1260$}}
\newcommand{\siliiasta}{\hbox{Si\,\textsc{ii}$^\ast$\,$\lambda1265$}}
\newcommand{\oi}{\hbox{O\,\textsc{i}\,$\lambda1302$}}

\newcommand{\siliib}{\hbox{Si\,\textsc{ii}\,$\lambda1304$}}

\newcommand{\ciib}{\hbox{C\,\textsc{ii}\,$\lambda1335$}}
\newcommand{\ov}{\hbox{O\,\textsc{v}\,$\lambda1371$}}
\newcommand{\silivd}{\hbox{Si\,\textsc{iv}\,$\lambda\lambda1394,1403$}}
\newcommand{\silivt}{\hbox{Si\,\textsc{iv}\,$\lambda1400$}}
\newcommand{\siliic}{\hbox{Si\,\textsc{ii}\,$\lambda1527$}}
\newcommand{\siliidast}{\hbox{Si\,\textsc{ii}$^\ast$\,$\lambda1533$}}
\newcommand{\civd}{\hbox{C\,\textsc{iv}\,$\lambda\lambda1548,1551$}}
\newcommand{\civ}{\hbox{C\,\textsc{iv}\,$\lambda1550$}}
\newcommand{\feiia}{\hbox{Fe\,\textsc{ii}\,$\lambda1608$}}
\newcommand{\heii}{\hbox{He\,\textsc{ii}\,$\lambda1640$}}
\newcommand{\oiiid}{\hbox{O\,\textsc{iii}]$\lambda\lambda1661,1666$}}

\newcommand{\alii}{\hbox{Al\,\textsc{ii}\,$\lambda1671$}}
\newcommand{\niv}{\hbox{N\,\textsc{iv}\,$\lambda1719$}}
\newcommand{\feiilinea}{\hbox{Fe\,\textsc{ii}\,$\lambda1786$}}
\newcommand{\siliie}{\hbox{Si\,\textsc{ii}\,$\lambda1808$}}
\newcommand{\aliii}{\hbox{Al\,\textsc{iii}\,$\lambda\lambda1855,1863$}}
\newcommand{\ciiid}{\hbox{[C\,\textsc{iii}]$\lambda1907$+C\,\textsc{iii}]$\lambda1909$}}

\newcommand{\feiib}{\hbox{Fe\,\textsc{ii}\,$\lambda2344$}}
\newcommand{\feiie}{\hbox{Fe\,\textsc{ii}\,$\lambda2374$}}
\newcommand{\feiic}{\hbox{Fe\,\textsc{ii}\,$\lambda2383$}}
\newcommand{\feiibla}{\hbox{Fe\,\textsc{ii}\,$\lambda\lambda2389,2396$}}
\newcommand{\siliiid}{\hbox{[Si\,\textsc{iii}]$\lambda1883$+Si\,\textsc{iii}]$\lambda1892$}}

\newcommand{\oiiib}{\hbox{[O\,\textsc{iii}]$\lambda2321$}}
\newcommand{\oiib}{\hbox{[O\,\textsc{ii}]\,$\lambda2471$}}
\newcommand{\feiilineb}{\hbox{Fe\,\textsc{ii}\,$\lambda2400$}}

\newcommand{\feiid}{\hbox{Fe\,\textsc{ii}\,$\lambda2587$}}
\newcommand{\feiilinec}{\hbox{Fe\,\textsc{ii}\,$\lambda2600$}}
\newcommand{\mniiline}{\hbox{Mn\,\textsc{ii}\,$\lambda2606$}}
\newcommand{\feiiblb}{\hbox{Fe\,\textsc{ii}\,$\lambda2612$--$\lambda2632$}}
\newcommand{\mgiiline}{\hbox{Mg\,\textsc{ii}\,$\lambda\lambda2796,2804$}}
\newcommand{\heilinea}{\hbox{He\,\textsc{i}\,$\lambda2829$}}
\newcommand{\ciilinea}{\hbox{C\,\textsc{ii}\,$\lambda2838$}}
\newcommand{\mgiline}{\hbox{Mg\,\textsc{i}\,$\lambda2852$}}
\newcommand{\heilineb}{\hbox{He\,\textsc{i}\,$\lambda2945$}}
\newcommand{\feivlined}{\hbox{Fe\,\textsc{iv}\,$\lambda3096$}}
\newcommand{\heilinec}{\hbox{He\,\textsc{i}\,$\lambda3121$}}
\newcommand{\heilined}{\hbox{He\,\textsc{i}\,$\lambda3188$}}

\newcommand{\lhei}{\hbox{He\,\textsc{i}}}
\newcommand{\lsili}{\hbox{Si\,\textsc{i}}}
\newcommand{\lsilii}{\hbox{Si\,\textsc{ii}}}
\newcommand{\lsiliiast}{\hbox{Si\,\textsc{ii}$^\ast$}}
\newcommand{\lsiliii}{\hbox{Si\,\textsc{iii}}}
\newcommand{\lsiliv}{\hbox{Si\,\textsc{iv}}}
\newcommand{\loi}{\hbox{O\,\textsc{i}}}
\newcommand{\loiii}{\hbox{O\,\textsc{iii}]}}

\newcommand{\lni}{\hbox{N\,\textsc{i}}}
\newcommand{\lniv}{\hbox{N\,\textsc{iv}}}
\newcommand{\lfei}{\hbox{Fe\,\textsc{i}}}
\newcommand{\lfeii}{\hbox{Fe\,\textsc{ii}}}
\newcommand{\lfeiii}{\hbox{Fe\,\textsc{iii}}}
\newcommand{\lfeiv}{\hbox{Fe\,\textsc{iv}}}
\newcommand{\lfev}{\hbox{Fe\,\textsc{v}}}
\newcommand{\lci}{\hbox{C\,\textsc{ i}}}
\newcommand{\lciast}{\hbox{C\,\textsc{i}$^\ast$}}
\newcommand{\lcii}{\hbox{C\,\textsc{ii}}}
\newcommand{\lciii}{\hbox{C\,\textsc{iii}}}
\newcommand{\lciv}{\hbox{C\,\textsc{iv}}}
\newcommand{\lali}{\hbox{Al\,\textsc{i}}}
\newcommand{\lalii}{\hbox{Al\,\textsc{ii}}}
\newcommand{\laliii}{\hbox{Al\,\textsc{iii}}}
\newcommand{\lcoi}{\hbox{Co\,\textsc{i}}}
\newcommand{\lcoii}{\hbox{Co\,\textsc{ii}}}
\newcommand{\lcri}{\hbox{Cr\,\textsc{i}}}
\newcommand{\lcrii}{\hbox{Cr\,\textsc{ii}}}
\newcommand{\lnii}{\hbox{Ni\,\textsc{i}}}
\newcommand{\lniii}{\hbox{Ni\,\textsc{ii}}}
\newcommand{\lmgi}{\hbox{Mg\,\textsc{i}}}
\newcommand{\lmgii}{\hbox{Mg\,\textsc{ii}}}
\newcommand{\lmni}{\hbox{Mn\,\textsc{i}}}
\newcommand{\lmnii}{\hbox{Mn\,\textsc{ii}}}
\newcommand{\lznii}{\hbox{Zn\,\textsc{ii}}}

\newcommand{\bl}{\hbox{Bl\;}}
\newcommand{\siliv}{\hbox{Si\,\textsc{iv}\;1397}}
\newcommand{\fe}{\hbox{Fe\;1453}}
\newcommand{\civa}{\hbox{C\,\textsc{iv}$^\mathrm{a}$}}
\newcommand{\civc}{\hbox{C\,\textsc{iv}$^\mathrm{c}$}}
\newcommand{\cive}{\hbox{C\,\textsc{iv}$^\mathrm{e}$}}
\newcommand{\feii}{\hbox{Fe\,\textsc{ii}\;}}
\newcommand{\mgii}{\hbox{Mg\,\textsc{ii}\;2800}}
\newcommand{\mgi}{\hbox{Mg\,\textsc{i}\;2852}}
\newcommand{\mgw}{\hbox{Mg\,wide}}
\newcommand{\fei}{\hbox{Fe\,\textsc{i}\;3000}}

\title[Ultraviolet-line diagnostics of stars and the ISM]{Modelling ultraviolet-line diagnostics of stars, the ionized and the neutral interstellar medium in star-forming galaxies}
\author[A. Vidal-Garc\'ia et al.]
  {A. Vidal-Garc\'ia$^{1}$\thanks{E-mail: vidal@iap.fr}, S. Charlot$^{1}$, G. Bruzual$^{2}$ and I. Hubeny$^{3}$
\\
\\
$^{1}$Sorbonne Universit\'es, UPMC-CNRS, UMR7095, Institut d'Astrophysique de Paris, F-75014, Paris, France\\
$^{2}$Instituto de Radioastronom{\'i}a y Astrof{\'i}sica, UNAM, Campus Morelia, Michoacan, M{\'e}xico, C.P. 58089, M{\'e}xico\\
$^{3}$Department of Astronomy and Steward Observatory, University of Arizona, Tucson, AZ 85721, USA
}

\date{Accepted XXX. Received YYY; in original form ZZZ}

\pubyear{2016}

\begin{document}
\label{firstpage}
\pagerange{\pageref{firstpage}--\pageref{lastpage}}
\maketitle

\begin{abstract}
We combine state-of-the-art models for the production of stellar radiation and its transfer through the interstellar medium (ISM) to investigate ultraviolet-line diagnostics of stars, the ionized and the neutral ISM in star-forming galaxies. We start by assessing the reliability of our stellar population synthesis modelling by fitting absorption-line indices in the ISM-free ultraviolet spectra of 10 Large-Magellanic-Cloud clusters. In doing so, we find that neglecting  stochastic sampling of the stellar initial mass function in these young ($\sim10$--100\,Myr), low-mass clusters affects negligibly ultraviolet-based age and metallicity estimates but can lead to significant overestimates of stellar mass. Then, we proceed and develop a simple approach, based on an idealized description of the main features of the ISM, to compute in a physically consistent way the combined influence of nebular emission and interstellar absorption on ultraviolet spectra of star-forming galaxies. Our model accounts for the transfer of radiation through the ionized interiors and outer neutral envelopes of short-lived stellar birth clouds, as well as for radiative transfer through a diffuse intercloud medium. We use this approach to explore the entangled signatures of stars, the ionized and the neutral ISM in ultraviolet spectra of star-forming galaxies. We find that, aside from a few notable exceptions, most standard ultraviolet indices defined in the spectra of ISM-free stellar populations are prone to significant contamination by the ISM, which increases with metallicity. We also identify several nebular-emission and interstellar-absorption features, which stand out as particularly clean tracers of the different phases of the ISM.
\end{abstract}


\section{Introduction}\label{sec:intro}

The ultraviolet spectral energy distributions of star-forming galaxies exhibit numerous spectral signatures of stars and gas in the ionized and neutral interstellar medium (ISM). The ability to interpret these features in the spectra of star-forming galaxies has received increasing interest with the possibility, offered by new-generation ground-based telescope and the future {\it James Webb Space Telescope} ({\it JWST}), to sample the rest-frame ultraviolet emission from large samples of young galaxies out to the epoch of cosmic reionization.   

Several major observational studies have focused on the interpretation of ultraviolet galaxy spectra in terms of constraints on stellar and interstellar properties. Most studies of low-redshift galaxies are based on data from the {\it International Ultraviolet Explorer} ({\it IUE}), the {\it Hubble Space Telescope} ({\it HST}), the {\it Hopkins Ultraviolet Telescope} ({\it HUT}) and the {\it Far Ultraviolet Spectroscopic Explorer} ({\it FUSE}), while those of high-redshift galaxies rely on deep ground-based infrared spectroscopy.  Ultraviolet absorption lines from stellar photospheres and winds often used to characterise young stellar populations, chemical enrichment and stellar initial mass function (IMF) in galaxies include the prominent \silivd\ and \civd\ lines, but also \lyb, \ovid, \ciia\ and \nvd\  \citep[e.g.,][]{GonzaDel1998,Pettini2000,Mehlert2002,Mehlert2006,Halliday2008}. \cite{Fanelli1992} define around 20 absorption-line indices at wavelengths in the range $1200\lesssim\lambda\lesssim3200\,\AA$ in the {\it IUE} spectra of $218$ nearby stars. \cite{Leitherer2011} propose a complementary set of 12 indices over a similar wavelength range based on {\it HST} observations of 28 local starburst and star-forming galaxies, which also includes interstellar absorption lines. In fact, interstellar absorption lines are routinely used to characterise the chemical composition and large-scale kinematics of the ionized and neutral ISM in star-forming galaxies, among which \siliia, \oi, \siliib, \ciib, \silivd, \siliic, and \civd, but also \feiia,  \alii, \feiib\ and  \feiic\ \citep[e.g.,][]{Pettini2000,Pettini2002,Sembach2000,Shapley2003,Steidel2010,Lebouteiller2013,James2014}. The physical conditions in the ionized gas can also be traced using the luminosity of emission lines, such as \lya, \civd, \heii, \oiiid, \siliiid\ and \ciiid\ \citep[e.g.,][]{Shapley2003,Erb2010,Christensen2012,Stark2014,Stark2015a,Sobral2015}. So far, however, combined constraints on the neutral and ionized components of the ISM derived from the ultraviolet emission of a galaxy have generally been performed using independent (and hence potentially inconsistent) estimates of the chemical compositions of both components \citep[e.g.,][]{Thuan2002,Thuan2005,Aloisi2003,Lebouteiller2004,Lebouteiller2009,Lee2004,Lebouteiller2013}.

On the theoretical front, the ultraviolet spectral modelling of galaxies has been the scene of important progress over the past two decades. Early spectral synthesis models relied on libraries of observed {\it IUE} and {\it HST} spectra of O and B stars in the Milky Way, the Small and the Large Magellanic Clouds \citep[LMC; e.g.,][]{Leitherer1999,Leitherer2001}. The restricted range in stellar (in particular, metallicity and wind) parameters probed by such spectra triggered the development of increasingly sophisticated libraries of synthetic spectra \citep[e.g.,][see section~2 of \citealt{Leitherer2010} for a description of the different types of model atmosphere calculations]{Hauschildt1999,Hillier1999,Pauldrach2001,Lanz2003,Lanz2007,Hamann2004,Martins2005,Puls2005,RodrigMerin2005}. These theoretical libraries have enabled the exploration of the dependence of selected ultraviolet spectral indices on stellar effective temperature, gravity, metallicity and, for massive stars, wind parameters (controlling P-Cygni line profiles). \citet{Chavez2007,Chavez2009} identify 17 mid-ultraviolet spectral diagnostics in the wavelength range $1900\lesssim\lambda\lesssim3200\,\AA$, which they exploit to constrain the ages and metallicities of evolved Galactic globular clusters. Other studies have focused on the exploration of a small number of stellar metallicity diagnostics of star-forming galaxies at far-ultraviolet wavelengths, over the range $900\lesssim\lambda\lesssim2100\,$\AA\ \citep[][see also \citealt{Faisst2015}]{Rix2004,Sommariva2012}. A challenge in the application of these metallicity diagnostics to the interpretation of observed spectra is the required careful determination of the far-ultraviolet continuum in a region riddled with absorption features. This difficulty is avoided when appealing to indices defined by two `pseudo-continuum' bandpasses flanking a central bandpass \citep[e.g.,][]{Fanelli1992,Chavez2007}. \citet{Maraston2009} compare in this way the strengths of the \citet{Fanelli1992} indices predicted by their models for young stellar populations (based on the stellar spectral library of \citealt{RodrigMerin2005}) with those measured in the {\it IUE} spectra of 10 LMC star clusters observed by \citet{Cassatella1987}. \citet{Maraston2009} however do not account for the fact that these clusters have small stellar masses, between about $3\times10^3$ and $7\times10^4\Msun$ \citep{Mackey2003,McLaughlin2005}, and hence, that stochastic IMF sampling can severely affect integrated spectral properties \citep[e.g.,][]{Fouesneau2012}. A more general limitation of the use of ultraviolet spectral diagnostics to interpret observed galaxy spectra is that no systematic, quantitative estimate of the contamination of these diagnostics by interstellar emission and absorption has been performed yet.

In this paper, we use a combination of state-of-the-art models for the production of radiation by young stellar populations and its transfer through the ionized and the neutral ISM to investigate diagnostics of these three components in ultraviolet spectra of star-forming galaxies. We start by using the latest version of the \citet{Bruzual2003} stellar population synthesis code (Charlot \& Bruzual, in preparation; which incorporates the ultraviolet spectral libraries of \citealt{Lanz2003,Lanz2007,Hamann2004,RodrigMerin2005,Leitherer2010}) to investigate the dependence of the \citet{Fanelli1992} ultraviolet spectral indices on age, metallicity and integrated stellar mass, for simple (i.e. instantaneous-burst) stellar populations (SSPs). We demonstrate the ability of the models to reproduce observations and assess the extent to which accounting for stochastic IMF sampling can change the constraints on age, metallicity and stellar mass derived from the {\it IUE} spectra of the \citet{Cassatella1987} clusters. On these grounds, we develop a simple approach to compute in a physically consistent way the combined influence of nebular emission and interstellar absorption on the ultraviolet spectra of star-forming galaxies. We achieve this through an idealized description of the main features of the ISM (inspired from the dust model of \citealt{CharlotFall2000}) and by appealing to a combination of the photoionization code \cloudy\ (version 13.3; \citealt{Ferland2013}) with the spectral synthesis code \synspec\footnote{\url{http://nova.astro.umd.edu/Synspec49/synspec.html}} \citep[e.g.,][]{Hubeny2011}, which allows the computation of interstellar-line strengths based on the ionization structure solved by \cloudy\ (in practice, this combination is performed via the program \cloudspec\ of \citealt{Hubeny2000}; see also \citealt{Heap2001}). We use this approach to investigate the ultraviolet spectral features individually most sensitive to young stars, the ionized and the neutral ISM. We find that, aside from a few notable exceptions, most standard ultraviolet indices defined in the spectra of ISM-free stellar populations can potentially suffer from significant contamination by the ISM, which increases with metallicity. We also identify several nebular-emission and interstellar-absorption features, which stand out as particularly clean tracers of the different phases of the ISM. Beyond an a posteriori justification of the main spectral diagnostics useful to characterise young stars, the ionized and the neutral ISM, the models presented in this paper provide a means of simulating and interpreting in a versatile and physically consistent way the entangled contributions by these three components to the ultraviolet emission from star-forming galaxies. 

We present the model we adopt to compute the ultraviolet emission from young stellar populations in Section~\ref{sec:stelpops}, where we investigate the dependence of the \citet{Fanelli1992} ultraviolet spectral indices on age, metallicity and integrated stellar mass for SSPs. In Section~\ref{sec:uvinterpret}, we use this model to interpret the {\it IUE} spectra of the 10 LMC star clusters observed by \citet{Cassatella1987} and assess potential biases in age, metallicity and stellar-mass estimates introduced through the neglect of stochastic IMF sampling. We present our approach to model the influence of nebular emission and interstellar absorption on the ultraviolet spectra of star-forming galaxies in Section~\ref{influenceism}. We analyse in detail two representative models of star-forming galaxies: a young, metal-poor galaxy and a more mature, metal-rich galaxy. We use these models to identify ultraviolet spectral indices individually most sensitive to stars, the ionized and the neutral ISM. Our conclusions are summarised in Section~\ref{conclu}.

\section{Ultraviolet signatures of young stellar populations}\label{sec:stelpops}

In this section, we start by describing the main features of the stellar population synthesis code we adopt to compute ultraviolet spectral signatures of young stellar populations (section~\ref{sec:popsyn}). Then, we briefly review the main properties of the \citet{Fanelli1992} ultraviolet spectral indices (Section~\ref{sec:uvindices}). We examine the dependence of index strengths on age, metallicity and integrated stellar mass for simple stellar populations (Section~\ref{sec:modidx}), along with the dependence of ultraviolet, optical and near-infrared broadband magnitudes on integrated stellar mass (Section~\ref{sec:modmag}).

\subsection{Stellar population synthesis modelling}\label{sec:popsyn}

We adopt the latest version of the \citet{Bruzual2003} stellar population synthesis code (Charlot \& Bruzual, in preparation; see also \citealt{Wofford2016}) to compute emission from stellar populations of ages between $10^4$\,yr and 13.8\,Gyr at wavelengths between 5.6\,\AA\ and 3.6\,cm, for metallicities in the range $0.0001 \leq Z \leq 0.04$ (assuming scaled-solar heavy-element abundance ratios at all metallicities). This version of the code incorporates updated stellar evolutionary tracks computed with the PARSEC code of \citet{Bressan2012} for stars with initial masses up to 350\,\Msun\ \citep{Chen2015}, as well as the recent prescription by \citet{Marigo2013} for the evolution of thermally pulsing asymptotic-giant-branch (AGB) stars. The present-day solar metallicity in these calculations is taken to be $Z_\odot=0.01524$ (the zero-age main sequence solar metallicity being $Z^{0}_{\odot}= 0.01774$; see \citealt{Bressan2012}). We note that the inclusion of very low-metallicity, massive stars is important to investigate the properties of primordial stellar populations \citep{Bromm2011}. 

These evolutionary tracks are combined with various stellar spectral libraries to describe the properties of stars of different effective temperatures, luminosities, surface gravities, metallicities and mass-loss rates in the Hertzsprung-Russell diagram. For the most part (see adjustments below), the spectra of O stars hotter than 27,500\,K and B stars hotter than 15,000\,K are taken from the TLUSTY grid of metal line-blanketed, non-local thermodynamic equilibrium (non-LTE), plane-parallel, hydrostatic models of \citet[][see also \citealt{Lanz2003,Lanz2007}]{Hubeny1995}. The spectra of cooler stars are taken from the library of line-blanketed, LTE,  plane-parallel, hydrostatic models of \citet{Martins2005}, extended at wavelengths shorter than 3000\,\AA\ using similar models from the UVBLUE library of \citet{RodrigMerin2005}. At wavelengths in the range $3525\lesssim\lambda\lesssim7500\,\AA$, the spectra of stars with effective temperatures in the range $2800\lesssim T_{\rm eff}\lesssim 36,000\,$K are taken from the observational MILES library of \citet{Sanchez2006}. At wavelengths in the range $900\lesssim\lambda\lesssim3000\,\AA$, the spectra of main-sequence stars with effective temperatures in the range $17,000\lesssim T_{\rm eff}\lesssim 60,000\,$K are taken from the theoretical library of \citet{Leitherer2010}, computed using the WM-basic code of \citet{Pauldrach2001} for line-blanketed, non-LTE, spherically extended models including radiation driven winds. Finally, for Wolf-Rayet stars, the spectra are taken from the (high-resolution version of the) PoWR library of line-blanketed, non-LTE, spherically expanding models of \citet[][see also \citealt{Graefener2002, Hamann2003,Hamann2006,Sander2012,Hainich2014}]{Hamann2004}. The inclusion of the \citet{Leitherer2010} and \citet{Hamann2004} spectral libraries enables the modelling of P-Cygni line profiles originating from winds of massive OB and Wolf-Rayet stars in the integrated ultraviolet spectra of young stellar populations (for example, for the \nv, \silivt\ and \civ\ lines; e.g. \citealt{Walborn1984}). For completeness, the spectra of the much fainter, hot post-AGB stars are taken from the library of line-blanketed, non-LTE, plane-parallel, hydrostatic models of \citet{Rauch2002}. In Appendix~\ref{app:comparison}, we show how the predictions of this model in the age and metallicity ranges relevant to the present study compare with those of the original \citet{Bruzual2003} stellar population synthesis code.

\subsection{Ultraviolet spectral indices}\label{sec:uvindices}

To investigate the ultraviolet properties of young stellar populations, we appeal to the set of spectral indices originally defined by \citet{Fanelli1992} in the {\it IUE} spectra of 218 nearby stars spanning spectral type from O through K and iron abundances (available for only 94 stars) in the range $-1.12\lesssim\mathrm{[Fe/H]}\lesssim0.56$. We focus on the 19 absorption-line indices defined by means of a central bandpass flanked by two pseudo-continuum bandpasses.\footnote{The use of a pseudo-continuum in the definition of these indices is dictated by the $\sim 6 \AA$ resolution of {\it IUE} spectra \citep{Boggess1978}, which does not enable reliable measurements of the true continuum.} This includes 11 far-ultraviolet indices, with central wavelengths in the range $1300\lesssim\lambda\lesssim1850\,$\AA, and 8 mid-ultraviolet indices, with central wavelengths in the range $2400\lesssim\lambda\lesssim3100\,$\AA. Following \cite{Trager1998}, we compute the equivalent width of an absorption-line index in a spectral energy distribution defined by a flux per unit wavelength $F_{\lambda}$ as 
\begin{equation}
EW=\int_{\lambda_{1}}^{\lambda_{2}} d\lambda\,\left( 1- \frac{F_{\lambda}}{F_{c,\lambda}}\right)\,,
\end{equation}
where $\lambda_{1}$ and $\lambda_{2}$ are the wavelength limits of the central feature bandpass and $F_{c,\lambda}$ is the pseudo-continuum flux per unit wavelength, defined through a linear interpolation between the average fluxes in the blue and red flanking bandpasses.

\begin{table*}
\caption{Definition of 19 ultraviolet spectral indices in terms of a central bandpass flanked by two pseudo-continuum bandpasses \citep[from tables~4A and 4B of][]{Fanelli1992}. The rightmost column indicates the main atomic species responsible for each feature \citep{Fanelli1992,Chavez2007,Leitherer2011}.}
\begin{threeparttable}
\centering
\footnotesize\setlength{\tabcolsep}{7pt}
\begin{tabular*}{0.85\textwidth}{l c c c r l}
\toprule
Name\tnote{a} & Blue bandpass & Central bandpass & Red bandpass & $\Delta\mathrm{(EW/\AA)}$\tnote{b} & Features\tnote{c} \\
\midrule
\bl1302 & 1270--1290 & 1292--1312 & 1345--1365 & $-0.87$ \ \ \ \  & \lsilii, \lsiliii, \lciii, \loi\ \\
\siliv\  & 1345--1365 & 1387--1407 & 1475--1495 & $-0.47$ \ \ \ \  & \lsiliv\  \\
\bl1425 & 1345--1365 & 1415--1435 & 1475--1495 & $0.00$ \ \ \ \  & \lsiliii, \lfev, \lcii, \lciii\ \\
\fe\ & 1345--1365 & 1440--1466 & 1475--1495 & $0.00$ \ \ \ \  & \lniii, \lcoii\ \\
\civa\  & 1500--1520 & 1530--1550 & 1577--1597 & $-0.41$ \ \ \ \  & \lsiliiast, \lciv\ \\
\civc\ & 1500--1520 & 1540--1560 & 1577--1597 & $-0.65$ \ \ \ \  & \lciv\ \\
\cive\ & 1500--1520 & 1550--1570 & 1577--1597 & $-0.23$ \ \ \ \  & \lciv\  \\
\bl1617 & 1577--1597 & 1604--1630 & 1685--1705 & $0.00$ \ \ \ \  & \lfeii, \lciii\ \\
\bl1664 & 1577--1597 & 1651--1677 & 1685--1705 & $-0.64$ \ \ \ \  & \lci, \lciast, \loiii, \lfev, \lalii\ \\
\bl1719 & 1685--1705 & 1709--1729 & 1803--1823 & $0.06$ \ \ \ \  & \lniii, \lfeiv, \lniv, \lsiliv\ \\
\bl1853 & 1803--1823 & 1838--1868 & 1885--1915 & $-0.11$ \ \ \ \  & \lsili, \lalii, \laliii\ \\
\feii2402 & 2285--2325 & 2382--2422 & 2432--2458 & $-0.87$ \ \ \ \  & \lfei, \lfeii, \lcoi\ \\
\bl2538 & 2432--2458 & 2520--2556 & 2562--2588 & $0.55$ \ \ \ \  & \lfei, \lfeii, \lmgi, \lcri, \lni\ \\
\feii2609 & 2562--2588 & 2596--2622 & 2647--2673 & $-0.56$ \ \ \ \  & \lfei, \lfeii, \lmnii\ \\
\mgii\ & 2762--2782 & 2784--2814 & 2818--2838 & $-2.24$ \ \ \ \  & \lmgii, \lfei, \lmni\ \\
\mgi\ & 2818--2838 & 2839--2865 & 2906--2936 & $-0.48$ \ \ \ \  & \lmgi, \lfei, \lfeii, \lcrii\ \\
\mgw\ & 2470--2670 & 2670--2870 & 2930--3130 & $-2.72$ \ \ \ \  & \lmgi, \lmgii, \lfei, \lfeii, \lmni, \lcri, \lcrii\ \\
\fei\ & 2906--2936 & 2965--3025 & 3031--3051 & $0.00$ \ \ \ \  & \lfei, \lfeii, \lcri, \lnii\ \\
\bl3096 & 3031--3051 & 3086--3106 & 3115--3155 & $0.00$ \ \ \ \  & \lfei, \lali, \lnii, \lmgi\ \\
\bottomrule
\end{tabular*}
\begin{tablenotes}
\item [a] Indices originally qualified as blends of several species are labelled with `Bl'.\\
\item [b] Term to be added to the measured index strength to correct a posteriori for contamination by Milky-Way ISM absorption lines (Section~\ref{sec:sample} and Appendix~\ref{app:mwcorr}). A zero entry indicates that no  strong ISM absorption line affects the bandpasses defining this index.\\
\item [c] An asterisk indicates a fine-structure transition.
\end{tablenotes} 
\end{threeparttable}
\label{tab:tabindices}
\end{table*}

For reference, we list in Table ~\ref{tab:tabindices} the definitions by \citet{Fanelli1992} of the 19 ultraviolet absorption-line indices used in this work, along with an indication of the atomic species thought to contribute most to each feature \citep[according to, in particular,][]{Fanelli1992,Chavez2007,Maraston2009,Leitherer2011}. It is worth recalling that the three index definitions for the \lciv\ line, which exhibits a strong P-Cygni profile in O-type stars \citep{Walborn1984}, are centred on absorption feature (\civa), the central wavelength (\civc) and emission feature (\cive) of that line. We refer the reader to the original study of \citet[][see also \citealt{Chavez2007}]{Fanelli1992} for a more detailed description of the properties of these indices and of the dependence of their strengths on stellar spectral type and luminosity class. In general, far-ultraviolet indices tend to be dominated by hot O- and B-type stars, while mid-ultraviolet indices tend to be stronger in cooler, A- to K-type stars.

\subsection{Dependence of index strength on age, metallicity and integrated stellar mass}\label{sec:modidx}

\begin{figure*}
\begin{center}
\resizebox{\hsize}{!}{\includegraphics{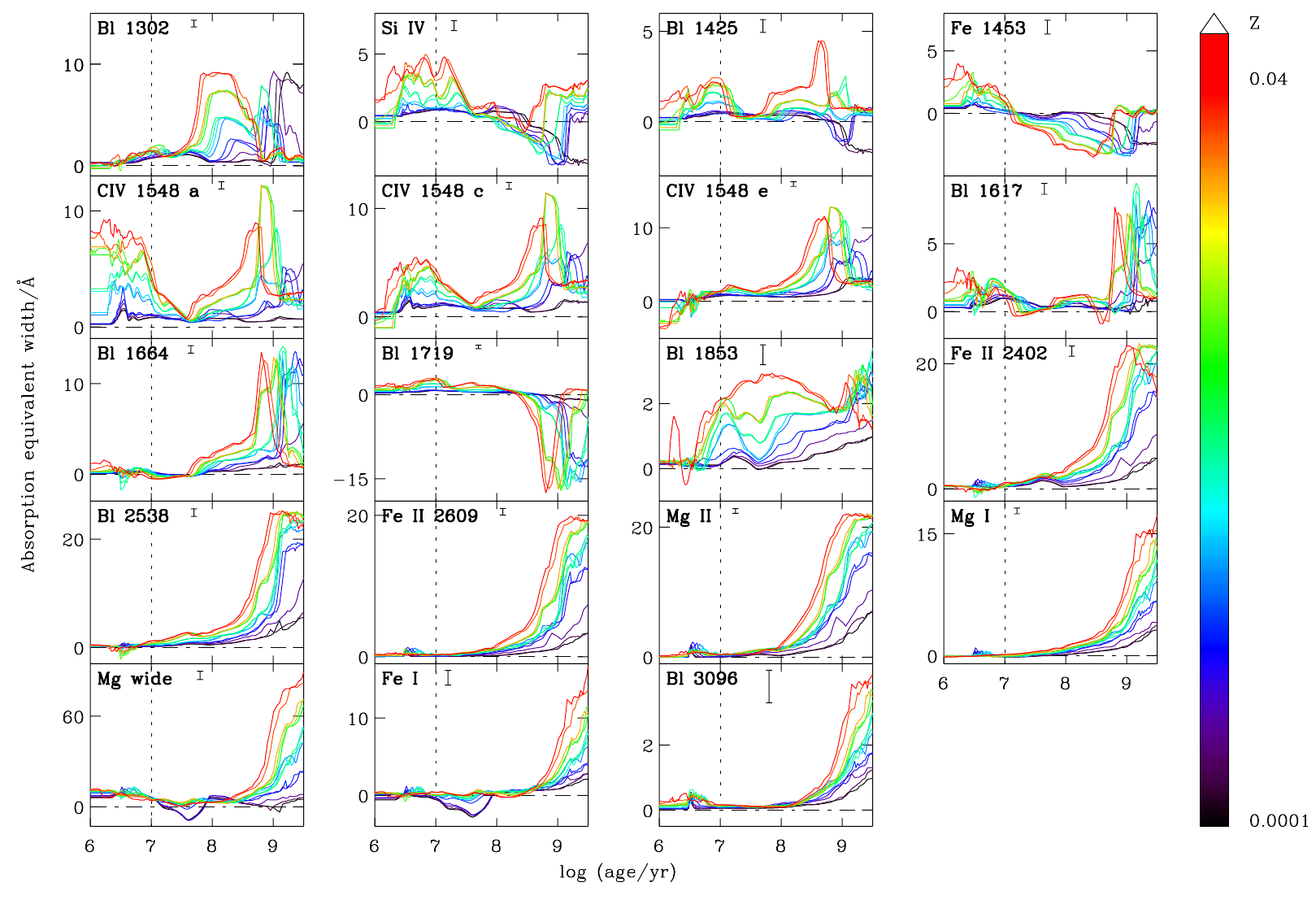}}
\end{center}
\caption{Strengths of the 19 \citet{Fanelli1992} ultraviolet spectral indices defined in Table~\ref{tab:tabindices} plotted against age, for SSPs with a smoothly sampled \citet{Chabrier2003} IMF and metallicities $Z=0.0001$, 0.0002, 0.0005, 0.001, 0.002, 0.004, 0.006, 0.008, 0.010, 0.014, 0.017, 0.02, 0.03 and 0.04, colour-coded as indicated on the right-hand scale (for clarity, the ordinate scale in each panel has been adjusted to reflect the dynamic range spanned by the strength of the corresponding index). Also shown next to the index name in each panel is the typical measurement error in the {\it IUE} spectra of the LMC clusters observed by \citet{Cassatella1987}. In each panel, the dotted vertical line marks the age of 10\,Myr, before which nebular emission can strongly affect index strengths (see text).}
\label{fig:indmet}
\end{figure*}

\citet{Chavez2007} and \citet{Maraston2009} have investigated the dependence of the \citet{Fanelli1992} index strengths on stellar effective temperature, gravity and metallicity, as well as, for integrated SSP spectra, stellar population age and metallicity. It is useful to start by examining the predictions of the new stellar population synthesis models described in Section~\ref{sec:stelpops} for the dependence of index strength on SSP age and metallicity. Fig.~\ref{fig:indmet} shows the time evolution of the strengths of the 19 ultraviolet spectral indices in Table~\ref{tab:tabindices}, for SSPs with 14 different metallicities in the range $ 0.0001 \leqslant Z \leqslant 0.04$ and a smoothly sampled \citet{Chabrier2003} IMF. Also indicated in each panel is the typical measurement error in the corresponding index strength in the {\it IUE} spectra of the LMC star clusters observed by \citet[][see Section~\ref{sec:sample} below]{Cassatella1987}. Since star clusters generally form in dense molecular clouds, which dissipate on a timescale of about 10\,Myr \citep[e.g.][]{Murray2010,Murray2011}, stellar absorption lines of SSPs younger than this limit are expected to be strongly contaminated by nebular emission from the H\,\textsc{ii} regions carved by newly-born stars within the clouds (Section~\ref{sec:ismmodels}). We therefore focus for the moment on the evolution of ultraviolet index strengths at ages greater than 10\,Myr (i.e. to the right of the dotted vertical line) in Fig.~\ref{fig:indmet}. 

The results of Fig.~\ref{fig:indmet} confirm those obtained using previous models by \citet{Chavez2007} and \citet{Maraston2009}, in that the strengths of most absorption-line indices tend to increase with metallicity -- because of the stronger abundance of absorbing metals --  with a few exceptions in some age ranges. The most notable exception is \fe, for which an inversion in the dependence of index strength on metallicity occurs at an age around $10\,$Myr. This is associated with a transition from positive to negative equivalent widths, induced by the development of a strong absorption blend affecting the red pseudo-continuum bandpass of this index. A similar absorption of the pseudo-continuum flux is also responsible for the negative equivalent widths of, e.g., \siliv, \bl1425 and \bl1719 at ages around 1\,Gyr. A main feature of Fig.~\ref{fig:indmet} is that, overall, the age range during which the dependence of an index strength on metallicity is the strongest tends to increase with wavelength, from ages between roughly $3\times10^7$ and $3\times10^8\,$yr for \bl1302 to ages greater than about 1\,Gyr for \bl3096. This is because, as the stellar population ages, stars of progressively lower effective temperature dominate the integrated ultraviolet emission, implying a shift in strong absorption features from the far to the mid ultraviolet (Section~\ref{sec:uvindices}). It is also important to note that, at ages greater than about 1\,Gyr, the ultraviolet emission from an SSP is dominated by hot post-AGB stars, whose contribution is orders of magnitude fainter than that of massive stars at young ages \citep[see, e.g., fig.~9 of][]{Bruzual2003}. 

\begin{figure*}
\begin{center}
\resizebox{\hsize}{!}{\includegraphics{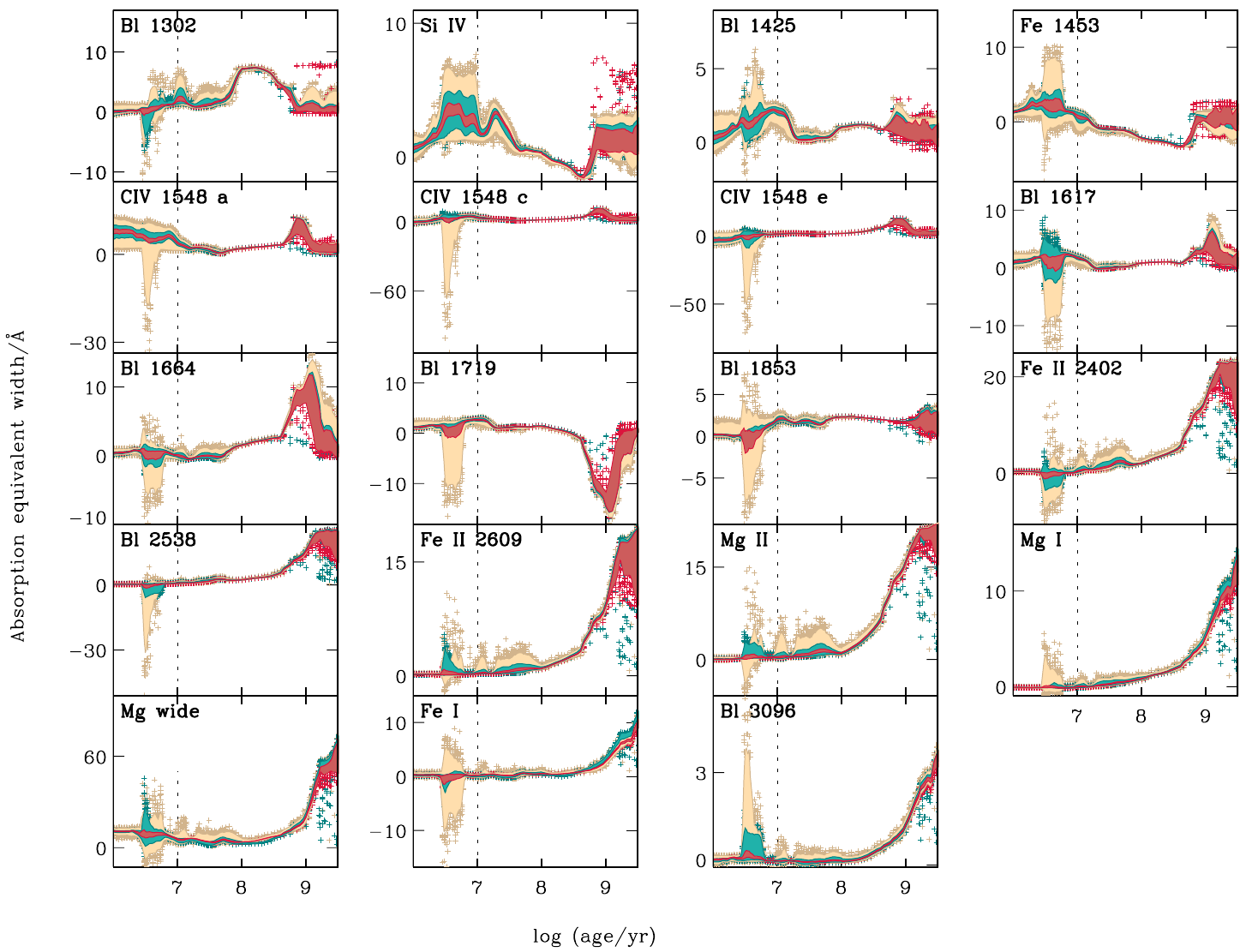}}
\end{center}
\caption{Strengths of the 19 \citet{Fanelli1992} ultraviolet spectral indices defined in Table~\ref{tab:tabindices} plotted against age, for SSPs with a stochastically sampled \citet{Chabrier2003} IMF and integrated stellar masses $\Mcl=10^3\Msun$ (cream), $10^4\Msun$ (teal blue) and $10^5\Msun$ (indian red), for the metallicity $Z=0.017$ (for clarity, the ordinate scale in each panel has been adjusted to reflect the dynamic range spanned by the strength of the corresponding index). In each panel, the filled area shows the range spanned by 99 per cent of 1100 SSP realisations with these three target stellar masses at each age, while crosses show the remaining 1 per cent of most distant outliers. The dotted vertical line has the same meaning as in Fig.~\ref{fig:indmet}.}
\label{fig:ind017}
\end{figure*}

As mentioned in Section~\ref{sec:intro}, an important issue when appealing to SSP models to interpret observations of individual star clusters is that, for integrated stellar masses less than about $10^5\,$\Msun, stochastic IMF sampling can severely affect the integrated spectral properties of the clusters \citep[e.g.,][]{Bruzual2003,Fouesneau2012}. We investigate this by computing the ultraviolet spectral properties of stellar populations with different integrated stellar masses, using an approach similar to that described by \citet[][inspired from \citealt{Santos1997}]{Bruzual2002}. Given an age and a metallicity, this consists in drawing stars randomly from the IMF until a chosen integrated SSP mass is reached.\footnote{In building up the integrated stellar mass of an SSP at a given age, we account for the loss of mass returned to the ISM by evolved stars in the form of winds and supernova explosions (this was not considered by \citealt{Santos1997} and \citealt{Bruzual2002}).} Fig.~\ref{fig:ind017} shows the dispersion in the strengths of the 19 \citet{Fanelli1992} indices for SSPs with integrated stellar masses $\Mcl=10^3\Msun$ (cream), $10^4\Msun$ (teal blue) and $10^5\Msun$ (indian red) and a stochastically sampled \citet{Chabrier2003} IMF, at ages between 1\,Myr and 3\,Gyr, for a fixed metallicity $Z=0.017$. This was obtained by performing, at each age, 1100 realisations of SSPs for each of these three target stellar masses. We note that, because of the large dispersion in some age ranges, the ordinate scale in a given panel in Fig.~\ref{fig:ind017} generally spans a wider dynamic range than in the corresponding panel in Fig.~\ref{fig:indmet}.

Fig.~\ref{fig:ind017} shows some important results. Firstly, as expected, the dispersion in index strength generally tends to increase from large to small SSP mass \Mcl, as the presence of a single massive bright star can affect more strongly the integrated emission from a low-mass than a high-mass stellar population.  A most remarkable result from Fig.~\ref{fig:ind017} is that, while the dispersion in index strength can be very large at ages below $\sim$10\,Myr and above $\sim$500\,Myr, it is generally more moderate at ages in between. This is because, at the youngest ages, the integrated ultraviolet light can be strongly influenced by the occasional presence of rare, very massive, hot, short-lived stars, while at ages greater than a few hundred million years, it can be strongly influenced by the emergence of more numerous, but extremely short-lived, hot post-AGB stars.\footnote{The evolution of post-AGB stars through the bright, hot phase giving rise to significant ultraviolet emission lasts between a few $\times10^3$ and several $\times10^4\,$yr, depending on star mass and metallicity \citep{VassiliadisWood1994}. This is so fast that, when drawing `only' 1100 models at each age,  the most massive, hottest PAGB stars are hardly sampled at ages around 1\,Gyr in the $\Mcl=10^5\Msun$ model and not sampled at all in the $10^4\Msun$ and $10^3\Msun$ models. This explains why, for example, only a few indian-red models (and no teal-blue nor cream model) exhibit large strengths of far-ultraviolet indices, such as \bl1302 and  \siliv, around $10^9\,$yr  in Fig.~\ref{fig:ind017}. As an experiment, we have checked that, when drawing $5\times10^5$ models at the age of $10^9\,$yr, 401 (243) indian-red, 59 (8) teal-blue and 15 (0) cream crosses reach the largest \siliv\ (\bl1302) index strengths.} At ages in the range from $\sim$10\,Myr to $\sim$500\,Myr, therefore, the impact of stochastic IMF sampling on age and metallicity estimates of star clusters from ultraviolet spectroscopy should be moderate. We find that this is even more true at metallicities lower than that of $Z=0.017$ adopted in Fig.~\ref{fig:ind017} (not shown), since, as illustrated by Fig.~\ref{fig:indmet}, index strengths tend to rise with metallicity. As we shall see in Section~\ref{sec:paramest}, despite this result, stochastic IMF sampling can affect stellar-mass estimates of star clusters more significantly than age and metallicity estimates at ages between $\sim$10\,Myr to $\sim$500\,Myr.

\subsection{Associated broadband magnitudes}\label{sec:modmag}

\begin{figure}\includegraphics[width=\columnwidth]{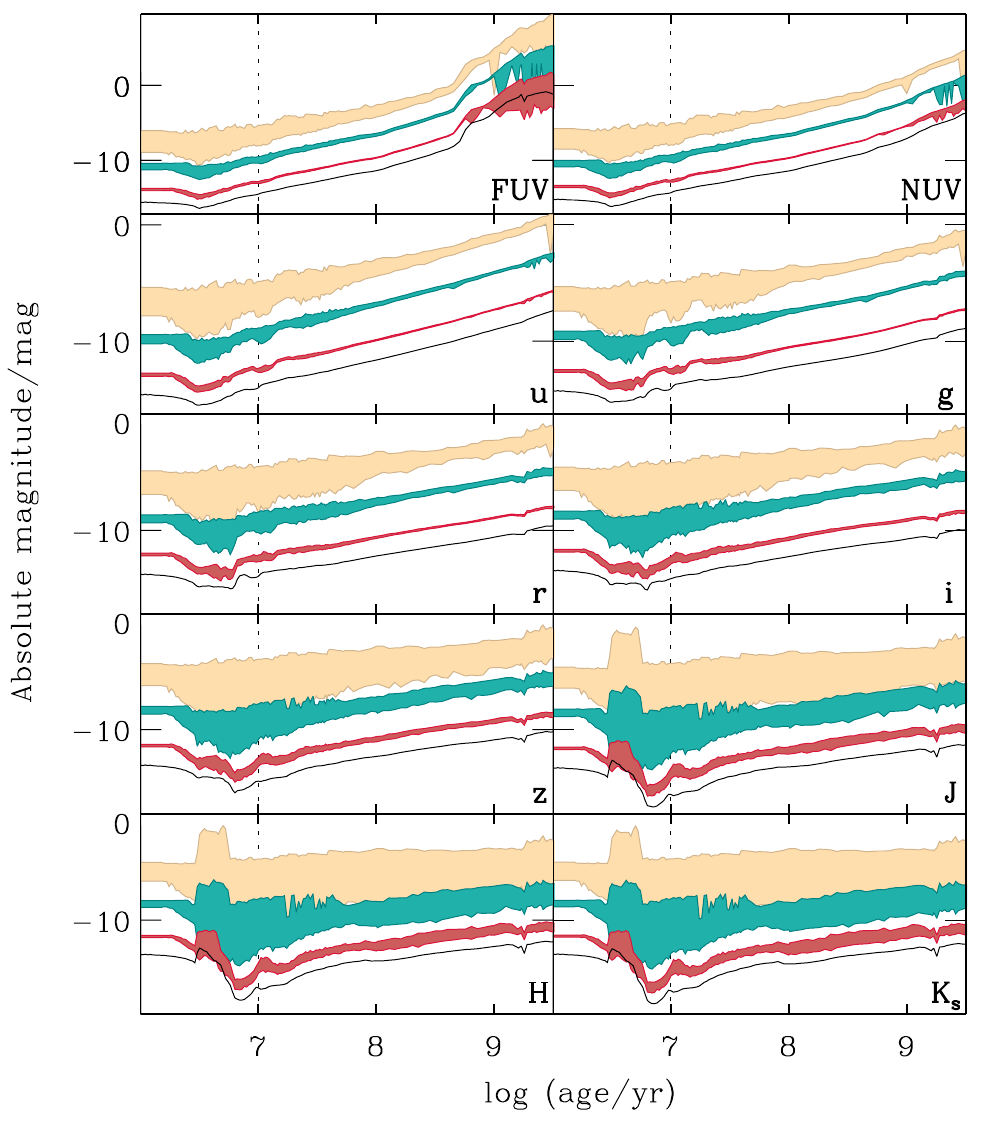}
\caption{Absolute magnitudes of the same models  as in Fig.~\ref{fig:ind017} [with integrated stellar masses $\Mcl=10^3\Msun$ (cream), $10^4\Msun$ (teal blue) and $10^5\Msun$ (indian red), for the metallicity $Z=0.017$], computed through the {\it GALEX} FUV and NUV, SDSS $ugriz$ and 2MASS $JHK_\textsc{s}$ filters ({\it GALEX} and SDSS magnitudes are on the AB system, while 2MASS ones are on the Vega system). For comparison, black curve shows the evolution of a $10^6$\,\Msun, $Z=0.017$ model with a smoothly sampled IMF.}
\label{fig:mag017}
\end{figure}

Mass estimates of observed stellar populations require an absolute flux measurement in addition to that of a spectral energy distribution. For this reason, we show in Fig.~\ref{fig:mag017} the broadband magnitudes of the same models as in Fig.~\ref{fig:ind017}, computed through the {\it Galaxy Evolution Explorer} ({\it GALEX}) FUV and NUV, Sloan Digital Sky Survey (SDSS) $ugriz$ and Two-Micron All-Sky Survey (2MASS) $JHK_\textsc{s}$ filters. Also shown for comparison in Fig.~\ref{fig:mag017} is the evolution of a $10^6$\,\Msun, $Z=0.017$ model with a smoothly sampled IMF (black curve). Again, as expected, the dispersion in integrated spectral properties induced by stochastic IMF sampling increases from the most massive (indian red) to the least massive (cream) model, the eventual presence of a single luminous, massive star having a larger differential effect on the integrated luminosity of a faint, $10^3$-\Msun\ stellar population than on that of a bright, $10^5$-\Msun\ one. Also, consistently with our findings in Fig.~\ref{fig:ind017}, the dispersion in ultraviolet magnitudes in Fig.~\ref{fig:mag017} is largest at ages below $\sim$10\,Myr and above $\sim$500\,Myr, although a difference with respect to ultraviolet spectral indices is that the dispersion in magnitude can still be substantial at ages in between, for small \Mcl. Fig.~\ref{fig:mag017} further shows that the effect of stochastic IMF sampling increases from ultraviolet to near-infrared magnitudes. This is because stars in nearly all mass ranges experience a bright, rapid, cool (red supergiant or AGB) phase at the end of their evolution, whose influence on integrated near-infrared spectral properties can be strongly affected by stochastic IMF sampling. We have also examined the predictions of models with different \Mcl\ similar to those in Fig.~\ref{fig:mag017}, but for metallicities lower than $Z=0.017$ (not shown). The properties of such models are qualitatively similar to those of the models in Fig.~\ref{fig:mag017}, except that low-metallicity models run at brighter magnitudes, as the deficiency of metals in stellar atmospheres reduces the ability for stars to cool and hence fade \citep[e.g.][]{Bressan2012}.

\section{Interpretation of ultraviolet star-cluster spectroscopy}\label{sec:uvinterpret}

In this section, we use the stellar population models presented in the previous section to interpret ultraviolet spectroscopic observations of a sample of young LMC star clusters. Our main goals are to assess the ability of the models to reproduce the strengths of the \citet{Fanelli1992} indices in observed cluster spectra and to quantify the influence of stochastic IMF sampling on age, metallicity and stellar-mass estimates.

\subsection{Observational sample}\label{sec:sample}

To assess the reliability of the ultraviolet spectral synthesis models present in Section~\ref{sec:stelpops}, we appeal to the {\it IUE} spectra of $10$ LMC globular clusters observed by \citet{Cassatella1987}, in which the strengths of the 19 \citet{Fanelli1992} indices listed in Table~\ref{tab:tabindices} were measured by \citet{Maraston2009}. Although \citet{Cassatella1987} originally corrected their spectra for reddening by dust in the Milky Way and the LMC (using the extinction laws of \citealt{Savage1979} and \citealt{Howarth1983}), these smooth corrections do not account for the potential contamination of the spectra by discrete ISM absorption lines arising from resonant ionic transitions \citep[see, e.g.,][]{SavagedB1979,SavagedB1981,Savage1997}. We account for this effect by correcting the index measurements in tables~B.1 and B.2 of \citet{Maraston2009}, which these authors performed in the reddening-corrected {\it IUE} spectra of \citet{Cassatella1987}, for the strongest interstellar absorption features, as described in Appendix~\ref{app:mwcorr}. In brief, since we do not know the column densities of different ionic species along the lines of sight to individual LMC clusters, we follow \citet[][see their table~3]{Leitherer2011} and adopt a mean correction based on the median equivalent widths of the 24 strongest Milky Way absorption lines measured in the wavelength range $1150\lesssim\lambda\lesssim3200\,\AA$ along 83 lines of sight by \citet{Savage2000}. Although this correction does not account for potential extra contamination by LMC absorption lines, we expect such a contribution to be moderate. This is because the \citet{Cassatella1987} clusters are old enough ($\gtrsim10\,$Myr; see Section~\ref{sec:paramest}) to have broken out of their parent molecular clouds \citep{Murray2010,Murray2011}, as also indicated by the absence of nebular emission lines in their spectra.

The resulting ISM correction term, $\Delta\mathrm{(EW/\AA)}$, is listed in Table~\ref{tab:tabindices} for each index (strong interstellar Mg\,\textsc{ii} absorption makes the corrections to the \mgw\ and \mgii\ indices particularly large). The final corrected index strengths and associated errors for all clusters are listed in Table~\ref{tab:idxcorr} of Appendix~\ref{app:mwcorr} (along with the adopted $V$-band magnitudes of the clusters). For reference, not including these corrections for interstellar line absorption would imply changes of typically 1 (5), 13 (25) and 10 (7) per cent, respectively, in the logarithmic estimates of age, metallicity and stellar-mass of the \citet{Cassatella1987} star clusters using the stochastic (smooth) models presented in Section~\ref{sec:paramest} below. 

\subsection{Model library}\label{sec:modelib}

\begin{figure*}
\begin{center}
\resizebox{\hsize}{!}{\includegraphics{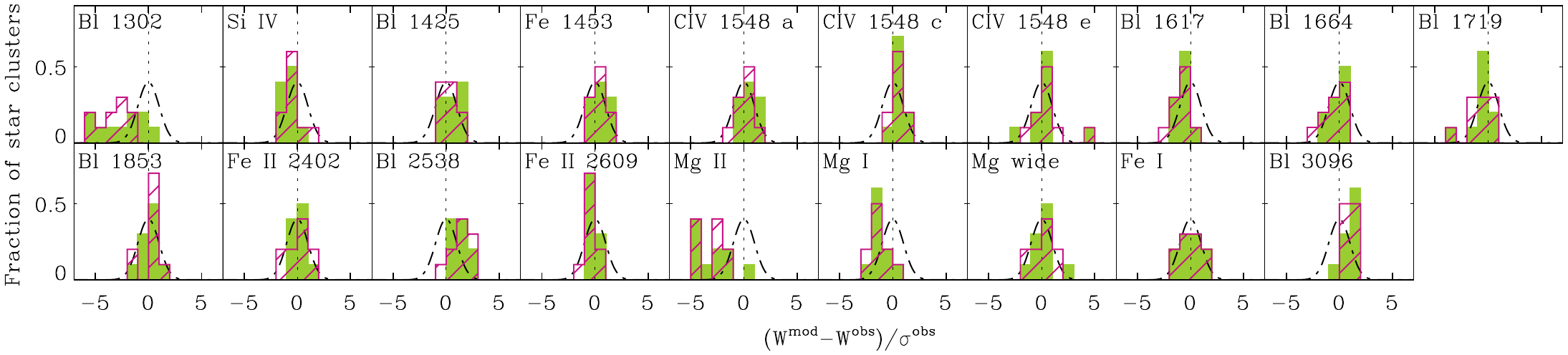}}
\end{center}
\caption{Distribution of the difference in index strength between best-fitting model and observed spectra, in units of the observational error, for the 10 clusters in the \citet{Cassatella1987} sample and the 19 \citet{Fanelli1992} ultraviolet spectral indices defined in Table~\ref{tab:tabindices} (as indicated). For each index, the filled (green) and hatched (violet) histograms show the distributions obtained using models with stochastically sampled and smoothly sampled IMFs, respectively, while the black dot-dashed line shows a reference Gaussian distribution with unit standard deviation.}
\label{fig:indicesR}
\end{figure*}

To interpret the observed ultraviolet index strengths of the \citet{Cassatella1987} star clusters, we use the models presented in Section~\ref{sec:stelpops} to build large libraries of SSP spectra for both a smoothly sampled and a stochastically sampled \citet{Chabrier2003} IMFs (we make the usual assumption that individual star clusters can be approximated as SSPs). Specifically,  we compute models at 14 metallicities, $Z=0.0001$, 0.0002, 0.0005, 0.001, 0.002, 0.004, 0.006, 0.008, 0.010, 0.014, 0.017, 0.020, 0.030 and 0.040, for 67 logarithmically spaced ages between 10\,Myr and 1\,Gyr. The spectra for a smoothly sampled IMF scale linearly with stellar mass \citep[e.g.,][]{Bruzual2003}. For a stochastically sampled IMF, at each of the 67 stellar ages of the grid, we compute 220 realizations of SSP spectra for 33 logarithmically spaced stellar masses between $2\times10^2\,\Msun$ and $3\times10^5\,\Msun$ (as described in Section~\ref{sec:modidx}).

It is important to start by evaluating the extent to which we can expect these models to reproduce the observed ultraviolet index strengths of the \citet{Cassatella1987} star clusters. In particular, the models assume scaled-solar heavy-element abundance ratios at all metallicities, while the relative abundances of different elements (such as the abundance ratio of $\alpha$ elements to iron-peak elements, $\alpha$/Fe) can vary from cluster to cluster in the LMC \citep[e.g.,][]{Colucci2012}. A simple way to assess the goodness-of-fit by a model $j$ of the observed index strengths in the ultraviolet spectrum of a given cluster is to compute the $\chi^2$ statistics,
\begin{equation}
\chi_j^2 = \sum _{i} \left(\frac{ W_{i}^j-W_{i}^{\rm obs}}{\sigma_{i}^{\rm obs} }\right)^{2}\,,
\label{eq:chi2}
\end{equation}
where the summation index $i$ runs over all spectral indices observed, $W_{i}^j$ and $W_{i}^{\rm obs}$ are the equivalent widths of index $i$ in the model and observed spectra, respectively, and $\sigma_{i}^{\rm obs}$ is the observational error. 

Fig.~\ref{fig:indicesR} shows the distribution of the difference in index strength between best-fitting model (corresponding to the minimum $\chi_j^2$ as computed using equation~\ref{eq:chi2}) and observed spectra, in units of the observational error, for the 10 clusters in the \citet{Cassatella1987} sample. Each panel corresponds to a different spectral index in Table~\ref{tab:tabindices}. In each case, the filled and hatched histograms show the distributions obtained using models with stochastically sampled and smoothly sampled IMFs, respectively, while the dot-dashed line shows a reference Gaussian distribution with unit standard deviation. As expected from the similarity between models with a smoothly sampled and a stochastically sampled IMFs at ages between 10\,Myr and 1\,Gyr in Figs.~\ref{fig:indmet} and \ref{fig:ind017}, the filled and hatched histograms are quite similar in all panels in Fig.~\ref{fig:indicesR}. The fact that most histograms in this figure fall within the reference Gaussian distribution further indicates that, globally, the models reproduce reasonably well the ultraviolet index strengths in the observed spectra. Two notable exceptions are the distributions for \bl1302 and \mgii, which display significant tails relative to a Gaussian distribution. For these features, the implied systematic larger absorption in the data relative to the models could arise from either an enhanced $\alpha$/Fe ratio in the \citet{Cassatella1987} clusters, or an underestimate of interstellar absorption, or both. To proceed with a meaningful comparison of models with data in the next paragraphs, we exclude \bl1302 and \mgii\ from our analysis.\footnote{\label{foot:criterion} For \bl1302 and \mgii, the median of the difference $W^{\rm mod}-W^{\rm obs}$ in Fig.~\ref{fig:indicesR} exceeds $1.5\,\sigma^{\rm obs}$ for models with a stochastically sampled IMF. We use this threshold to define `badly fitted indices' in Fig.~\ref{fig:indicesR} and keep the 17 other, better-fitted indices to pursue our analysis.}

\subsection{Age, metallicity and stellar-mass estimates}\label{sec:paramest}

\begin{figure*}
\begin{center}
\resizebox{\hsize}{!}{\includegraphics{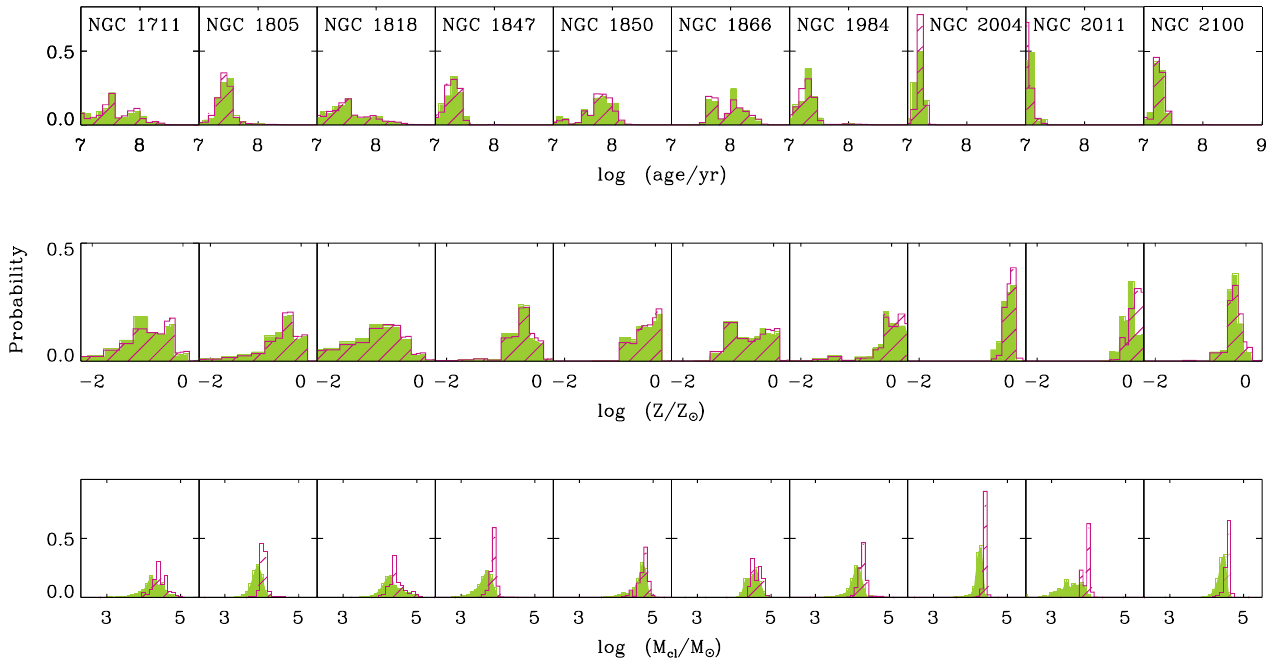}}
\end{center}
\caption{Posterior probability distributions of age (top row), metallicity (middle row) and stellar mass (bottom row) for the 10 clusters in the \citet{Cassatella1987} sample, obtained as described in Section~\ref{sec:paramest} using models with stochastically sampled (green filled histograms) and smoothly sampled (violet hatched histograms) IMFs.}
\label{fig:parampdf}
\end{figure*}

\renewcommand{\arraystretch}{1.4}
\begin{table*}
\caption{Median estimates of age, metallicity and stellar mass -- and associated 68 per cent central credible intervals -- for the 10 clusters in the \citet{Cassatella1987} sample, obtained as described in Section~\ref{sec:paramest} using models with stochastically sampled and smoothly sampled IMFs (the quantity $\chi ^{\rm 2,sto}_{\rm min}/\chi^{\rm 2,smo}_{\rm min}$ indicates the ratio of minimum $\chi_j^2$'s, as computed using equations~\ref{eq:chi2} and \ref{eq:chi2stoc}, between the two types of models). Literature values, along with ultraviolet-based age and metallicity estimates from \citet{Maraston2009}, are also reported for reference.}
\begin{threeparttable}

	\centering
	\begin{tabular}{llllllll}
\toprule						    

\multirow{2}{*}{Cluster\hskip0.5cm}  &  \multicolumn{3}{c}{This work: smoothly sampled IMF} &\multirow{2}{*}{\hskip0.1cm} 
& \multicolumn{3}{c}{This work: stochastically sampled IMF} \\     

\cmidrule{2-4}  \cmidrule{6-8}

&$\log ({\rm age/yr})$ &  $\log(Z/Z_{\odot})$ &  $\log ({\rm \Mcl/\Msun})$ &&$\log ({\rm age/yr})$ &  $\log(Z/Z_{\odot})$ &  $\log ({\rm \Mcl/\Msun})$ \\
\midrule

NGC 1711 & $7.51^{+0.44}_{-0.26}$ & $-0.45^{+0.27}_{-0.73}$ & $4.37^{+0.23}_{-0.07}$ && $7.49^{+0.46}_{-0.34}$ & $-0.57^{+0.39}_{-0.61}$ & $4.19^{+0.31}_{-0.19}$ \\
NGC 1805 & $7.41^{+0.14}_{-0.06}$ & $-0.27^{+0.31}_{-0.32}$ & $3.99^{+0.11}_{-0.09}$ && $7.43^{+0.12}_{-0.08}$ & $-0.29^{+0.34}_{-0.29}$ & $3.83^{+0.17}_{-0.13}$ \\
NGC 1818 & $7.47^{+0.48}_{-0.22}$ & $-0.63^{+0.45}_{-0.85}$ & $4.37^{+0.33}_{-0.07}$ && $7.46^{+0.49}_{-0.21}$ & $-0.73^{+0.55}_{-0.75}$ & $4.23^{+0.37}_{-0.13}$ \\
NGC 1847 & $7.28^{+0.17}_{-0.03}$ & $-0.26^{+0.30}_{-0.15}$ & $3.83^{+0.17}_{-0.03}$ && $7.28^{+0.17}_{-0.13}$ & $-0.28^{+0.33}_{-0.13}$ & $3.62^{+0.18}_{-0.22}$ \\
NGC 1850 & $7.76^{+0.19}_{-0.21}$ & $-0.12^{+0.24}_{-0.16}$ & $4.72^{+0.18}_{-0.12}$ && $7.76^{+0.29}_{-0.21}$ & $-0.18^{+0.29}_{-0.23}$ & $4.63^{+0.17}_{-0.23}$ \\
NGC 1866 & $7.93^{+0.32}_{-0.28}$ & $-0.26^{+0.31}_{-0.62}$ & $4.55^{+0.25}_{-0.05}$ && $7.97^{+0.28}_{-0.32}$ & $-0.28^{+0.33}_{-0.60}$ & $4.47^{+0.23}_{-0.17}$ \\
NGC 1984 & $7.27^{+0.18}_{-0.12}$ & $+0.05^{+0.24}_{-0.23}$ & $4.23^{+0.17}_{-0.03}$ && $7.29^{+0.16}_{-0.14}$ & $+0.01^{+0.28}_{-0.19}$ & $4.07^{+0.13}_{-0.17}$ \\
NGC 2004 & $7.20^{+0.15}_{-0.05}$ & $+0.03^{+0.09}_{-0.06}$ & $4.35^{+0.15}_{-0.05}$ && $7.18^{+0.17}_{-0.03}$ & $+0.00^{+0.12}_{-0.04}$ & $4.19^{+0.11}_{-0.09}$ \\
NGC 2011 & $7.03^{+0.11}_{-0.09}$  & $+0.18^{+0.24}_{-0.14}$ & $3.93^{+0.17}_{-0.13}$ && $7.06^{+0.19}_{-0.01}$ & $+0.06^{+0.23}_{-0.10}$  & $3.46^{+0.34}_{-0.26}$ \\
NGC 2100 & $7.24^{+0.11}_{-0.09}$  & $-0.23^{+0.19}_{-0.06}$ & $4.53^{+0.17}_{-0.03}$ && $7.24^{+0.11}_{-0.09}$ & $-0.25^{+0.21}_{-0.03}$ & $4.37^{+0.13}_{-0.07}$ \\

\midrule
						    
\multirow{2}{*}{Cluster\hskip0.8cm}	    & \multirow{2}{*}{$\chi ^{\rm 2,sto}_{\rm min}/\chi^{\rm 2,smo}_{\rm min}$\ \ } &  \multicolumn{3}{c}{Literature} &\multirow{2}{*}{\hskip0.cm} &\multicolumn{2}{c}{\citet{Maraston2009}} \\     

\cmidrule{3-5}  \cmidrule{7-8} 

& &$\log ({\rm age/yr})$ &  $\log(Z/Z_{\odot})$ &  $\log ({\rm \Mcl/\Msun})$ &&$\log ({\rm age/yr})$ &  $\log(Z/Z_{\odot})$  \\						    
						    
\midrule

NGC 1711 & \ \ 0.864 &  $7.70\pm0.05^{\rm a}$   & $-0.57\pm0.17^{\rm a}$      & $4.21^{+0.16}_{-0.16}$ $^{\rm h}$      && \ \ 7.50 & $-0.10$ \\
NGC 1805 & \ \ 0.909 &  $7.65\pm0.05^{\rm i}$    & $-0.25\pm0.25^{\rm b,e}$   & $3.52^{+0.13}_{-0.13}$ $^{\rm h}$      && \ \ 7.20 & $-0.33$ \\
NGC 1818 & \ \ 0.864 &  $7.40\pm0.30^{\rm b}$   & $-0.25\pm0.25^{\rm b,e,l}$ & $4.13^{+0.15}_{-0.14}$ $^{\rm h}$      && \ \ 7.50 & $-0.20$ \\
NGC 1847 & \ \ 0.918 &  $7.42\pm0.30^{\rm d}$   & $-0.40^{\rm h,j}$                 & $3.86^{+0.09}_{-0.10}$ $^{\rm h}$      && \ \ 6.85 & $-0.40$ \\
NGC 1850 & \ \ 0.995 &  $7.97\pm0.10^{\rm j}$    & $-0.12\pm0.20^{\rm f}$       & $4.87^{+0.13}_{-0.14}$ $^{\rm h,m}$  && \ \ 8.00 & $-0.10$ \\
NGC 1866 & \ \ 0.970 &  $8.25\pm0.05^{\rm k}$   & $-0.28^{12}$                       & $4.63^{+0.08}_{-0.08}$ $^{\rm h}$      && \ \ 8.00 &  $-0.50$ \\
NGC 1984 & \ \ 0.882 &  $7.06\pm0.30^{\rm d}$   & $-0.90\pm0.40^{\rm g}$      & $3.38^{+0.35}_{-0.28}$ $^{\rm h}$      && \ \ 7.10 &  $-0.10$ \\
NGC 2004 & \ \ 0.966 &  $7.30\pm0.20^{\rm c,j}$ & $-0.56\pm0.20^{\rm f,j}$      & $4.43^{+0.24}_{-0.23}$ $^{\rm h}$      && \ \ 7.20 &  $+0.00$   \\
NGC 2011 & \ \ 0.682 &  $6.99\pm0.30^{\rm d}$   & $-0.47\pm0.40^{\rm g}$       & $3.47^{+0.38}_{-0.32}$ $^{\rm h}$      && \ \ 6.70 &  $+0.00$   \\
NGC 2100 & \ \ 0.918 &  $7.31\pm0.04^{\rm j}$    & $-0.32\pm0.20^{\rm f,j}$      & $4.48^{+0.33}_{-0.30}$ $^{\rm h}$      && \ \ 7.35 &  $+0.20$   \\

\bottomrule
	\end{tabular}
\begin{tablenotes}
References:   $^{\rm a}$\cite{Dirsch2000}; $^{\rm b}$\cite{deGrijs2002};   $^{\rm c}$\cite{Elson1991};$^{\rm d}$\cite{ElsonFall1988};   $^{\rm e}$\cite{Johnson2001}; $^{\rm f}$\cite{JasniewiczThevenin1994};  $^{\rm g}$\cite{OlivaOriglia1998}; $^{\rm h}$\cite[masses estimated from {\it HST} optical photometry and mass-to-light ratios from SSP models by \citealt{FiocRocca1997}, adopting cluster ages and metallicities from the literature]{Mackey2003}; $^{\rm i}$\cite{Liu2009}; $^{\rm j}$\cite{Niederhofer2015} ; $^{\rm k}$\cite{Bastian2013} ; $^{\rm l}$\cite{Korn2000}; $^{\rm m}$ \cite{McLaughlin2005}.\\
\end{tablenotes}
\label{tab:agemetest}	
\end{threeparttable}
\end{table*}
\renewcommand{\arraystretch}{1.}

We use of the library of SSP models computed in the previous section to estimate the ages, metallicities and stellar masses of the \citet{Cassatella1987} star clusters on the basis of the 17 \citet{Fanelli1992} ultraviolet indices that can be reasonably well reproduced by the models (footnote~\ref{foot:criterion}). We adopt a standard Bayesian approach and compute the likelihood of an observed set of spectral indices $\bmath{W^{\rm obs}}$ given a model $j$ with parameters (age, metallicity and stellar mass) $\bmath{\Theta^j}$ as
\begin{equation}
P(\bmath{W^{\rm obs}} \mid \bmath{\Theta^j}) \propto \exp ( -\chi_j^2/2 )\,,
\label{eq:likely}
\end{equation}
where we have assumed that the observed index strengths can be modelled as a multi-variate Gaussian random variable, with mean given by the index strengths of the model with parameters $ \bmath{\Theta^j}$ and noise described by a diagonal covariance matrix \citep[e.g.,][]{Chevallard2016}. For models with a smoothly sampled IMF, the value of $\chi^2_j$ entering equation~\eqref{eq:likely} is that given in equation~\eqref{eq:chi2}. In this case, the posterior probability distribution of stellar mass for a given cluster can be derived from that of the mass-to-light ratio and the observed absolute $V$-band magnitude. For models with a stochastically sampled IMF, the luminosity does not scale linearly with stellar mass (Section~\ref{sec:modmag}), and the fit of the absolute $V$-band magnitude must be inserted in the definition of $\chi_j^2$, i.e.,
\begin{equation}
\chi_j^2 = \left(\frac{V^j-V^{\rm obs}}{\sigma_V^{\rm obs}}\right)^{2}+ \sum _{i} \left(\frac{ W_{i}^j-W_{i}^{\rm obs}}{\sigma_{i}^{\rm obs} }\right)^{2}\,.
\label{eq:chi2stoc}
\end{equation}
Here $V^j$ and $V^{\rm obs}$ are the model and observed absolute $V$-band magnitudes, respectively, $\sigma_V^{\rm obs}$ is the associated observational error and the other symbols have the same meaning as in equation~\eqref{eq:chi2}. The combination of the likelihood function in equation~\eqref{eq:likely} with the model library of Section~\ref{sec:modelib}, which assumes flat prior distributions of age, metallicity and stellar mass, allows us to compute the posterior probability distributions of these parameters for each cluster \citep[see, e.g., equation~3.2 of][]{Chevallard2016}.

Fig.~\ref{fig:parampdf} shows the posterior probability distributions of age (top row), metallicity (middle row) and stellar mass (bottom row) obtained in this way for the 10 clusters in the \citet{Cassatella1987} sample, using models with  stochastically sampled (filled histograms) and smoothly sampled (hatched histograms) IMFs. Table~\ref{tab:agemetest} lists the corresponding median estimates of age, metallicity and stellar mass -- and the associated 68 per cent central credible intervals -- along with previous age and metallicity estimates from the literature. In general, we find that the probability distributions in Fig.~\ref{fig:parampdf} are narrower for clusters with index strengths measured with larger signal-to-noise ratio (Table~\ref{tab:idxcorr}). The similarity of the constraints derived on age and metallicity using both types of IMF sampling is striking, although expected from Figs.~\ref{fig:indmet}, \ref{fig:ind017} and \ref{fig:indicesR}, given that all star clusters turn out to have ages in the range $10\lesssim t\lesssim100\,$Myr (and hence, in the favourable range between 10\,Myr and 1\,Gyr; see Section~\ref{sec:modidx} and Table~\ref{tab:agemetest}). Interestingly, models with a stochastically sampled IMF always fit the data better than those with a smoothly sampled IMF, as reflected by the minimum-$\chi^2$ ratios reported in Table~\ref{tab:agemetest}. This table also shows that the constraints derived here on cluster ages and metallicities (in the range $0.2\lesssim Z/Z_{\odot}\lesssim1.1$) generally agree, within the errors, with previous estimates based on different models and observables. 

\begin{figure}
\includegraphics[width=\columnwidth]{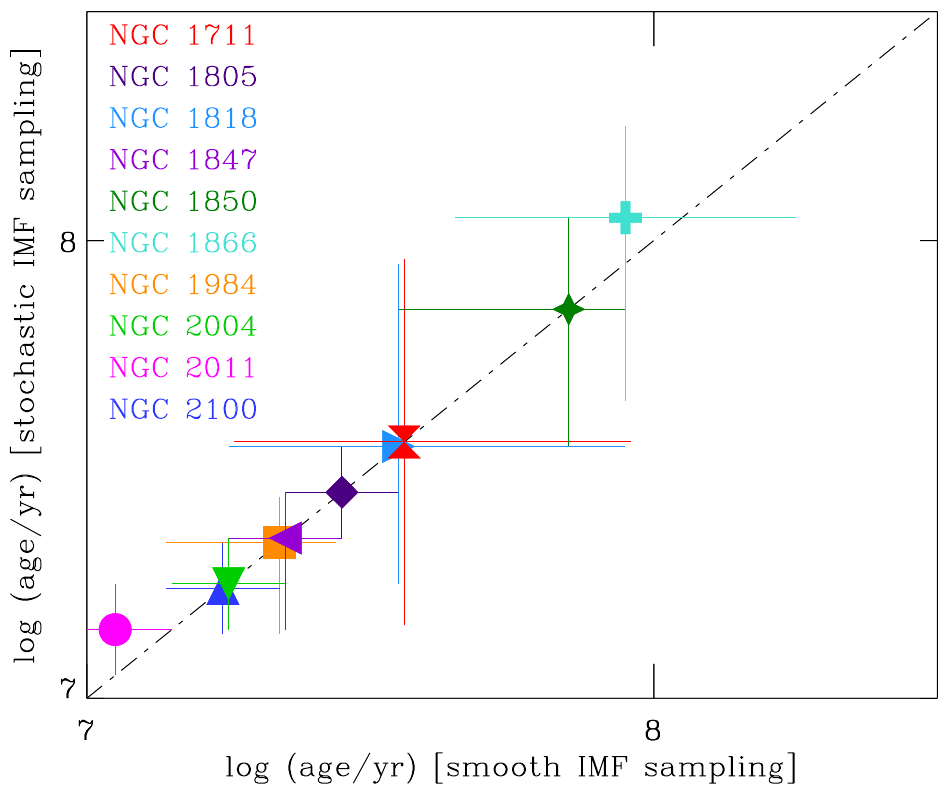}
\caption{Comparison of median estimates of age -- and associated 68 per cent central credible intervals -- obtained as described in Section~\ref{sec:paramest} using models with stochastically sampled and smoothly sampled IMFs, for the 10 clusters in the \citet{Cassatella1987} sample (colour-coded as indicated).}
\label{fig:agecomp}
\end{figure}

\begin{figure}\includegraphics[width=\columnwidth]{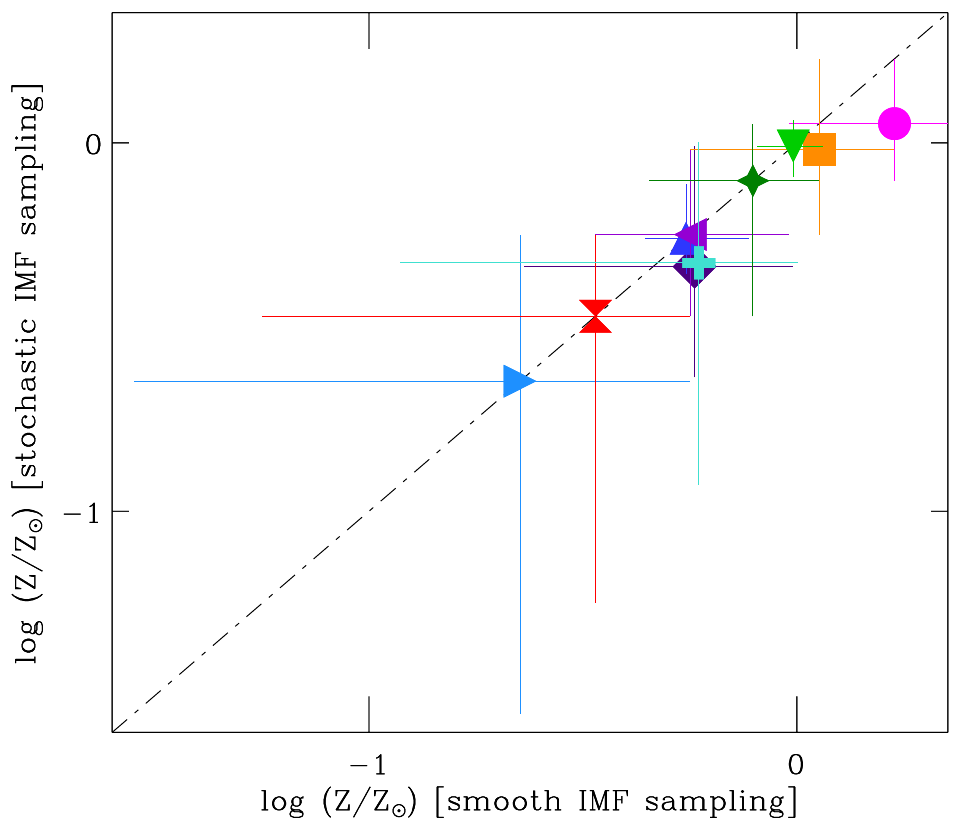}
\caption{Same as Fig.~\ref{fig:agecomp}, but for metallicity estimates.}
\label{fig:metcomp}
\end{figure}

\begin{figure}\includegraphics[width=\columnwidth]{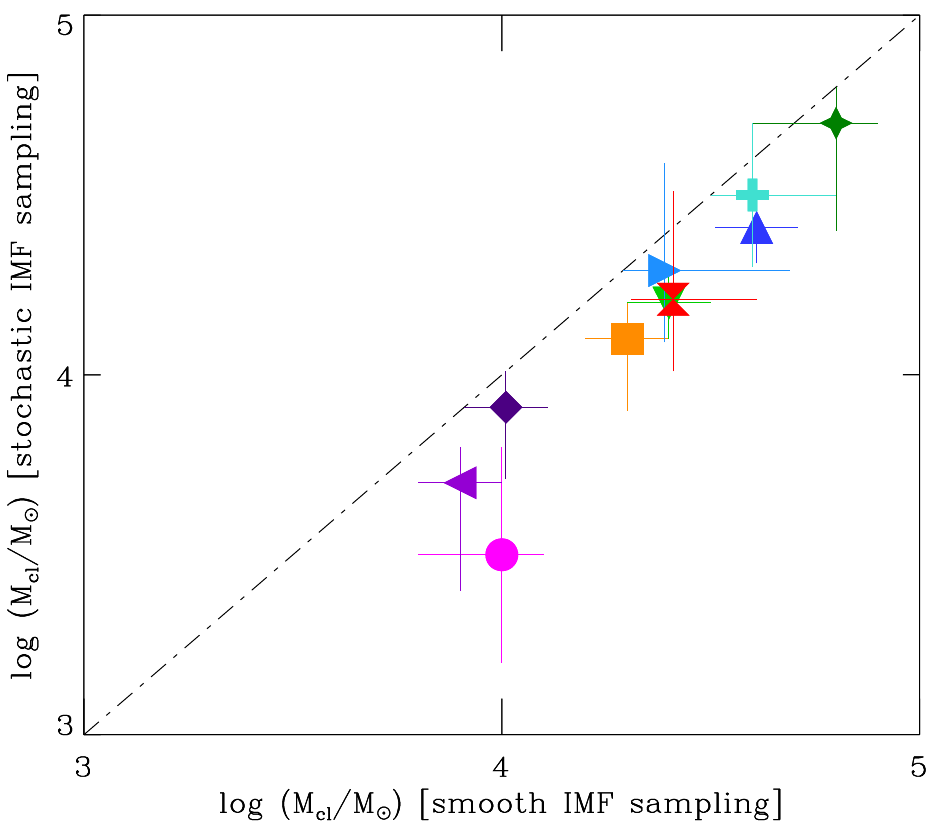}
\caption{Same as Fig.~\ref{fig:agecomp}, but for stellar-mass estimates.}
\label{fig:masscomp}
\end{figure}

\begin{figure}
\includegraphics[width=\columnwidth]{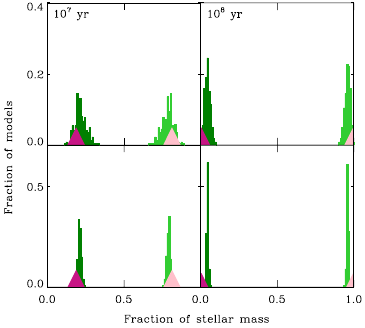}
\caption{Distribution of the fraction of integrated stellar mass contributed by stars more/less massive than 5\,\Msun\ (dark/light green) for four sets of 220 realisations of SSPs with a stochastically sampled \citet{Chabrier2003} IMF and target stellar masses of $10^3\,\Msun$ (top) and $10^4\,\Msun$ (bottom) at the ages of 10\,Myr (left) and 100\,Myr (right), for the metallicity $Z=0.008$. Dark/light pink triangles indicate the corresponding values for models with a smoothly sampled \citet{Chabrier2003} IMF.}
\label{fig:ml_mhimto}
\end{figure}

A notable result from Fig.~\ref{fig:parampdf} is the systematic lower stellar mass obtained when using models with a stochastically sampled IMF, as appropriate for star clusters with masses in the range $3\times10^3\lesssim \Mcl/M_{\odot}\lesssim4\times10^4$ (Section~\ref{sec:modidx} and Table~\ref{tab:agemetest}), relative to a smoothly sampled IMF. The difference reaches up to a factor of 3 in the case of the youngest cluster, NGC~2011. This result, which contrasts with those pertaining to age and metallicity estimates, is visualised otherwise in Figs.~\ref{fig:agecomp}--\ref{fig:masscomp}, in which we compare the median estimates of age, metallicity and stellar mass (and the associated 68 per cent central credible intervals) from models with stochastically sampled and smoothly sampled IMFs. While age and metallicity estimates from both types of models are in good general agreement (within the errors) in Figs~\ref{fig:agecomp} and \ref{fig:metcomp}, the systematic offset in stellar-mass estimates between stochastically and smoothly sampled IMFs appears clearly in Fig.~\ref{fig:masscomp}.   To investigate the origin of this effect, we plot in Fig.~\ref{fig:ml_mhimto} the contributions by stars more/less massive than 5\,\Msun\ \citep[roughly the limit between massive and intermediate-mass stars; e.g.][]{Bressan2012} to the integrated stellar mass of SSPs weighing $10^3$ and $10^4$\Msun, at the ages of 10 and 100\,Myr, for models with stochastically and smoothly sampled IMFs (in the former case, we show the results of 220 cluster realisations for each combination of stellar mass and age; see the figure caption for details). We adopt the metallicity $Z=0.008$, typical of the \citet{Cassatella1987} clusters in Table~\ref{tab:agemetest}. 

\begin{figure}
\includegraphics[width=\columnwidth]{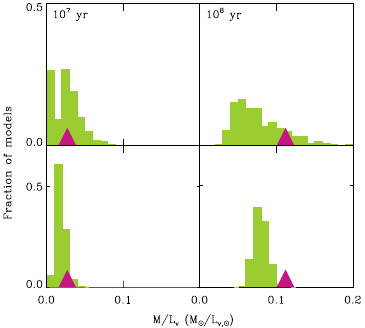}
\caption{Integrated mass-to-$V$-band luminosity ratio of the same model SSPs as in Fig.~\ref{fig:ml_mhimto}, with stochastically sampled (green) and smoothly sampled (pink) IMFs.}
\label{fig:ml_lhilto}
\end{figure}

Fig.~\ref{fig:ml_mhimto} shows that stars more massive than 5\,\Msun\ account typically for a larger fraction of the total cluster mass when the IMF is stochastically sampled than when it is smoothly sampled, for SSP masses of both $10^3$ and $10^4$\Msun\ and ages of both 10 and 100\,Myr. This is because, when a massive star is drawn in a stochastic model, it can account for a substantial fraction of the cluster mass, while models with a smoothly sampled IMF can contain `fractions' of massive stars in fixed proportion to the number of less massive stars.\footnote{For example, a $10^3$\Msun\ cluster with a smoothly sampled \citet{Chabrier2003} IMF truncated at 0.1 and 100\,\Msun\ contains 0.07 stars with masses between 90 and 100\,\Msun\ (and hence, a total of $\sim$993\,\Msun\ of lower-mass stars). In contrast, if a 100\,Mo star is drawn when stochastically sampling the IMF, only 900\,\Msun\ can be accounted by lower-mass stars.} This is also why models with smoothly sampled IMFs appear as single triangles in Fig.~\ref{fig:ml_mhimto}, while the histograms for models with stochastically sampled IMFs indicate that the mass fraction in stars more massive than 5\,\Msun\ depends on the actual stellar masses drawn in each cluster realisation. Since the mass-to-light ratio of massive stars is much smaller than that of low- and intermediate-mass stars, the results of Fig.~\ref{fig:ml_mhimto} have implications for the integrated mass-to-light ratio of young star clusters. This is shown in Fig.~\ref{fig:ml_lhilto}, where we plot the mass-to-$V$-band luminosity ratio of the same model star clusters as in Fig.~\ref{fig:ml_mhimto}. As anticipated, models with stochastically sampled IMFs have systematically lower mass-to-light ratio than those with smoothly sampled IMFs. This explains the difference in the masses of the \citet{Cassatella1987} star clusters retrieved using both types of models in Fig.~\ref{fig:masscomp} and Table~\ref{tab:agemetest}. We note that changing the upper mass limit of the IMF by a factor of a few would have no influence on the results of Figs.~\ref{fig:agecomp}--\ref{fig:masscomp}, since the turnoff mass of the youngest cluster (NGC~2011) is already as low as 18\,\Msun.

\begin{figure}
\includegraphics[width=\columnwidth]{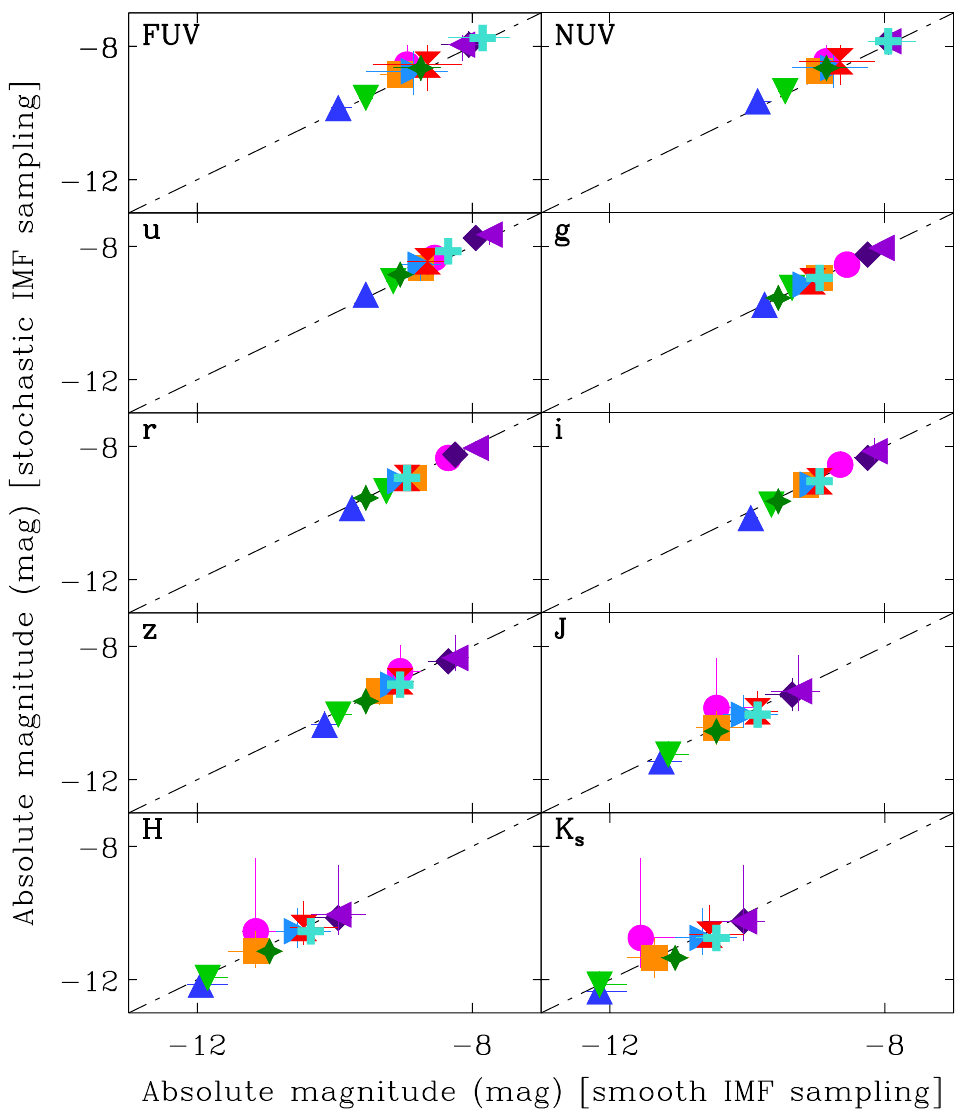}
\caption{Comparison of the median absolute magnitudes in the {\it GALEX} FUV and NUV, SDSS $ugriz$ and 2MASS $JHK_\textsc{s}$ filters  -- and associated 68 per cent central credible intervals -- predicted by the same models with stochastically and smoothly sampled IMFs as in Figs.~\ref{fig:agecomp}--\ref{fig:masscomp}.}
\label{fig:magcomp}
\end{figure}

It is also interesting to examine the difference in the absolute broadband magnitudes predicted at ultraviolet, optical and near-infrared wavelengths for the \citet{Cassatella1987} star clusters by the same models with stochastically and smoothly sampled IMFs as shown in Figs.~\ref{fig:agecomp}--\ref{fig:masscomp}. We present this in Fig.~\ref{fig:magcomp} for the magnitudes predicted in the {\it GALEX} FUV and NUV, SDSS $ugriz$ and 2MASS $JHK_\textsc{s}$ filters. Only for the youngest cluster, NGC~2011, is the difference substantial between the median magnitudes predicted using both types of IMF sampling,  particularly in the reddest bands but with a large uncertainty.

We conclude from this section that the stellar population models presented in Section~\ref{sec:stelpops} provide reasonable fits to the observed ultraviolet spectral signatures of young stellar populations at ages between 10\,Myr and 100\,Myr. At these ages, the neglect of stochastic variations in the number of massive stars hosted by individual star clusters does not have a strong influence on age and metallicity estimates. However, such a neglect can introduce a systematic bias in stellar-mass estimates. Given that the spectral evolution predicted at ages younger than 10\,Myr by the models presented in Section~\ref{sec:stelpops} has already been shown to provide reasonable fits of the nebular emission from observed galaxies \citep[e.g.,][see also Section~\ref{sec:nebem} below]{Stark2014,Stark2015a,Feltre2016}, we feel confident that these models represent a valuable means of exploring the rest-frame ultraviolet emission from star-forming galaxies. This is our aim in the remainder of this paper.

\section{Influence of the ISM on ultraviolet spectra of star-forming galaxies}\label{influenceism}

In this section we explore the impact of absorption and emission by the ISM on the ultraviolet spectra of star-forming galaxies. We consider three components: nebular emission from the photoionized ISM;  absorption by highly ionized species in the photoionized ISM; and absorption by weakly ionized species in the neutral ISM. We refer to the last two components globally as `interstellar absorption'. In Section~\ref{sec:ismmodels} below, we start by presenting our approach to describe nebular emission and interstellar absorption in a star-forming galaxy. Then, in Section~\ref{sec:idxselection}, we investigate the extent to which absorption and emission by the ISM can affect the strengths of the stellar absorption-line indices studied in Section~\ref{sec:stelpops}. We identify those ultraviolet spectral features individually most promising tracers of the properties of stars (Section~\ref{sec:idxstars}), nebular emission (Section~\ref{sec:nebem}) and interstellar absorption (Section~\ref{sec:ismreg}). Some strong, widely studied ultraviolet features turn out to often be mixtures of stellar and interstellar components. We describe those in Section~\ref{sec:mixed}, where we explore their dependence on metallicity, star formation history and upper mass cutoff of the IMF. 

\subsection{ISM modelling}\label{sec:ismmodels}

\subsubsection{Approach}\label{sec:ismmodelsap}

The influence of the ISM on the luminosity produced by stars in a galaxy can be accounted for by expressing the luminosity per unit wavelength $\lambda$ emerging at time $t$ from that galaxy as (using the `isochrone synthesis' technique of \citealt{CharlotBruzual1991})
\begin{equation}\label{eq:lumgal}
L_{\lambda}(t) = \int _{0}^{t}d\tprime\, \psi (t-\tprime) \, S_{\lambda}[\tprime, Z(t-\tprime)] \, T_{\lambda}(t,\tprime),
\end{equation}
where $\psi (t-\tprime)$ is the star-formation rate at time $t-\tprime$, $S_{\lambda}[\tprime, Z(t-\tprime)]$ the luminosity per unit wavelength per unit mass produced by a single stellar generation (SSP) of age $\tprime$ and metallicity $Z(t-\tprime)$ and $T_{\lambda}(t,\tprime)$ is the transmission function of the ISM. We compute the spectral evolution $S_{\lambda}[\tprime, Z(t-\tprime)]$ using the stellar population synthesis model described in Section~\ref{sec:popsyn}. At galactic stellar-mass scales, corresponding to stellar masses $\Mgal>10^{6}$\Msun, fluctuations in integrated spectral properties arising from stochastic sampling of the stellar IMF are no longer an issue \citep{Bruzual2003,Lancon2008}. Therefore, in all models in this section, we adopt a smoothly sampled \citet{Chabrier2003} IMF truncated at 0.1 and 100\,\Msun. The transmission function $T_{\lambda}(t,\tprime)$ in equation~\eqref{eq:lumgal} is defined as the fraction of the radiation produced at wavelength $\lambda$ at time $t$ by a population of stars of age $\tprime$ that is transferred through the ISM.

Following \citet{CharlotFall2000}, we express the transmission function of the ISM as the product of the transmission functions of stellar birth clouds (i.e. giant molecular clouds) and the intercloud medium  (i.e. diffuse ambient ISM). We assume for simplicity that transmission through the birth clouds depends only on SSP age \tprime, while transmission through the intercloud medium depends on the radiation field from the entire stellar population, of age $t$. We thus write\footnote{The function $T_{\lambda}^{\rm ICM}(t)$ in expression~\eqref{eq:transmission} replaces the function $T_{\lambda}^{\rm ISM}$ in the notation of \citet{CharlotFall2000}.}
\begin{equation}\label{eq:transmission}
T_{\lambda}(t,\tprime)= T_{\lambda}^{\rm BC}(\tprime) \, T_{\lambda}^{\rm ICM}(t).
\end{equation}
Furthermore, as in \citet{CharlotFall2000}, we assume that the birth clouds are all identical and consist of an inner \hii\ region ionized by young stars and bounded by an outer \hi\ region. We thus rewrite the transmission function of the birth clouds as
\begin{equation}\label{eq:transBC_a}
T_{\lambda}^{\rm BC}(\tprime)= T_{\lambda}^{\rm HII}(\tprime) \, T_{\lambda}^{\rm HI}(\tprime).
\end{equation}
We note that, with the assumption that the ionized regions are bounded by neutral material, the function $T_{\lambda}^{\rm HII}$ will be close to zero at wavelengths blueward of the H-Lyman limit and greater than unity at  wavelengths corresponding to emission lines. According to the stellar population synthesis model described in Section~\ref{sec:stelpops}, less than 0.1 per cent of H-ionizing photons are produced at ages greater than 10\,Myr by a single stellar generation. This is similar to the typical timescale for the dissipation of giant molecular clouds in star-forming galaxies \citep[e.g.,][]{Murray2010,Murray2011}. We therefore assume
\begin{equation}\label{eq:transBC_b}
T_{\lambda}^{\rm HII}(\tprime) =T_{\lambda}^{\rm HI}(\tprime)=T_{\lambda}^{\rm BC}(\tprime)=1\,, \hspace{3mm}  {\rm for} \hspace{3mm} \tprime > 10\,{\rm Myr} \,.
\end{equation}
We now need prescriptions to compute the functions $T_{\lambda}^{\rm HII}(\tprime)$ and $T_{\lambda}^{\rm HII}(\tprime)$ at earlier ages, along with $T_{\lambda}^{\rm ICM}(t)$.

Recently, \citet{gutkin_modelling_2016} computed the transmission function of ionized gas in star-forming galaxies [$T_{\lambda}^{+}(\tprime)$ in their notation], using the approach proposed by \citet[hereafter CL01]{ChaLon01}. This consists in combining a stellar population synthesis model with a standard photoionization code to describe the  galaxy-wide transfer of stellar radiation through ionized gas via a set of `effective' parameters. \citet{gutkin_modelling_2016} combine in this way the stellar population synthesis model described in Section~\ref{sec:stelpops} with the latest version of the photoionization code \cloudy\ (c13.03; described in \citealt{Ferland2013}). The link between the two codes is achieved through the time-dependent rate of ionizing photons produced by a typical star cluster ionizing an effective \hii\ region,
\begin{equation}\label{eq:rateions}
Q(\tprime)= \dfrac{M_\ast}{hc} \int _{0}^{\lambda_{\rm L}} d\lambda \, \lambda \, S_{\lambda}(\tprime),
\end{equation}
where $M_\ast$ is the mass of the ionizing star cluster, $h$ and $c$ are the Planck constant and the speed of light, $\lambda_{\rm L}=912\,\AA$ the wavelength at the Lyman limit and, for clarity, the dependence of $S_\lambda$ (and hence $Q$) on stellar metallicity $Z$ has been dropped. The radius of the Str\"omgren sphere ionized by this star cluster in gas with effective density \nh\ (assumed independent of \tprime) is given by
\begin{equation}\label{eq:stromgren}
R_{\rm S}^3(\tprime)={3Q(\tprime)}\slash{(4\pi \nh^2 \epsilon \alpha_{\rm B})}\,, 
\end{equation}
where $\epsilon$ is the volume-filling factor of the gas (i.e., the ratio of the volume-averaged hydrogen density to \nh) and $\alpha_{\rm B}$ the case-B hydrogen recombination coefficient. The ionization parameter (i.e., the dimensionless ratio of the number density of H-ionizing photons to \nh) at the Str\"omgren radius is then
\begin{equation}\label{eq:us}
\Us(\tprime) =Q(\tprime)/(4\pi R_{\rm S}^2 \nh{c})=\frac{\alpha_{\rm B}^{2/3}}{3c} \left[ \frac{3 Q(\tprime)\epsilon^{2} \nh}{4\pi} \right]^{1/3}.
\end{equation}
In this approach, for a given input spectral evolution of single stellar generation, $S_{\lambda}(\tprime)$, the galaxy-wide transmission function of the ionized gas can be computed by specifying the zero-age ionization parameter at the Str\"omgren radius,
\begin{equation}\label{eq:usdef}
\Us \equiv \Us(0),
\end{equation}
along with \nh\ and the abundances of the different metals and their depletions onto dust grains (see below). The effective star-cluster mass $M_\ast$ in equation~\eqref{eq:rateions} has no influence on the results other than that of imposing a maximum \Us\ at fixed \nh\ (corresponding to $\epsilon=1$; CL01). We note that \cloudy\ incorporates a full treatment of dust, including, in particular, absorption and scattering of photons, radiation pressure, photoelectric heating of the gas and collisional energy exchange between dust grains and the gas \citep[][see also \citealt{vanHoof2004}]{Ferland2013}.

The computations by \citet{gutkin_modelling_2016} of the transmission function $T_{\lambda}^{+}(\tprime)$ of ionized gas in star-forming galaxies cannot be used straightforwardly to represent the function $T_{\lambda}^{\rm HII}(\tprime)$ entering equation~\eqref{eq:transBC_a}. This is because the transmission function $T_{\lambda}^{+}(\tprime)$ calculated by \cloudy\ does not include {\em interstellar-line absorption} in the ionized gas, even though the code computes the full ionization structure of the nebula to produce the emission-line spectrum (of main interest to \citealt{gutkin_modelling_2016}). It is possible to account for interstellar-line absorption in the spectrum emerging from an \hii\ region by appealing to another code to exploit the ionization structure solved by \cloudy: the general spectrum synthesis program \synspec\ \citep[e.g.,][]{Hubeny2011}, which computes absorption signatures in the emergent spectrum, based on an exact radiative transfer solution for a specified structure of the medium (temperature, density and possibly atomic energy level populations) and a specified set of opacity sources (continua, atomic and molecular lines).\footnote{See \url{http://nova.astro.umd.edu/Synspec49/synspec.html}} The combination of \cloudy\ and \synspec\ can be achieved via an interactive program called \cloudspec\ \citep[][see also \citealt{Heap2001}]{Hubeny2000}, written in the Interactive Data Language (\idl). The \cloudspec\ program calls \synspec\ to solve the radiative transfer equation along the line of sight toward an ionizing source, based on the depth-dependent output of \cloudy, to compute the strengths and profiles of interstellar absorption lines. This has been used successfully to interpret in detail the observed ultraviolet spectrum of, for example, the metal-poor nearby star-forming galaxy I\,Zw\,18 \citep{Lebouteiller2013}. We note that, while \citet{gutkin_modelling_2016} stop their photoionization calculations at the edge of the Str\"omgren sphere, when the electron density falls below 1 per cent of \nh\ or if the temperature falls below 100\,K, the \cloudy\ calculations can be carried out further into the outer \hi\ envelopes of the clouds. This has little interest for the computation of emission-line luminosities, since attenuation by dust eventually associated with \hi\ can be included a posteriori (see, e.g., section~2.6 of \citealt{Chevallard2016}), but more interest for that of interstellar absorption lines.

In this paper, we use the \cloudspec\ wrapper of \cloudy\ and \synspec\ to compute in one go the transmission function of an effective (i.e. typical) birth cloud, $T_{\lambda}^{\rm BC}(\tprime)$, through both the inner \hii\ and outer \hi\ regions [i.e., we do not compute separately $T_{\lambda}^{\rm HII}(\tprime)$ and  $T_{\lambda}^{\rm HI}(\tprime)$ in equation~\ref{eq:transBC_b}]. We achieve this by adopting equations~\eqref{eq:rateions}--\eqref{eq:usdef} above to describe the gas photoionized by stars younger than 10\,Myr in terms of the effective zero-age ionization parameter at the Str\"omgren radius, \Us, gas density, \nh, and metal abundances and depletion factors. For the latter, we adopt the self-consistent, versatile prescription of \citet{gutkin_modelling_2016} to model in detail the influence of `gas-phase' and `interstellar' (i.e., total gas+dust phase) abundances on nebular emission. This prescription is based on the solar chemical abundances compiled by \citet{Bressan2012} from the work of \cite{Grevesse1998}, with updates from \citet[][see table~1 of \citealt{Bressan2012}]{Caffau2011}, and small adjustments for the solar nitrogen ($-0.15$\,dex) and oxygen ($+0.10$\,dex) abundances relative to the mean values quoted in table~5 of \citet[][see \citealt{gutkin_modelling_2016} for details]{Caffau2011}. The corresponding present-day solar (photospheric) metallicity is $\zsun=0.01524$, and the protosolar metallicity (i.e. before the effects of diffusion) $\zpsun=0.01774$. Both N and C are assumed to have primary and secondary nucleosynthetic components. The abundance of combined primary+secondary nitrogen is related to that of oxygen via equation~(11) of \citet{gutkin_modelling_2016}, while secondary carbon production is kept flexible via an adjustable \CO\ ratio. For reference, the solar \NO\ and \CO\ ratios in this prescription are $\NOsol=0.07$ and $\COsol=0.44$.\footnote{\label{foot:cno}This implies that, at solar metallicity, C, N and O represent respectively about 24, 4 and 54 per cent of all heavy elements by number (these values differ from those in footnote~3 of \citealt{gutkin_modelling_2016}, computed without accounting for the fine tuning of N and O abundances).} The depletion of heavy elements onto dust grains is computed using the default depletion factors of \cloudy, with updates from \citet[][see their table~2]{Groves2004} for C, Na, Al, Si, Cl, Ca and Ni and and from \citet[][see their table~1]{gutkin_modelling_2016} for O. The resulting dust-to-metal mass ratio for solar interstellar metallicity, $\zism=\zsun$, is $\xidsol=0.36$, with corresponding gas-phase abundances $12+\log\OHgassol=8.68$,  $\NOgassol=0.10$ and $\COgassol=0.31$. The dust-to-metal mass ratio, \xid, can be treated as an ajustable parameter, along with the interstellar metallicity, \zism. 

We use the above approach to compute the transmission function $T_{\lambda}^{\rm BC}(\tprime)$ for given  $S_{\lambda}(\tprime)$, \nh, \Us, \zism, \CO\ and \xid, tuning the \cloudy\ `stopping criterion' to allow calculations to expand beyond the \hii\ region, into the outer \hi\ envelope of a birth cloud. In the Milky Way, cold atomic gas organized in dense clouds, sheets and filaments has typical temperatures in the range 50--100\,K and densities in the range $1\lesssim\nh\lesssim10^3\,{\rm cm}^{-3}$ (e.g., \citealt{Ferriere2001}; this density range is similar to that of $1\leq\nh\leq10^4\,{\rm cm}^{-3}$ considered by \citealt{gutkin_modelling_2016} for their calculations of effective \hii\ regions). We therefore stop the \cloudy\ calculations when the kinetic temperature of the gas falls below 50\,K and take this to define the \hi\ envelope of a typical birth cloud. The corresponding \hi\ column density and dust attenuation optical depth depend on the other adjustable parameters of the model, i.e.,  $S_{\lambda}(\tprime)$, \nh, \Us, \zism, \CO\ and \xid. Following \citet{gutkin_modelling_2016}, we adopt the same metallicity for the ISM as for the ionizing stars, i.e., we set $\zism=Z$, and adopt spherical geometry for all models. We note that \cloudspec\ also allows adjustments of the velocity dispersion of the gas, $\sigma_v$, and the macroscopic velocity field as a function of radius, $v(r)$.

Photons emerging from stellar birth clouds, and photons emitted by stars older than the typical dissipation time of a birth cloud (i.e. $\sim$10\,Myr; equation~\ref{eq:transBC_b}), must propagate through the diffuse intercloud component of the ISM before they escape from the galaxy. This is accounted for by the transmission function $T_{\lambda}^{\rm ICM}(t)$ in equation~\eqref{eq:transmission}. Little ionizing radiation is striking this medium since, in our model, the birth clouds are ionization bounded, while less than 0.1 per cent of H-ionizing photons are produced at ages greater than 10\,Myr by an SSP. Before proceeding further, it is important to stress that the galaxy-wide intercloud medium considered here should not be mistaken for the `diffuse ionized gas' observed to contribute about 20--50 per cent of the total H-Balmer-line emission in nearby spiral and irregular galaxies, which appears to be spatially correlated with \hii\ regions and ionized, like these, by massive stars \citep[e.g.,][and references therein; see also \citealt{gutkin_modelling_2016} and references therein]{Haffner2009}. In our model, this diffuse ionized gas is subsumed in the \hii-region component described by the effective, galaxy-wide parameters entering equations~\eqref{eq:rateions}--\eqref{eq:usdef} (see also CL01). Instead, the intercloud medium refers to the warm, largely neutral gas filling much of the volume near the midplane of disc galaxies like the Milky Way \citep[e.g.,][]{Ferriere2001}. 

We appeal once more to \cloudspec\ to compute the transmission function $T_{\lambda}^{\rm ICM}(t)$ of this component. Since our aim here is to illustrate the signatures of diffuse interstellar absorption in ultraviolet galaxy spectra rather than compute an exhaustive grid of models encompassing wide ranges of parameters, we fix the hydrogen density of the intercloud medium at $\nhicm=0.3\,{\rm cm}^{-3}$, roughly the mean density of the warm Milky-Way ISM \citep[e.g.,][]{Ferriere2001}. Furthermore, for simplicity, we adopt the same interstellar metallicity, dust-to-metal mass ratio and \CO\ ratio for this component as for the birth clouds (\zism, \xid\ and \CO). The weakness of the ionizing radiation in the intercloud medium argues against a parametrization of the transmission function in terms of the zero-age ionization parameter via a flexible volume-filling factor, as was appropriate for the birth clouds (equations~\ref{eq:us}--\ref{eq:usdef}). Instead, we fix the volume-filling factor of the intercloud medium at $\epsilon=1$ and choose as main adjustable parameter the typical \hi\ column density seen by photons in the intercloud medium, \Nhicm, which we use as stopping criterion for \cloudy. The radiation  striking this medium is the sum of that emerging from the birth clouds and that produced by stars older than 10\,Myr. A subtlety arises from the fact that these sources are distributed throughout the galaxy, hence the radiation striking the intercloud medium in any location is more dilute than if all the sources were concentrated in a single point near that location. We wish to account for this effect in a simple way, without having to introduce geometric dilution parameters. We achieve this by taking the energy density of the interstellar radiation field of the Milky Way at $\lambda_{1500}=1500\,\AA$ as reference and assuming that this quantity scales linearly with star formation rate. Specifically, we require that the energy density $u_{1500}(t)$ at time $t$ at $\lambda_{1500}$ of a galaxy with current star formation rate $\psi(t)={\rm 1\,\Msun\,yr}^{-1}$, roughly equal to that of the Milky Way \citep[e.g.,][]{Robitaille2010}, be that of the local interstellar radiation field, $u_{1500}^{\rm MW}\approx3\times10^{-16}\,{\rm erg\,cm}^{-3}\AA^{-1}$ \citep[e.g.,][]{Porter2005,Maciel2013}, and write
\begin{equation}\label{eq:nrgdens}
u_{1500}(t)=3\times10^{-16}\,[\psi(t)/{\rm 1\,\Msun\,yr}^{-1}]\,{\rm erg\,cm}^{-3}\AA^{-1}\,.
\end{equation}
In practice, we meet the above condition by adjusting the inner radius of the ionized nebula in \cloudy, $r_{\rm in}$, such that
\begin{equation}\label{eq:lumtonrg}
\frac{L_{1500}^{\rm BC}+L_{1500,\ast}^{\rm ICM}(t)}{4\pi r_{\rm in}^2c}=u_{1500}(t),
\end{equation}
where $L_{1500}^{\rm BC}+L_{1500,\ast}^{\rm ICM}(t)$ is the 1500-\AA\ luminosity emitted by the birth clouds and stars older than 10\,Myr at time $t$  (equations~\ref{eq:lumbc} and \ref{eq:lumstaricm} below). The gas velocity dispersion and macroscopic velocity field in the intercloud medium, $\sigma_v^{\rm ICM}$ and $v^{\rm ICM}(r)$, can be defined independently of those in the birth clouds.

\begin{table}
\caption{Main adjustable parameters of stars and the ISM in the model of star-forming galaxies presented in Section~\ref{influenceism}. For the ISM, the parameters are `effective' ones describing galaxy-wide properties of gas and dust (see text for details).}
\begin{threeparttable}
	\centering
	\begin{tabular*}{0.45\textwidth}{l l}

\toprule
 Parameter  & Physical meaning  \\
\midrule
\multicolumn{2}{c}{Stars} \\
\midrule
$\psi(t)$        & Star formation rate as a function of time\\
$Z(t)$               &  Stellar metallicity as a function of time\\
\mup\            & Upper mass cutoff of the IMF\tnote{\it a} \\
\midrule
\multicolumn{2}{c}{Stellar birth clouds} \\
\midrule
\nh\  & Hydrogen number density \\
\Us\ &  Zero-age ionization parameter\tnote{\it b} \\
\zism\ & Interstellar metallicity [$\zism=Z(t)$ by default]\tnote{\it c} \\
$\xi_{d}$ &  Dust-to-metal mass ratio \\
\CO\  &  Carbon-to-oxygen abundance ratio \\
$\sigma_v$ &  Velocity dispersion of the gas \\
$v(r)$ & Macroscopic velocity field of the gas \\
\midrule
\multicolumn{2}{c}{Diffuse intercloud medium\tnote{\it d}} \\
\midrule
\nhicm\  & Hydrogen number density \\ 
\Nhicm\  & \hi\ column density \\ 
$\sigma_v^{\rm ICM}$ &  Velocity dispersion of the gas \\
$v^{\rm ICM}(r)$ & Macroscopic velocity field of the gas \\

\bottomrule
\end{tabular*}
\begin{tablenotes}
\item [{\it a}] The default IMF is that of \citet{Chabrier2003}.\\
\item [{\it b}] At the Str\"omgren radius, as defined by equations~\eqref{eq:us}--\eqref{eq:usdef}.\\
\item [{\it c}] By default, the interstellar metallicity in the birth clouds is taken to be the same as that of the youngest ionizing stars.\\
\item [{\it d}] By default, \zism, \xid\ and \CO\ in the intercloud medium are taken to be the same as those in the birth clouds.\\
\end{tablenotes}
\end{threeparttable}
\label{tab:summaryparams}
\end{table}

Table~\ref{tab:summaryparams} summarizes the main adjustable parameters of the stars and ISM in our model.  While idealized, this parametrization provides a means of exploring in a physically consistent way those features of the ISM that are expected to have the strongest influence on the ultraviolet spectra of star-forming galaxies. In practice, we compute the luminosity per unit wavelength emerging at time $t$ from a star-forming galaxy (equation~\ref{eq:lumgal}) as
\begin{equation}\label{eq:lumgaltrue}
L_{\lambda}(t) = T_{\lambda}^{\rm ICM}(t)\left[\,L_{\lambda}^{\rm BC}+L_{\lambda,\ast}^{\rm ICM}(t)\right],
\end{equation}
where
\begin{equation}\label{eq:lumbc}
L_{\lambda}^{\rm BC}=\int _{0}^{\rm 10\,Myr}d\tprime\, \psi (t-\tprime) \, S_{\lambda}[\tprime, Z(t-\tprime)]\,T_{\lambda}^{\rm BC}(\tprime)
\end{equation}
is the luminosity emerging from the birth clouds and
\begin{equation}\label{eq:lumstaricm}
L_{\lambda,\ast}^{\rm ICM}(t)=\int _{\rm 10\,Myr}^{t}d\tprime\, \psi (t-\tprime) \, S_{\lambda}[\tprime, Z(t-\tprime)]
\end{equation}
is the emission from stars older than 10\,Myr.

\subsubsection{Examples of model spectra}\label{sec:ismmodelsex}

\begin{figure*}
\includegraphics{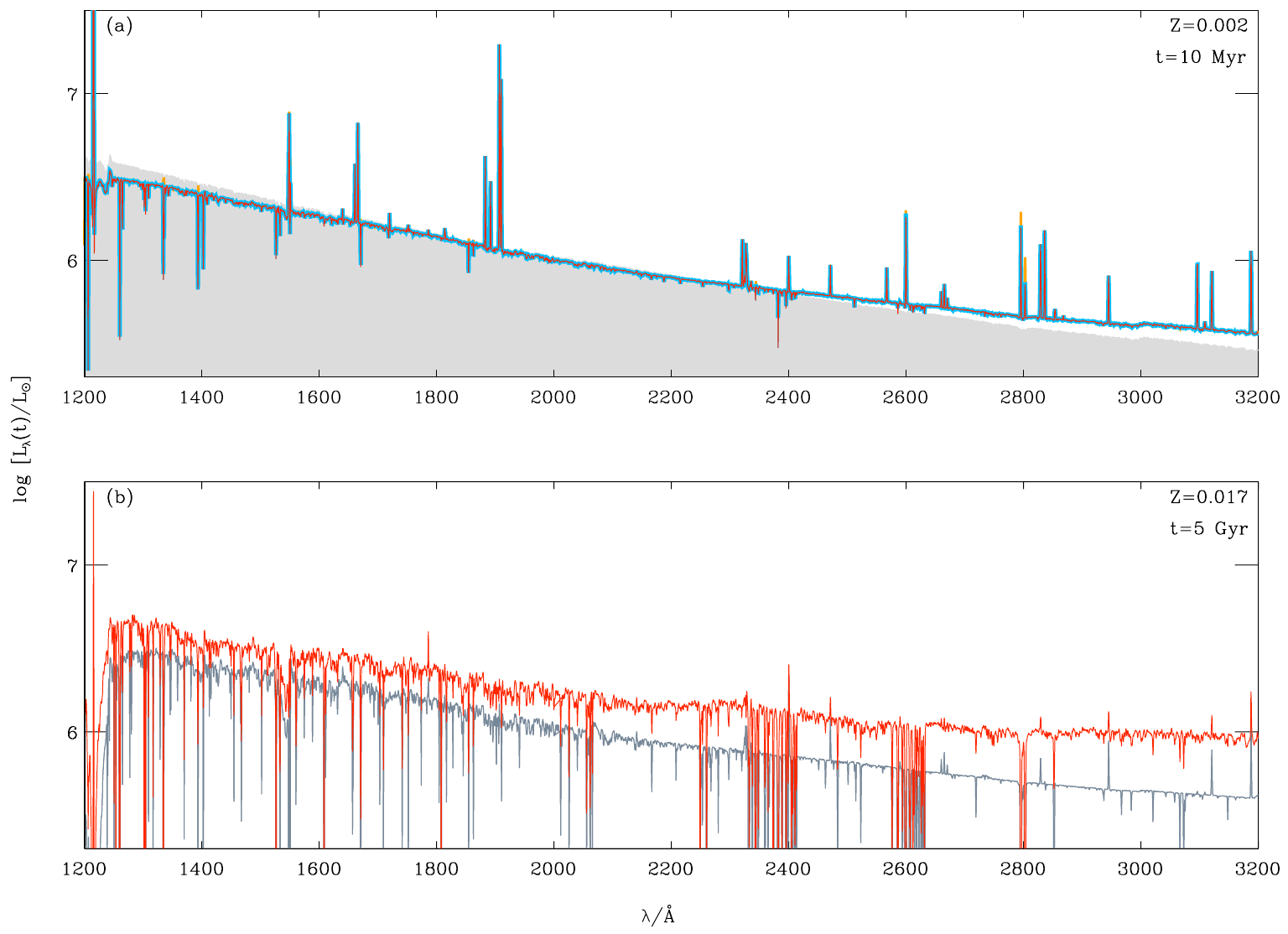}
\caption{Examples of ultraviolet spectra of star-forming galaxies computed using the model described in Section~\ref{sec:ismmodelsap}: (a) low-metallicity ($Z=0.002$), young galaxy model with constant star formation $\psi(t)=1\,\Msun{\rm yr}^{-1}$ for $t=10\,$Myr. The different spectra shown are that produced by stars (solid grey swath), the spectrum output by \cloudy\ when used in a standard way to compute the transmission function of ionized gas (gold curve), the same spectrum after accounting with \cloudspec\ for interstellar-line absorption in the ionized gas (thick light-blue curve) and the spectrum emerging from the outer \hi\ boundaries of birth clouds (red curve); (b) metal-rich ($Z=0.017$), mature galaxy model with constant star formation $\psi(t)=1\,\Msun{\rm yr}^{-1}$ for $t=5\,$Gyr and an intercloud medium. The two spectra shown are the contribution by stellar birth clouds to the emergent luminosity (grey curve) and the total luminosity emerging from the galaxy (in red). See text Section~\ref{sec:ismmodelsex} for more details about the ISM parameters of these models.}
\label{fig:oldyoung}
\end{figure*}

Fig.~\ref{fig:oldyoung} shows two examples of ultraviolet spectra of star-forming galaxies computed using the model described in Section~\ref{sec:ismmodelsap} above, corresponding to a young ($t=10\,$Myr), low-metallicity ($Z=0.002$) galaxy (Fig.~\ref{fig:oldyoung}a) and a mature ($t=5\,$Gyr), more metal-rich ($Z=0.017$) galaxy (Fig.~\ref{fig:oldyoung}b). Since all the stars in the young-galaxy model are still embedded in their birth clouds, we do not include any intercloud medium in this case. For simplicity, in both models, we adopt constant star formation rate, an IMF upper mass cutoff  $\mup=100\,\Msun$, and we assume that all stars in the galaxy have the same metallicity $Z$. In our approach, the absolute star formation rate must be specified only to determine the rate of ionizing photons striking the intercloud medium in the mature-galaxy model, where we take $\psi(t)=1\,\Msun{\rm yr}^{-1}$. For the ISM, we adopt in both models $\xid=0.3$, $\CO=\COsol$, $\sigma_v=50\,{\rm km\,s}^{-1}$  and static birth clouds ($v=0$) with $\nh=100\,{\rm cm}^{-3}$. This value of $\sigma_v$ is intermediate between the typical velocity dispersion of ionized gas in nearby dwarf galaxies \citep[$\sim$20--30\,km\,s$^{-1}$;][]{Moiseev2015} and in the Milky-Way ISM \citep[$\sim$50--110\,km\,s$^{-1}$;][see also \citealt{Leitherer2011}]{Savage2000}. For reference, the spectral resolution of the stellar population synthesis model of Section~\ref{sec:stelpops} ranges roughly from 300\,km\,s$^{-1}$ (full width at half maximum) at $\lambda=1000\,\AA$ (resolving power $R\approx1000$) to 100\,km\,s$^{-1}$ at $\lambda=3000\,\AA$ ($R\approx3000$). We also need to specify the zero-age ionization parameter of the birth clouds. We adopt $\log\Us=-2.0$ for the young-galaxy model, i.e. the typical value observed by \citet{Stark2014} in four low-metallicity dwarf star-forming galaxies at redshift $z\sim2$, and $\log\Us=-3.5$ for the mature-galaxy model. This is consistent with the dependence of \Us\ on \zism\ identified by Carton et al. (in preparation; see equation~25 of \citealt{Chevallard2016}) in a large sample of SDSS galaxy spectra.

To isolate the effect of each model component on the predicted spectrum of the young-galaxy model in Fig.~\ref{fig:oldyoung}a, we show four spectra: the spectrum produced by stars (solid grey swath);  the spectrum output by \cloudy\ when used in a standard way to compute the transmission function of ionized gas [gold curve; this corresponds to, for example, the transmission function $T_{\lambda}^{+}(\tprime)$ computed by \citealt{gutkin_modelling_2016}]; the same spectrum after accounting with \cloudspec\ for interstellar-line absorption in the ionized gas [thick light-blue curve; this corresponds to the transmission function $T_{\lambda}^{\rm HII}(\tprime)$ in equation~\ref{eq:transBC_a}]; and the spectrum emerging from the outer \hi\ boundaries of birth clouds [red curve; corresponding to the full transmission function $T_{\lambda}^{\rm BC}(\tprime)$ in equation~\ref{eq:transBC_a} and computed using equation~\ref{eq:lumbc}]. The three spectra transferred through the ionized gas lie above the input stellar population spectrum at wavelengths $\lambda\gtrsim2500\,$\AA\ in Fig.~\ref{fig:oldyoung}a. This is because of the contribution by  Balmer-recombination continuum photons produced in the nebula. At shorter wavelengths, the transferred radiation is fainter than that of the input stellar population because of absorption by dust, whose optical depth increases blueward, from $\tau_\lambda\approx0.2$ at $\lambda=3200\,$\AA\ to $\tau_\lambda\approx0.4$ at $\lambda=1200\,$\AA. We note that, for the adopted high $\log\Us=-2.0$, the bulk of this absorption occurs in the ionized interiors of birth clouds. This is because the hydrogen column density of the ionized region scales roughly as $\epsilon\nh R_{\rm S}\propto\Us$; hence the dust optical depth scales also as \Us\ at fixed dust-to-gas ratio $\xid\zism$ (see also \citealt{Mathis1986}; CL01). For reference, in this low-metallicity, young-galaxy model, the \hii\ and \hi\ column densities of a new birth cloud (i.e. corresponding to an SSP age $\tprime=0$ in equation~\ref{eq:transBC_a}) are $\sim3\times10^{21}{\rm cm}^{-2}$ and $\sim5\times10^{19}{\rm cm}^{-2}$, respectively, implying a much thinner \hi\ envelope (out to a kinetic gas temperature of 50\,K) than the interior \hii\ region.

A most remarkable feature of Fig.~\ref{fig:oldyoung}a is the prominence of interstellar absorption lines arising from the ionized region, as revealed by the post-processing with \synspec\ of  \cloudy\ output using the \cloudspec\ tool (thick light-blue curve). Among the deepest absorption lines are those corresponding to the \siliia, \oi, \siliib, \ciib, \silivd, \siliic, \siliidast, \civd\ and \alii\ transitions. Accounting for this absorption is important, as it can influence the net emission luminosity of lines emerging from the ionized gas, as shown by the difference between the gold and light-blue spectra at the wavelengths of, for example, the \ciib, \silivd\ and \mgiiline\ lines in Fig.~\ref{fig:oldyoung}a. The difference between the light-blue and red spectra further shows that, in this example, only a modest contribution to interstellar absorption arises from the thin neutral envelopes of birth clouds. Aside from \lya, the strongest emission lines in the emergent spectrum are \civd, \oiiid, \siliiid\ and mainly \ciiid, with equivalent widths of 2.5, 2.8, 2.0 and 21.3\,\AA, respectively.

In the case of the mature-galaxy model, we must also specify the typical \hi\ column density seen by photons in the intercloud medium. We adopt $\Nhicm=5\times10^{20}{\rm cm}^{-2}$, typical of the \hi\ column densities observed through the discs of nearby spiral galaxies \citep[e.g.,][]{Warmels1988}.  Furthermore, for simplicity, we assume that gas in the intercloud medium is static  ($v^{\rm ICM}=0$), with a velocity dispersion $\sigma_v^{\rm ICM}=\sigma_v=50\,{\rm km\,s}^{-1}$. In Fig.~\ref{fig:oldyoung}b, we show two spectra for this model:  the contribution by stellar birth clouds to the emergent luminosity [grey curve; this corresponds to the term $T_{\lambda}^{\rm ICM}(t)\,L_{\lambda}^{\rm BC}$ in equation~\eqref{eq:lumgaltrue}]; and the total luminosity emerging from the galaxy [in red; corresponding to the quantity $L_{\lambda}(t)$ in equation~\eqref{eq:lumgaltrue}]. The birth clouds in this high-metallicity, low-ionization ($\log\Us=-3.5$) model have zero-age \hii\ and \hi\ column densities of $\sim$ $1\times10^{20}{\rm cm}^{-2}$ and $\sim$ $2\times10^{20}{\rm cm}^{-2}$, respectively. This combination yields a dust absorption optical depth similar to that found for the young-galaxy model above (with $\tau_\lambda\approx0.3$ at $\lambda_{1500}$), but with most of the absorption now occurring in the thick neutral envelopes of the clouds. From Fig.~\ref{fig:oldyoung}b, we find that the birth clouds contribute from nearly 80 per cent of the total ultraviolet emission of the galaxy at 1500\,\AA, to less than 50 per cent at 3200\,\AA. The emission lines in the emergent spectrum are much weaker than in the young-galaxy model of Fig.~\ref{fig:oldyoung}a. This is partly because of the higher \zism\ and lower \Us\ \citep[e.g.,][]{gutkin_modelling_2016}, but also because of strong interstellar absorption. 

The total emergent spectrum of the mature-galaxy model in Fig.~\ref{fig:oldyoung}b exhibits much stronger absorption features than that of the young-galaxy model in Fig.~\ref{fig:oldyoung}a. This is the case for all the interstellar lines mentioned previously, but also additional ones, such as for example \feiia, \siliie\ and several clusters of metallic lines (involving \lcoii, \lcrii, \lfeii, \lmgi, \lmnii, \lznii\ and other species) around 2050, 2400 and 2600\,\AA. While most of the absorption in high-ionization lines, such as \nvd\ and \civd, occurs in the hot interiors of the birth clouds, the intercloud medium dominates absorption in low-ionization lines, such as \oi, \siliib, \ciib, \feiia, \alii, \siliie, a blend of \lznii, \lcrii, \lcoii\ and \lmgi\ lines at $\lambda=2026\,\AA$, \feiic, \feiid, \feiilinec\ and \mgiiline. For other lines, such as \siliia\ and \mgiline, the neutral envelopes of the birth clouds can contribute up to 30 per cent of the total absorption. We note that interstellar absorption suppresses entirely emission in the \feiilinec\ and \mgiiline\ lines in the emergent spectrum of this model. The few emission lines standing out of the continuum in Fig.~\ref{fig:oldyoung}b are \feiilinea\ (`UV 191' multiplet, with an equivalent width of only 0.44\,\AA), \feiilineb, \oiib, \heilinea, \heilineb, \heilinec\ and \heilined. We also note the deep \lya\ absorption feature at 1216\,\AA, which arises from the resonant scattering of line photons through neutral hydrogen both in the birth-cloud envelopes and in the intercloud medium. For reference, the dust absorption optical depth in the intercloud medium is similar to that in the birth clouds (with $\tau_\lambda\approx0.4$ at $\lambda_{1500}$). The warm, largely neutral intercloud medium produces only 1.2 per cent of the total H\,$\alpha$ emission from the galaxy (the total emergent H\,$\alpha$ equivalent width being of the order of 130\,\AA).

Hence, the simple approach outlined in Section~\ref{sec:ismmodelsap} to model the influence of the ISM on starlight, while idealized, provides a unique means of exploring in a quantitative and physically consistent way the competing effects of stellar absorption, nebular emission and interstellar absorption in star-forming galaxy spectra. The prominent emission- and absorption-line features identified in the example ultraviolet spectra of a young, low-metallicity galaxy and a mature, more metal-rich galaxy in Fig.~\ref{fig:oldyoung} are commonly observed in high-quality spectra of star-forming galaxies at various redshifts \citep[e.g.,][]{Pettini2000,Shapley2003,Lebouteiller2013,Lefevre2013,James2014,Stark2014,Patricio2016}. In the following subsections, we use this simple yet physically consistent model to identify those ultraviolet spectral features individually most sensitive to the properties of stars, the ionized and the neutral ISM in star-forming galaxies.

\subsection{Ultraviolet tracers of stars and the ISM}\label{sec:idxselection}

The modeling approach presented in Section~\ref{sec:ismmodels} above offers a valuable means of identifying the features most sensitive to stars, the ionized and the neutral ISM in the spectra of star-forming galaxies. In Section~\ref{sec:idxstars} below, we start by exploring the influence of the ISM on ultraviolet spectral indices commonly used to constrain the ages and metallicities of stellar populations. Then, we use our model to identify potentially good independent tracers of nebular emission (Section~\ref{sec:nebem}) and interstellar absorption (Section~\ref{sec:ismreg}). We also investigate the complex dependence of well-studied lines, such as \ov, \silivd\ (hereafter simply \silivt), \civd\ (hereafter simply \civ), \heii, and \niv, on stellar absorption, nebular emission and interstellar absorption.

\begin{figure*}\includegraphics{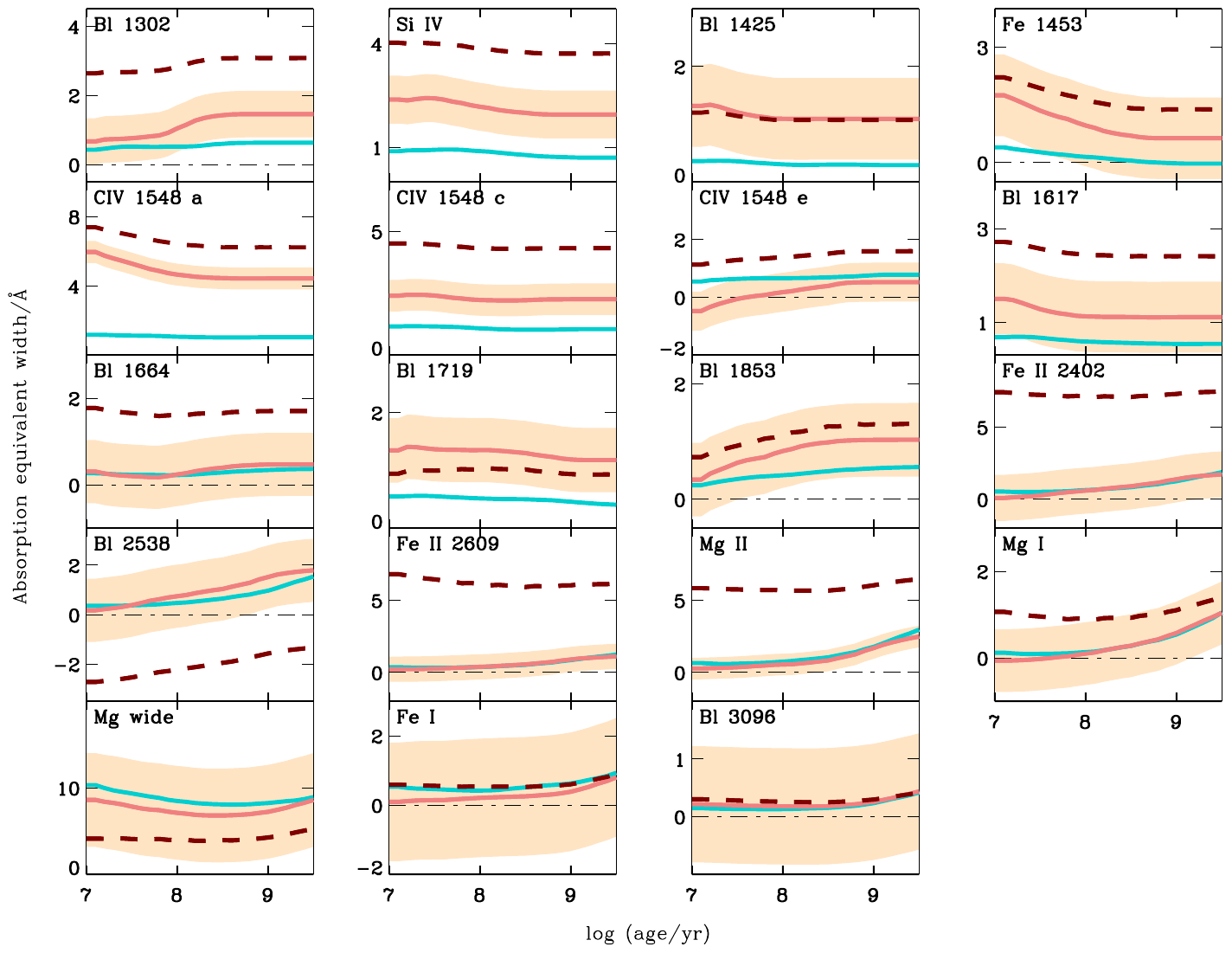}
\caption{Strengths of the 19 \citet{Fanelli1992} ultraviolet spectral indices defined in Table~\ref{tab:tabindices} plotted against age, for stellar populations with constant star formation rate and metallicities $Z=0.002$ (teal blue curve) and 0.017 (salmon curve). The filled cream area around the $Z=0.017$ model indicates the median index measurement errors in the {\it IUE} spectra of the LMC clusters observed by \citet{Cassatella1987}. The brown dashed curve shows the effect of accounting for nebular emission and interstellar absorption in the $Z=0.017$ model (see text Section~\ref{sec:idxselection} for details).}
\label{fig:indicesaffage}
\end{figure*}

\subsubsection{Features tracing young stars}\label{sec:idxstars}

Fig.~\ref{fig:indicesaffage} shows the strengths of the 19 \citet{Fanelli1992} indices studied in Sections~\ref{sec:stelpops} and \ref{sec:uvinterpret} (Fig.~\ref{fig:indmet}) measured in the emergent spectrum of the mature-galaxy model of Section~\ref{sec:ismmodelsex} at ages between $1\times10^7$ and $3\times10^9\,$yr (brown dashed curve in each panel). Also shown for comparison in each panel is the index strength measured in the pure stellar population spectrum, i.e., before including the effects of nebular emission and interstellar absorption (salmon curve). For most indices, the brown curve lies above the salmon one in Fig.~\ref{fig:indicesaffage}, indicating that interstellar absorption in the central index bandpass deepens the stellar absorption feature. For others, such as \bl1719, \bl2538 and \mgw, the behaviour is opposite and caused in general, in this example, by contamination of a pseudo-continuum bandpass by interstellar absorption rather than of the central bandpass by nebular emission. A way to gauge the impact of ISM contamination on the interpretation of ultraviolet stellar absorption features is to compare the offsets between the brown and salmon curves in Fig.~\ref{fig:indicesaffage} to typical uncertainties in index-strength measurements. In each panel, the filled cream area around the solid salmon curve shows the median measurement uncertainty from Table~\ref{tab:ews} for the 10 LMC star clusters observed by \citet{Cassatella1987}. Indices for which the salmon curve lies within the cream area may therefore be considered a priori as those whose interpretation will be least affected by ISM contamination. In this example, these include \bl1425, \fe, \bl1719, \bl1853, \mgw, \fei\ and \bl3096. In contrast, \lciv-based indices (\civa, \civc, \cive), \feii2402, \bl2538, \feii2609 and \mgii\ appear to be most highly affected by ISM contamination.

It is also of interest to compare the impact of ISM contamination and that of stellar metallicity on the index strengths in Fig.~\ref{fig:indicesaffage}. This is illustrated by the teal-blue curve, which shows in each panel the index strength measured in the pure stellar population spectrum of a model with constant star formation rate and metallicity $Z=0.002$. For most indices, the difference between the teal-blue and salmon curves is smaller than that between the salmon and brown curves, indicating that ISM contamination can affect the index strength more than a change by a factor of 8.5 in metallicity. When this is not the case, the change in index strength induced by such a change in  metallicity is generally smaller than the typical observational error (cream area in Fig.~\ref{fig:indicesaffage}; as in the case of \fe\ and \bl1617). Only for \bl1425, \civa\ and \bl1719 does the effect of metallicity seem measurable and stronger than that of ISM contamination. Among those, \bl1425 and \bl1719 appear to be the least sensitive to ISM contamination, and hence, the most useful to trace stellar population properties.

\begin{figure}\includegraphics[width=\columnwidth]{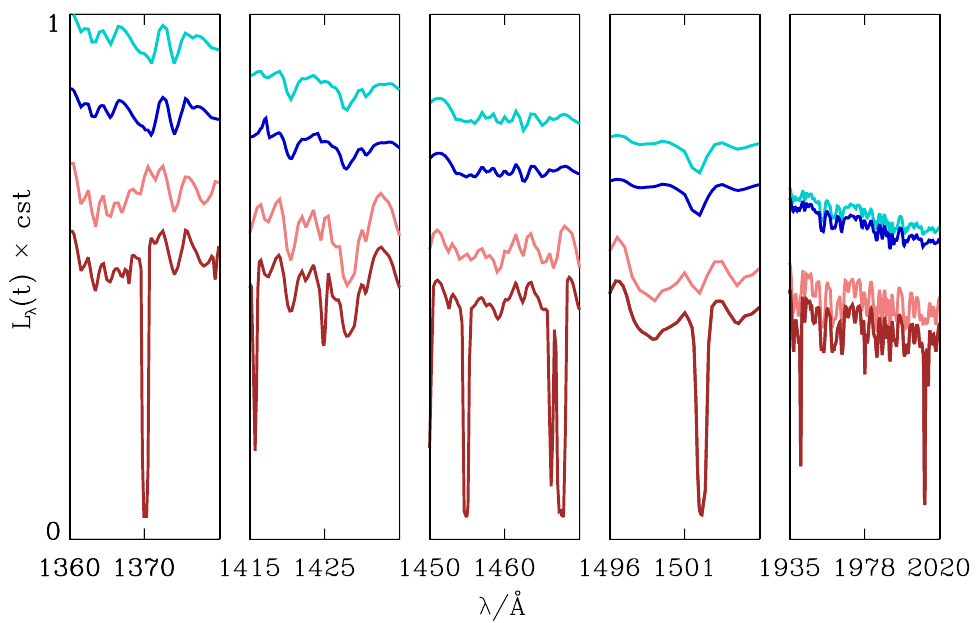}
\caption{Emergent spectra of the young- and mature-galaxy models of Fig.~\ref{fig:oldyoung} (dark-blue and brown curves, respectively), along with the input stellar population spectra of these models (light-blue and salmon curves), in the five spectral windows defining the metallicity-sensitive indices `1370', `1425', `1460', `1501' and `1978' \citep{Leitherer2001,Rix2004,Sommariva2012}.}\label{fig:sommariva}
\end{figure}

An alternative to the \citet{Fanelli1992} set of indices was proposed by \citet[][see also \citealt{Rix2004,Sommariva2012}]{Leitherer2001}, who explored metallicity indicators in the ultraviolet spectra of stellar population synthesis models computed using the Starburst99 code \citep{Leitherer1999}. Unlike \citet{Fanelli1992}, who define indices by measuring the fluxes in a central bandpass and two pseudo-continuum bandpasses in a spectrum, \citet{Leitherer2001} firstly normalize the spectrum to unit continuum through the division by a spline curve fitted to sections identified as free of stellar absorption lines in the model spectra. Then, they define line indices as the equivalent widths integrated in specific spectral windows of the normalized spectrum. Metallicity-sensitive indices defined in this way include the `1370' (1360--1380\,\AA) and `1425' (1415--1435\,\AA) indices of \citet{Leitherer2001}, the `1978' (1935--2020\,\AA) index of \citet{Rix2004} and the `1460' (1450--1470\,\AA) and `1501' (1496--1506\,\AA) indices of \citet{Sommariva2012}. Fig.~\ref{fig:sommariva} shows the emergent spectra of the young- and mature-galaxy models of Section~\ref{sec:ismmodelsex} (dark-blue and brown curves, respectively), along with the input stellar population spectra of these models (light-blue and salmon curves), in the five spectral windows defining these metallicity-sensitive indices. The low-metallicity, young galaxy model shows no strong nebular-emission nor interstellar-absorption line in any of the windows. However, interstellar absorption can contaminate the indices at high metallicity, most notably Ni\,\textsc{ii}\,$\lambda1370$ for 1370, Ni\,\textsc{ii}\,$\lambda1416$, Co\,\textsc{ii}\,$\lambda1425$ and S\,\textsc{i}\,$\lambda1425$ for 1425, Ni\,\textsc{ii}\,$\lambda1455$, Co\,\textsc{ii}\,$\lambda1466$ and Ni\,\textsc{ii}\,$\lambda\lambda1467,1468$ for 1460, Ni\,\textsc{ii}\,$\lambda1502$ for 1501 and Co\,\textsc{ii}\,$\lambda1941$, Si\,\textsc{i}\,$\lambda1978$ and Co\,\textsc{ii}\,$\lambda2012$ for 1978. In the example of Fig.~\ref{fig:sommariva}, contamination appears to be least severe for 1425, confirming our previous finding with regard to the \bl1425 index of \citet{Fanelli1992}. For the other indices, the potential influence of the ISM should be kept in mind when interpreting index strengths measured in observed galaxy spectra, especially at high metallicity.

\subsubsection{Features tracing nebular emission}\label{sec:nebem}

\begin{figure*}\includegraphics{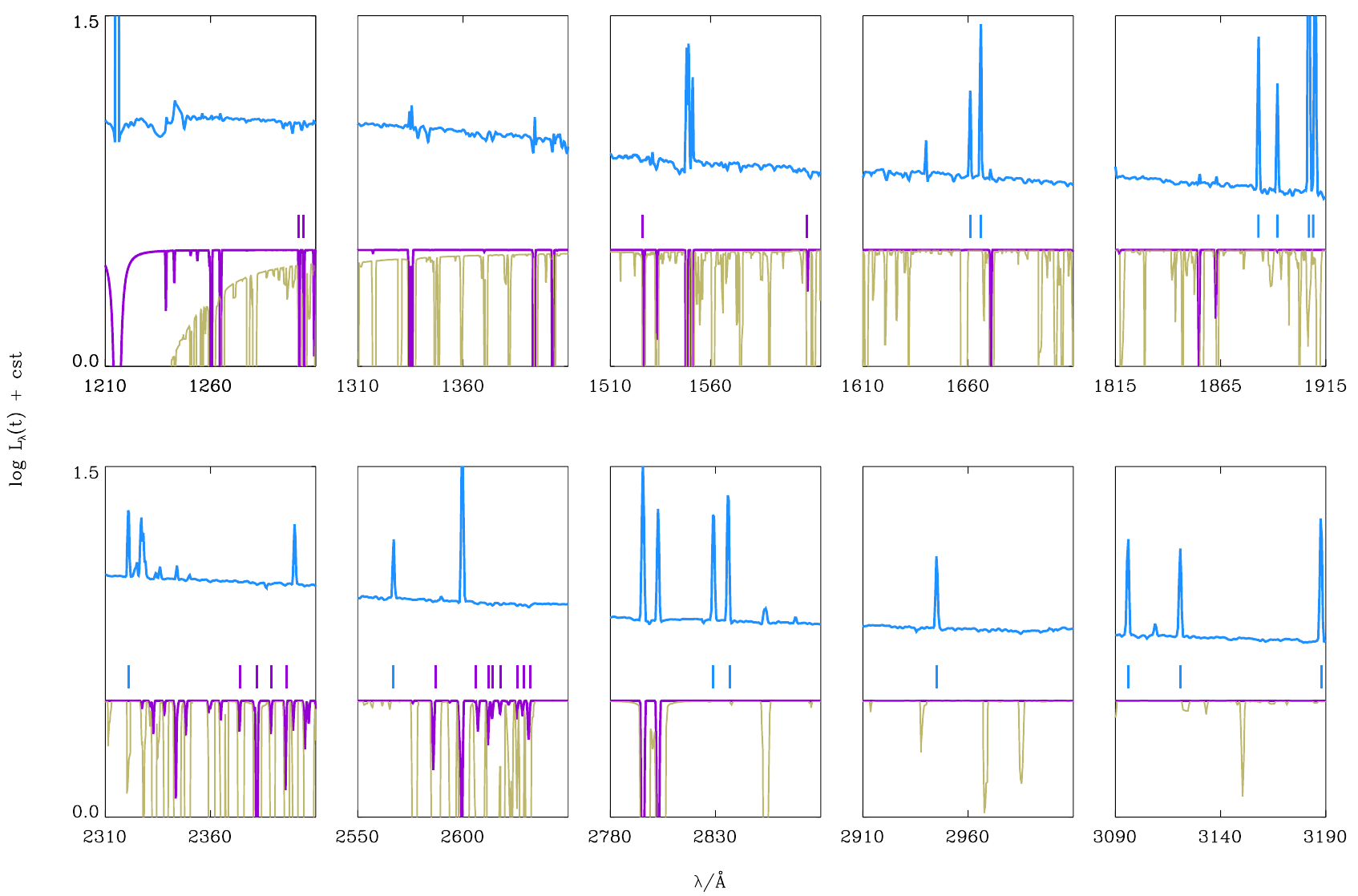}
\caption{Focus on spectral sections exhibiting the strongest features (marked by vertical bars) in the emission component (blue curve; corresponding to the gold curve in Fig.~\ref{fig:oldyoung}a) and the transmission function of the ISM (purple curve) in the 10\,Myr-old, $Z=0.002$ young-galaxy model of Fig.~\ref{fig:oldyoung}a. Also shown for reference is the transmission function of the ISM in the mature-galaxy model of Fig.~\ref{fig:oldyoung}b (olive-green curve).}
\label{fig:contamination}
\end{figure*}

We now wish to identify those nebular emission lines least contaminated by stellar and interstellar absorption in the ultraviolet spectrum of a star-forming galaxy. To this end, we plot in Fig.~\ref{fig:contamination} the emission component of the 10\,Myr-old galaxy spectrum of Fig.~\ref{fig:oldyoung}a (blue curve; corresponding to the gold curve in Fig.~\ref{fig:oldyoung}a), along with the transmission function of the absorbing ISM in this low-metallicity model (purple curve). Also shown for reference in Fig.~\ref{fig:contamination} is the transmission function of the ISM in the mature-galaxy model of Fig.~\ref{fig:oldyoung}b (olive-green curve), which unveils the locations of many more absorption features appearing at high metallicity. We choose to focus on the strongest emission and absorption features and do not show in Fig.~\ref{fig:contamination} the wavelength intervals 1410--1510, 1710--1815, 1915--2310, 2410-2550, 2650--2780, 2880--2910 and 3010--3080\,\AA, where only a few weak lines appear in the young-galaxy spectrum of Fig.~\ref{fig:oldyoung}a.

A simple comparison of the blue and purple spectra in Fig.~\ref{fig:contamination} reveals the strongest emission lines expected to be least sensitive to interstellar absorption. These are marked by blue vertical bars and reported in Table~\ref{tab:summarylines}. These features include emission lines routinely observed in the ultraviolet spectra of high-redshift star-forming galaxies, such as \oiiid, \siliiid\ and \ciiid\ \citep{Shapley2003,Erb2010,Christensen2012,Stark2014,Stark2015a}. We note the appearance at high metallicity of an absorption feature (dominated by C\,\textsc{i}) just blueward of O\,\textsc{iii}]\,$\lambda1661$, one (dominated by Ti\,\textsc{ii}) underneath C\,\textsc{iii}]\,$\lambda1907$ and another (dominated by Ni\,\textsc{i}) underneath \oiiib, but these ultraviolet emission lines are anyway expected to be intrinsically much weaker, and hence difficult to detect, at high metallicity \citep{gutkin_modelling_2016}. We refer the reader to \citet{gutkin_modelling_2016} for a comprehensive study on the dependence of the luminosities of the emission lines in Table~\ref{tab:summarylines} on the parameters of stars and stellar birth clouds in Table~\ref{tab:summaryparams}. These and other ultraviolet emission lines can also help discriminate between star formation and nuclear activity in galaxies \citep{Feltre2016}. We further note that none of the emission lines listed in Table~\ref{tab:summarylines} is strongly affected by absorption originating in stellar winds and photospheres \citep[e.g.,][]{Leitherer2011}.

\begin{table}
\caption{Uncontaminated tracers of nebular emission and interstellar absorption in the ultraviolet spectra of star-forming galaxies.}
\begin{threeparttable}
	\centering
	\begin{tabular*}{0.47\textwidth}{c}
\toprule
{Nebular emission lines} \\
\midrule
\oiiid; \,\, \siliiid; \\
\ciiid; \,\, \oiiib; \,\, Fe\,\textsc{iv}\,$\lambda2567$; \\
Fe\,\textsc{iv}\,$\lambda2829$+\heilinea; \ciilinea; \,\, \heilineb; \,\, \\
\feivlined; \,\, \heilinec; \,\, \heilined \\
\midrule
{Interstellar absorption lines} \\
\midrule
\oi; \,\, \siliib; \,\, \siliic; \\
\feiia; \,\, \feiie; \,\, \feiic; \,\, \feiibla; \\
 \feiid; \,\, \mniiline; \,\, \feiiblb\ blend\\
\bottomrule
\end{tabular*}
\end{threeparttable}
\label{tab:summarylines}
\end{table}

\subsubsection{Features tracing interstellar absorption}\label{sec:ismreg}

Fig.~\ref{fig:contamination} is also useful to identify the strongest interstellar absorption lines expected to be least sensitive to contamination by nebular emission and stellar absorption in the ultraviolet spectrum of a star-forming galaxy. These are marked by purple vertical bars and also reported in Table~\ref{tab:summarylines}. Among these, \oi, \siliib, \siliic, \feiia, \feiie\ and \feiic\ have all been observed in the spectra of high-redshift star-forming galaxies \citep{Pettini2002,Shapley2003,Erb2010,Christensen2012,Stark2014,Stark2015a}. We do not include in this list \siliia, \siliiasta, \ciib, \siliidast, \alii, \aliii, \feiilinec, \mgiiline\ and \mgiline, which are also observed at high redshift but can be contaminated by nebular emission. Moreover, the strong \silivt\  and \civ\ lines, which would be valuable tracers of the ionized ISM, are contaminated by stellar-wind features (these lines are discussed further in the next subsection). We note that \citet[][their table~7]{Leitherer2011} propose useful definitions of the ultraviolet spectral indices centred on \oi+\siliib, \siliic, \feiia\ and \feiie\ in Table~\ref{tab:summarylines}, based on a study of stellar and ISM features in the spectra of 28 local starburst and star-forming galaxies. 

\subsubsection{Important composite features}\label{sec:mixed}

\begin{figure}\includegraphics[width=\columnwidth]{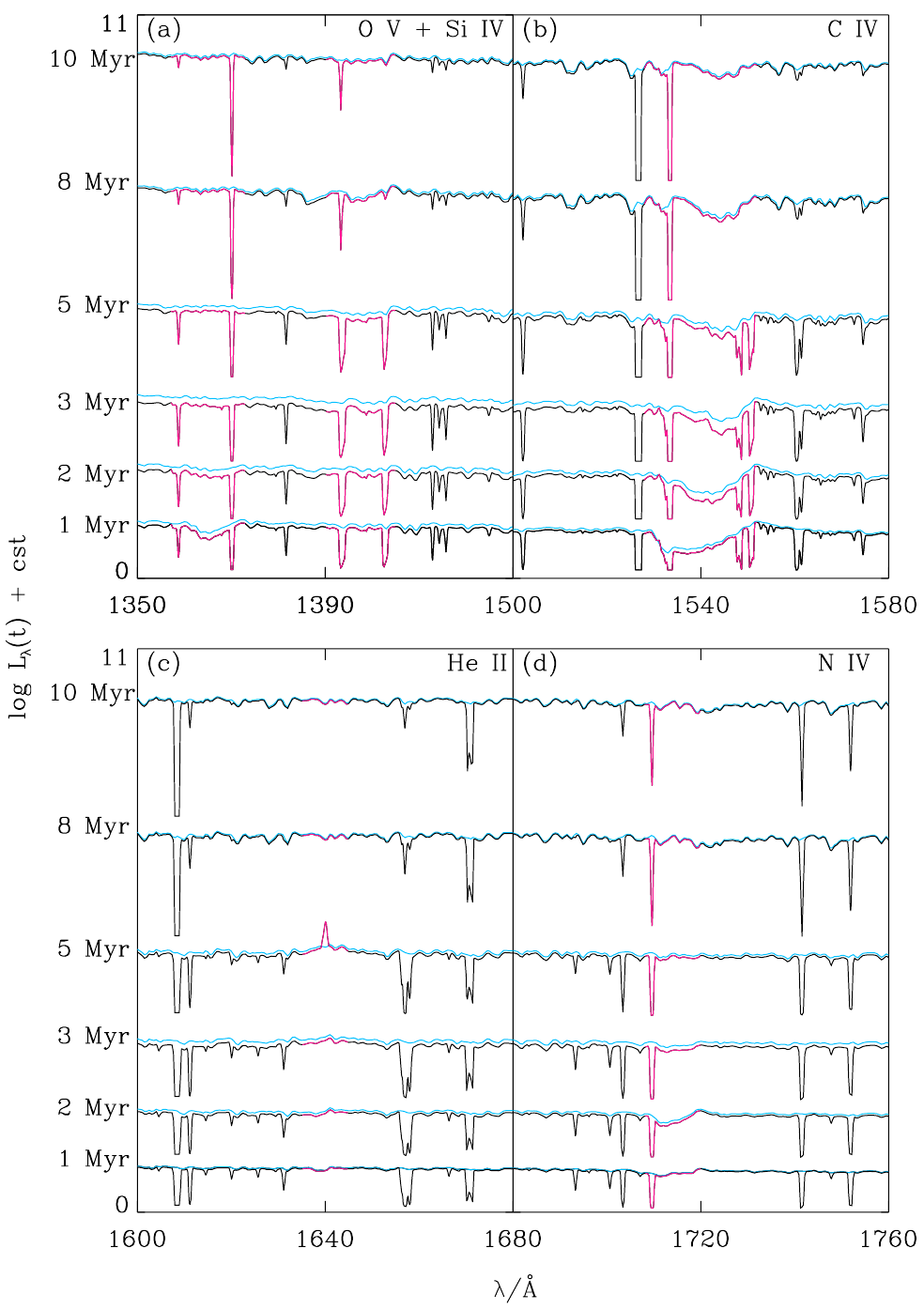}
\caption{Dependence of selected features on SSP age $\tprime$ in the spectrum emerging from a stellar birth cloud with metallicity $Z=\zism=0.017$, for $\tprime=1$, 2, 3, 5, 8 and 10\,Myr (as indicated). The birth-cloud parameters are the same as in the mature-galaxy model of Fig.~\ref{fig:oldyoung}b. In each panel and for each \tprime, the light-blue curve shows the input SSP spectrum and the black one the emergent spectrum, with the features of interest highlighted in pink: (a) \ov+\silivt; (b) \civ; (c) \heii; and (d) \niv.}
\label{fig:mixedlines_ssp}
\end{figure}

\begin{figure}\includegraphics[width=\columnwidth]{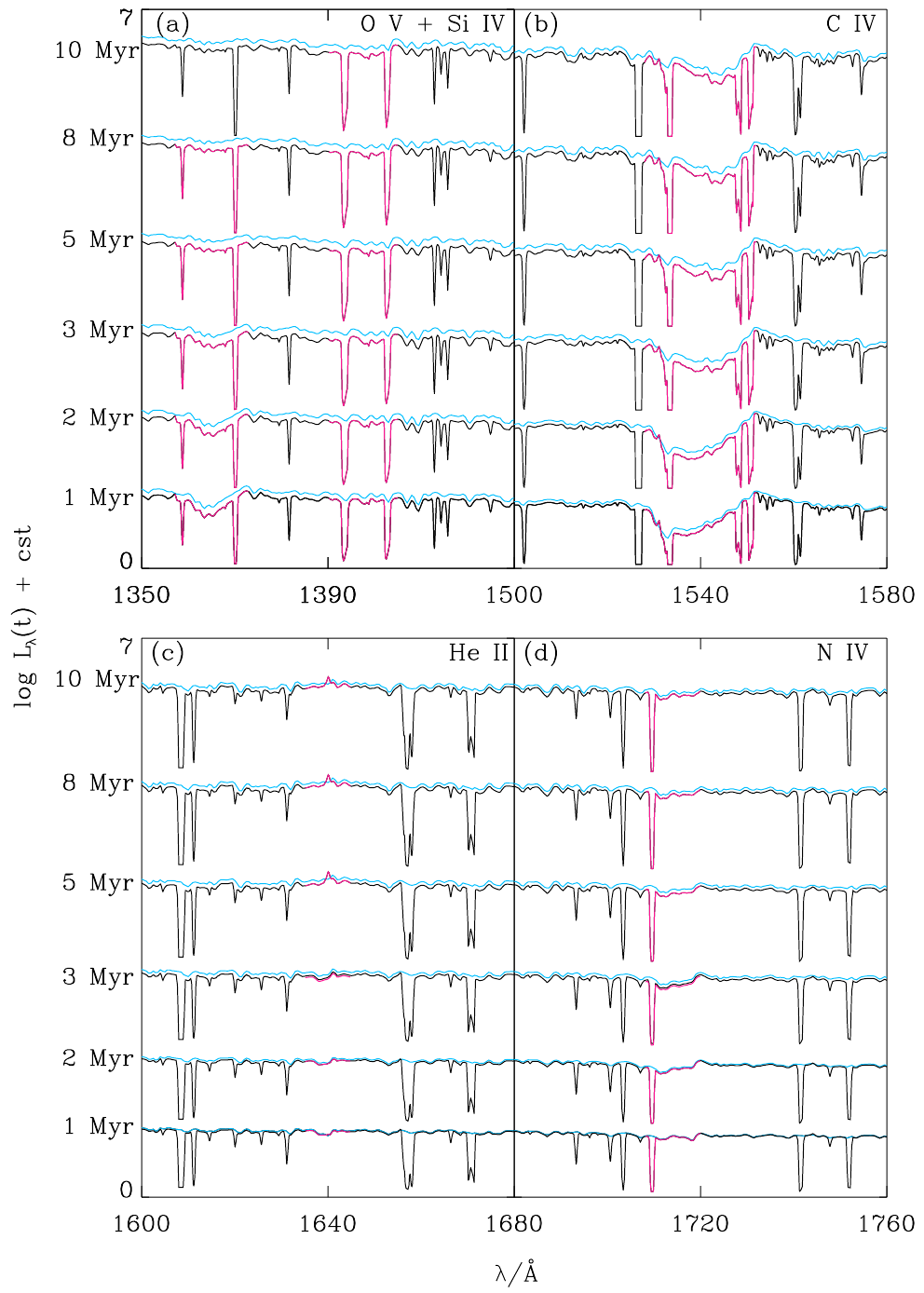}
\caption{Same as Fig.~\ref{fig:mixedlines_ssp}, but for a young-galaxy model with constant star formation rate and $Z=0.017$.}
\label{fig:mixedlines_csf}
\end{figure}

\begin{figure}\includegraphics[width=\columnwidth]{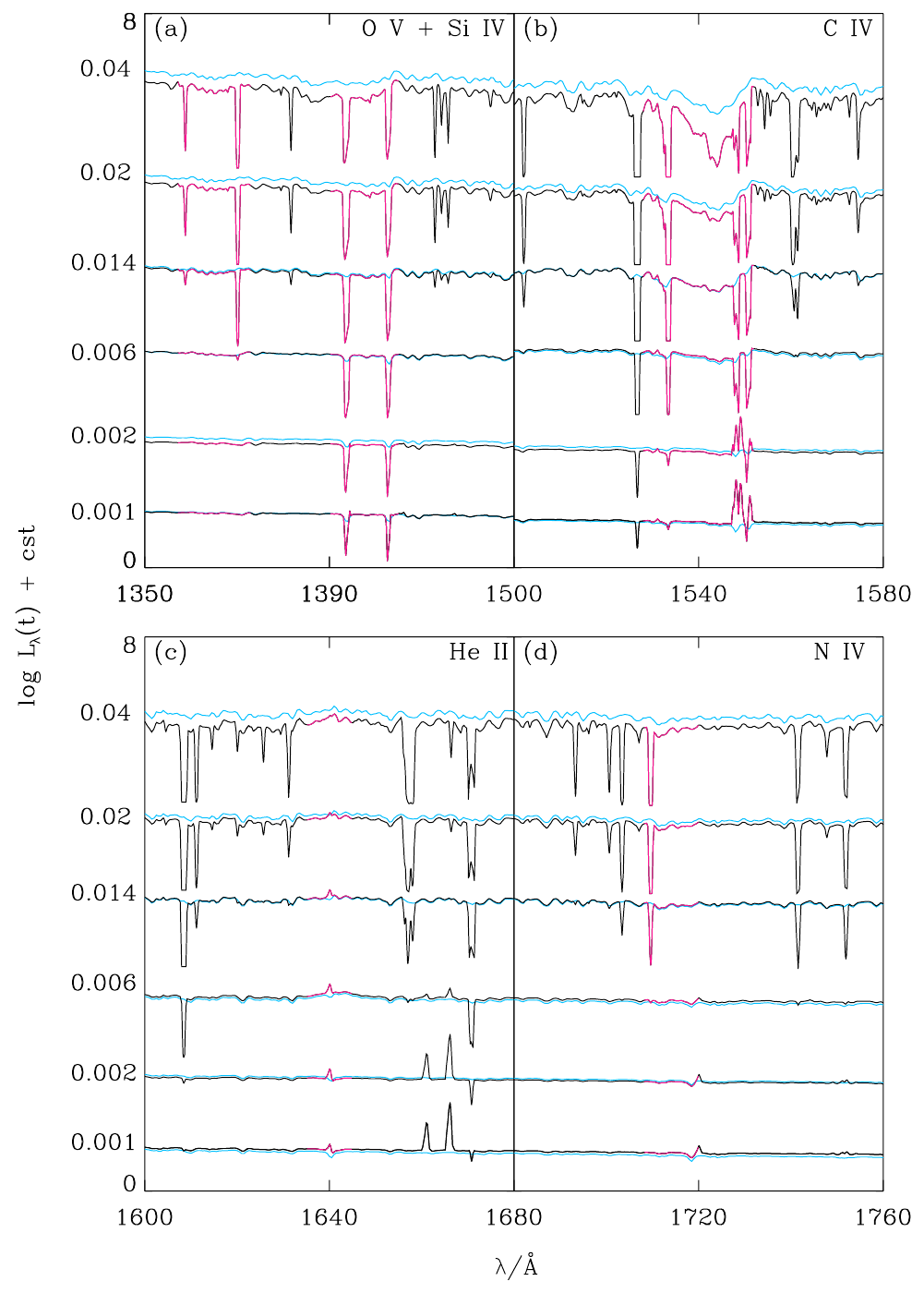}
\caption{Analog of Fig.~\ref{fig:mixedlines_csf} for young-galaxy models with constant star formation rate and metallicities $Z=0.001$, 0.002, 0.006, 0.014, 0.02 and 0.04, at fixed age $\tprime=10\,$Myr.}
\label{fig:mixedlines_met}
\end{figure}

\begin{figure}\includegraphics[width=\columnwidth]{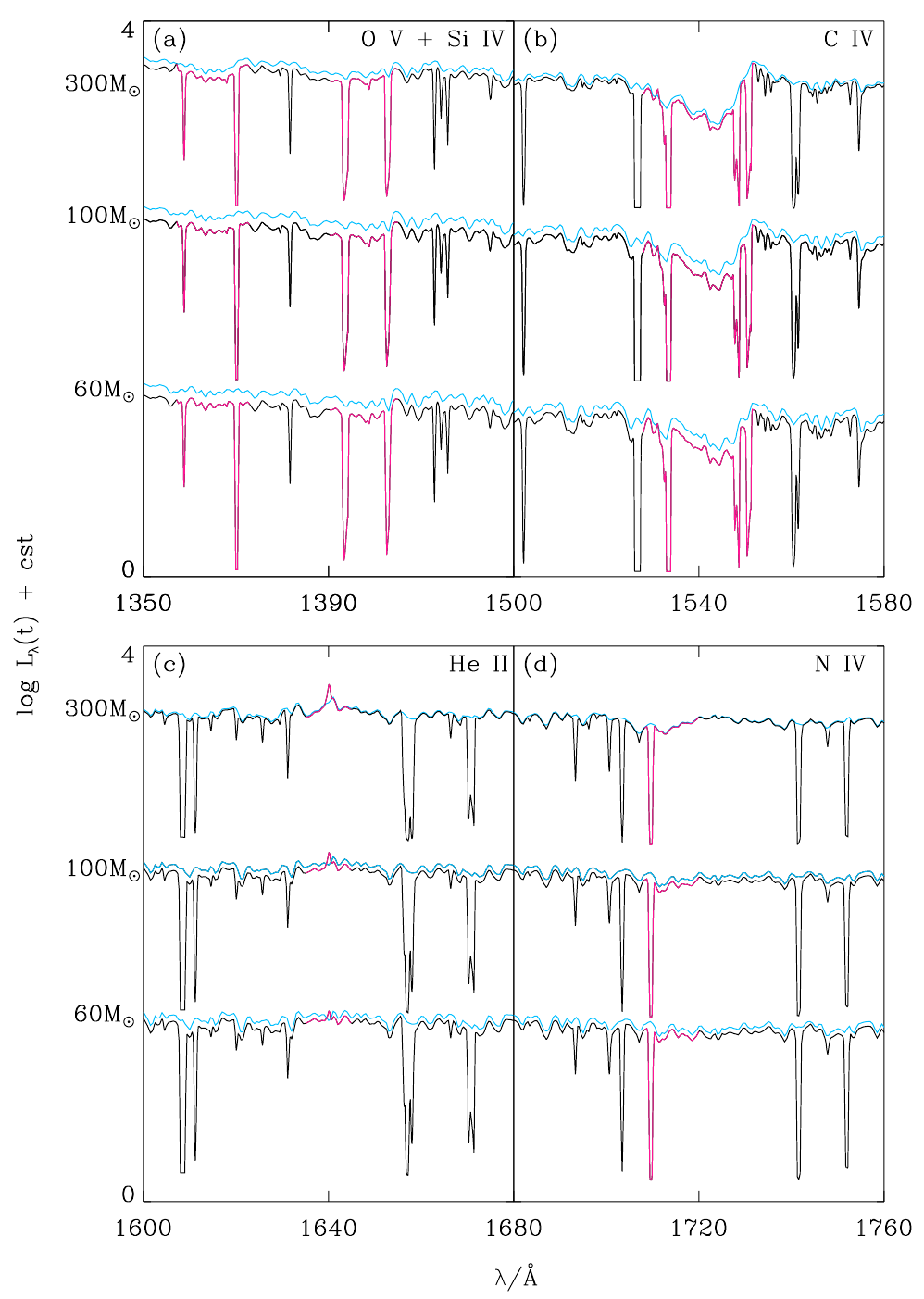}
\caption{Analog of Fig.~\ref{fig:mixedlines_csf} for young-galaxy models with constant star formation rate and IMF upper mass cutoffs $\mup=60$, 100 and 300\,\Msun, at fixed age $\tprime=10\,$Myr.}
\label{fig:mixedlines_imf}
\end{figure}

Some of the most prominent, widely studied features in the ultraviolet spectra of star-forming galaxy are blends of stellar (wind or photospheric) absorption, nebular emission and interstellar absorption. Our model may also be used to illustrate the dependence of these features on star formation and ISM parameters. In Fig.~\ref{fig:mixedlines_ssp}, we show zooms around \ov, \silivt, \civ, \heii\ and \niv\ in the spectrum emerging from a stellar birth cloud of metallicity $Z=\zism=0.017$ (chosen high enough to exhibit prominent metallic lines), at ages between 1 and 10\,Myr (the birth-cloud parameters are the same as in the mature-galaxy model of Section~\ref{sec:ismmodelsex}). This corresponds to the spectral energy distribution $S_{\lambda}(\tprime)\,T_{\lambda}^{\rm BC}(\tprime)$ in the notation of Section~\ref{sec:ismmodelsap}, for $1\leq\tprime\leq10\,$Myr. In each panel, we show for each age the input stellar population spectrum, $S_{\lambda}(\tprime)$ (in light blue), together with the emergent spectrum (in black, with the features of interest highlighted in pink). The continuum level in one spectrum relative to the other depends on the competing effects of nebular continuum radiation (most important at 1\,Myr) and attenuation by dust (dominating at 3\,Myr), the effects of both declining with age. The dependence of stellar features on SSP age \tprime\ in Fig.~\ref{fig:mixedlines_ssp} confirms previous findings \citep[e.g.,][see also \citealt{Wofford2011} for \ov]{Leitherer1995}. For example, the shape of \civ\ evolves a strong P-Cygni profile characteristic of O-type star winds at early ages to a profile characteristic of B-type star photospheres after 8\,Myr (Fig.~\ref{fig:mixedlines_ssp}b). For \silivt, the P-Cygni profile is apparent only at ages between 3 and 8\,Myr, consistent with the fact that this line traces high ratios of supergiant-to-main sequence stars (Fig.~\ref{fig:mixedlines_ssp}a; see \citealt{Walborn1984,Leitherer1995}). Instead, the P-Cygni profiles of \ov\ and \niv\ are prominent only at ages younger than about 3\,Myr, as these lines, which originate from excited energy levels (unlike \silivt\ and \civ), trace the dense winds of the most luminous O and Wolf-Rayet stars (Figs~\ref{fig:mixedlines_ssp}a,d). Finally, the broad but weak \heii\ emission is formed primarily in fast, dense winds of Wolf-Rayet stars and therefore most visible when these stars are more numerous relative to main-sequence stars, i.e., at SSP ages between roughly 3 and 5\,Myr (Fig.~\ref{fig:mixedlines_ssp}c).

The novelty in Fig.~\ref{fig:mixedlines_ssp} relative to previous studies is the ability with our model to compute in a physically consistent way the impact of nebular emission and interstellar absorption on these stellar features. At solar metallicity, nebular emission is weak in the ultraviolet (Fig.~\ref{fig:oldyoung}), and the most prominent features arise from interstellar absorption. This includes for example contamination of stellar \ov\ by interstellar Ni\,\textsc{ii}\,$\lambda1370$ at ages up to 10\,Myr, and that of stellar \silivt\ by interstellar lines of the same element at early ages (in the ionized interior of the birth cloud) and by Ni\,\textsc{ii}\,$\lambda1393$ up to older ages (in the outer neutral envelope of the cloud; Fig.~\ref{fig:mixedlines_ssp}a). The most prominent interstellar lines affecting \civ\ are those of the element itself, which are firstly in emission and then in absorption until they disappear at ages beyond 8\,Myr, and absorption by \siliidast, which remains up to 10\,Myr (Fig.~\ref{fig:mixedlines_ssp}b). \heii\ does not suffer from any significant interstellar absorption, and contamination by nebular emission is limited to  the brief phase when hot Wolf-Rayet stars are present, between 3 and 5 \,Myr (Fig.~\ref{fig:mixedlines_ssp}c). In the case of \niv, contamination arises mainly from the off-centred Ni\,\textsc{ii}\,$\lambda1710$ line, justifying a posteriori our finding in Section~\ref{sec:idxstars} that the  \bl1719 index of \citet{Fanelli1992} may be a useful (although not ideal) tracer of stellar population properties in star-forming galaxies (Fig.~\ref{fig:mixedlines_ssp}d).

Fig.~\ref{fig:mixedlines_csf} shows the analog of Fig.~\ref{fig:mixedlines_ssp} for a young-galaxy model with constant star formation rate and $Z=0.017$ (the spectrum at age 10\,Myr corresponds to the quantity $L_{\lambda}^{\rm BC}$ defined by equation~\ref{eq:lumbc}). The transient signatures of evolved supergiant and Wolf-Rayet stars identified in the spectral evolution of an SSP in Fig.~\ref{fig:mixedlines_ssp}, such as the P-Cygni profile of \silivt\ and stellar and nebular \heii\ emission, are weaker when the main sequence is constantly replenished in O-type stars (Figs~\ref{fig:mixedlines_csf}a,c). In contrast, features found also in O-type stars, such as \ov, \niv\ and especially \civ, are more pronounced and longer-lived when star formation is maintained at a constant rate (Figs~\ref{fig:mixedlines_csf}a,b,d). Perhaps the most noticeable feature in Fig.~\ref{fig:mixedlines_csf} is the steady strength of both high- and low-ionization interstellar absorption lines associated with a steady population of birth clouds at constant star formation rate. This implies a steady contamination of strong stellar lines by neighbouring interstellar lines, in the way described above. Also, in this example, attenuation by dust dominates over nebular recombination radiation at all ages, making the emergent continuum fainter than the input stellar population spectrum.

In Fig.~\ref{fig:mixedlines_met}, we show how the radiation emerging from a 10\,Myr-old galaxy with constant star formation rate depends on metallicity, for $0.001\leq Z\leq0.04$. We adopt $\log\Us=-2.0$ for $Z=0.001$ and 0.002, $\log\Us=-3.0$  for $Z=0.008$ and $\log\Us=-3.5$  for $Z=0.014,$ 0.02 and 0.04 (see Section~\ref{sec:ismmodelsex}; the spectrum for $Z=0.002$ is the same as that shown by the red curve in Fig.~\ref{fig:oldyoung}a). The strength of stellar-wind features increases from low to high $Z$, as expected from the associated enhanced efficiency of radiation pressure on metal lines \citep[e.g.,][]{Vink2001}. Lowering $Z$ also leads to a harder ionizing spectrum (since metal-poor stars evolve at higher effective temperatures than metal-rich ones; e.g., fig.~15 of \citealt{Bressan2012}). Only at the lowest metallicities considered here, $Z\lesssim0.002$, does nebular recombination radiation make the emergent spectrum slightly brighter than the input stellar-population spectrum, and some nebular lines stand out of the continuum, such as \civ\ (notched by a small interstellar absorption component; Fig.~\ref{fig:mixedlines_met}b), \heii\ and \oiiid\ (Fig.~\ref{fig:mixedlines_met}c; the small emission line just blueward of \niv\ in Fig.~\ref{fig:mixedlines_met}d is in reality the sum of S\,\textsc{iii}]\,$\lambda\lambda1713,1729$, which \cloudy\ places at 1720\,\AA). Interstellar absorption develops mainly at high metallicity, although significant \silivt\ and even \siliic\ absorption arises at $Z\lesssim0.002$ (Figs~\ref{fig:mixedlines_met}a,b). Fig.~\ref{fig:mixedlines_met}b further shows how, for the ISM parameters chosen here, \siliic, \siliidast\ and \feiia\ progressively saturate as metallicity increases.

Fig.~\ref{fig:mixedlines_imf} illustrates the influence of the upper mass cutoff of the IMF on the radiation emerging from a 10\,Myr-old galaxy with constant star formation rate and $Z=0.017$ (for $\log\Us=-3.5$, as in Fig.~\ref{fig:mixedlines_csf}). We show models with $\mup=60$, 100 (the standard value adopted throughout this paper) and $300\,\Msun$. Increasing \mup\ from 60 to 300\,\Msun\ strengthens all wind-line signatures from hot luminous stars, since more massive stars evolve at higher effective temperatures and luminosities than lower-mass stars. This also makes the ionizing spectrum of the stellar population harder, and hence, recombination-continuum and \heii-line emission stronger. Fig.~\ref{fig:mixedlines_imf} further shows that, in general, interstellar absorption features do not depend sensitively on \mup.

\section{Conclusions}\label{conclu}

We have used a combination of state-of-the-art models for the production of stellar radiation and its transfer through the ISM to investigate ultraviolet-line diagnostics of stars, the ionized and the neutral ISM in star-forming galaxies. We started by assessing the ability for the latest version of the \citet[][see also \citealt{Wofford2016}]{Bruzual2003} stellar population synthesis code to reproduce pure stellar absorption features in the ultraviolet spectra of ISM-free star clusters in the LMC, from the sample of \citet{Cassatella1987}. The fact that these clusters have small stellar masses, between about $3\times10^3$ and $4\times10^4\,\Msun$, forces one to consider the effects of stochastic IMF sampling on their integrated spectral properties \citep[e.g.,][]{Fouesneau2012}. We find that the stellar population models provide reasonable fits to the observed strengths of 17 ultraviolet indices defined by \citet{Fanelli1992} in the spectra of these clusters. In the corresponding age range, between 10\,Myr and 100\,Myr, the neglect of stochastic variations in the number of massive stars hosted by individual star clusters does not have a strong influence on ultraviolet index-based age and metallicity estimates. However, neglecting this effect can lead to systematic overestimates of stellar-mass, up to a factor of $\sim$3 for the cluster sample considered here.

Based on this success in reproducing stellar absorption features in ultraviolet spectra of young stellar populations, we have developed a new approach to model the combined influence of nebular emission and interstellar absorption on the ultraviolet spectra of star-forming galaxies. This approach builds on an idealized description of the main features of the ISM, inspired from the dust model of \citet{CharlotFall2000}. This accounts for the ionization of \hii\ regions in the interiors of the dense clouds in which stars form and the dissipation of these clouds on a timescale of typically 10\,Myr \citep[e.g.][]{Murray2010,Murray2011}. Photons emerging from stellar birth clouds and those produced by longer-lived stars must transfer through the intercloud medium on their way out of the galaxy. The combination of state-of-the-art stellar population synthesis and photoionization codes in this framework has already proven valuable to model and interpret the ultraviolet and optical nebular emission from star-forming galaxies (CL01, see also \citealt{gutkin_modelling_2016} and references therein). Here, we have expanded on these calculations to incorporate line absorption in the ionized and neutral ISM of a star-forming galaxy, by appealing to the spectrum synthesis code \synspec\ \citep[e.g.,][]{Hubeny2011}. This can be invoked via the program \cloudspec\ \citep[][see also \citealt{Heap2001}]{Hubeny2000} to compute strengths of interstellar absorption lines based on the ionization structure solved by \cloudy.

The main physical parameters of the ISM that need to be specified to compute absorption-line strengths in this context are the gas density, ionization parameter (linked to the stellar ionizing radiation), metal abundances and depletion on to dust grains. A simple parametrization of the birth clouds in terms of these physical quantities has been proposed by CL01. This was recently improved by \citet{gutkin_modelling_2016} to incorporate a versatile, fully consistent treatment of metals and dust allowing one to relate nebular emission to both gas-phase and dust-phase metal enrichment, over a wide range of chemical compositions. We adopt this parametrization to compute simultaneously nebular emission and interstellar-line absorption in the ionized interiors and outer \hi\ envelopes of birth clouds using \cloudspec. Since the birth clouds in this idealized approach are assumed to be ionization bounded, the parameters above should be regarded as `effective' ones describing the global physical conditions of the dense gas surrounding young stars throughout the galaxy (see CL01). In our model, we also account for interstellar-line absorption in the diffuse, intercloud medium. This is achieved by relating the 1500-\AA\ energy density of the interstellar radiation field to the star formation rate and by specifying the typical \hi\ column density seen by photons in the intercloud medium. 

We have used this approach to explore, in a physically consistent way, the competing effects of stellar absorption, nebular emission and interstellar absorption in ultraviolet spectra of star-forming galaxies. To this end, we have studied in detail the model spectra of a young, metal-poor galaxy and a more mature, metal-rich galaxy to identify the cleanest ultraviolet tracers of young stars, nebular emission and interstellar absorption. We find that most standard ultraviolet indices defined in the spectra of ISM-free stellar populations \citep[e.g.,][]{Fanelli1992,Leitherer2001,Rix2004,Sommariva2012} can potentially suffer from significant contamination by the ISM, which increases with metallicity. Two notable exceptions are spectral regions around 1425 and 1719\,\AA, for which contamination remains weak even at high metallicity. For the other indices, the potential influence of the ISM should be kept in mind when constraining young-star properties from observed spectra. We also identify 11 nebular-emission features, between \oiiid\ and \heilined, and 10 interstellar-absorption features, between \oi\ and the \feiiblb\ blend, which stand out as clean tracers of the different phases of the ISM (Table~\ref{tab:summarylines}). Finally, we illustrate how our model allows one to explore the complex dependence of prominent, widely studied features, such as \ov, \silivt, \civ, \heii\ and \niv, on stellar and ISM properties. 

The results presented in this paper are only examples of the potential of our approach to investigate in a physically consistent way the entangled signatures of stars, the ionized and the neutral ISM in ultraviolet spectra of star-forming galaxies. The ability to conduct such investigations opens the door to quantitative studies of the influence of the different adjustable model parameters, designed to capture the main features of stellar populations and the ISM, on emission- and absorption-line strengths. Also, while we have considered only static interstellar media in this paper, our model allows the exploration of kinematic signatures of gas infall and outflows on the profiles of lines with different ionization potentials tracing different phases of the ISM. Studies of this kind are the key to better understand the interplay between gas infall, star formation, metal and dust production and enrichment of the circumgalactic medium from ultraviolet spectroscopy of star-forming galaxies.

\section*{Acknowledgements}

We thank the anonymous referee for helpful comments. We also thank P.~Petitjean for enlightening discussions about ultraviolet interstellar features and A.~Feltre, A.~Wofford and the entire NEOGAL team for valuable advice. AVG, SC and GB acknowledge support from the ERC via an Advanced Grant under grant agreement no. 321323-NEOGAL. GB acknowledges support from the National Autonomous University of M\'exico (UNAM) through grant PAPIIT IG100115. This research has made use of the SIMBAD database, operated at CDS, Strasbourg, France \citep{2000A&amp;AS..143....9W}.

\bibliographystyle{mnras} 
\input{paper_UV_nobold.bbl}

\appendix

\section{Comparison of stellar population synthesis code versions}\label{app:comparison}

\begin{figure*}
\begin{center}
\resizebox{\hsize}{!}{\includegraphics{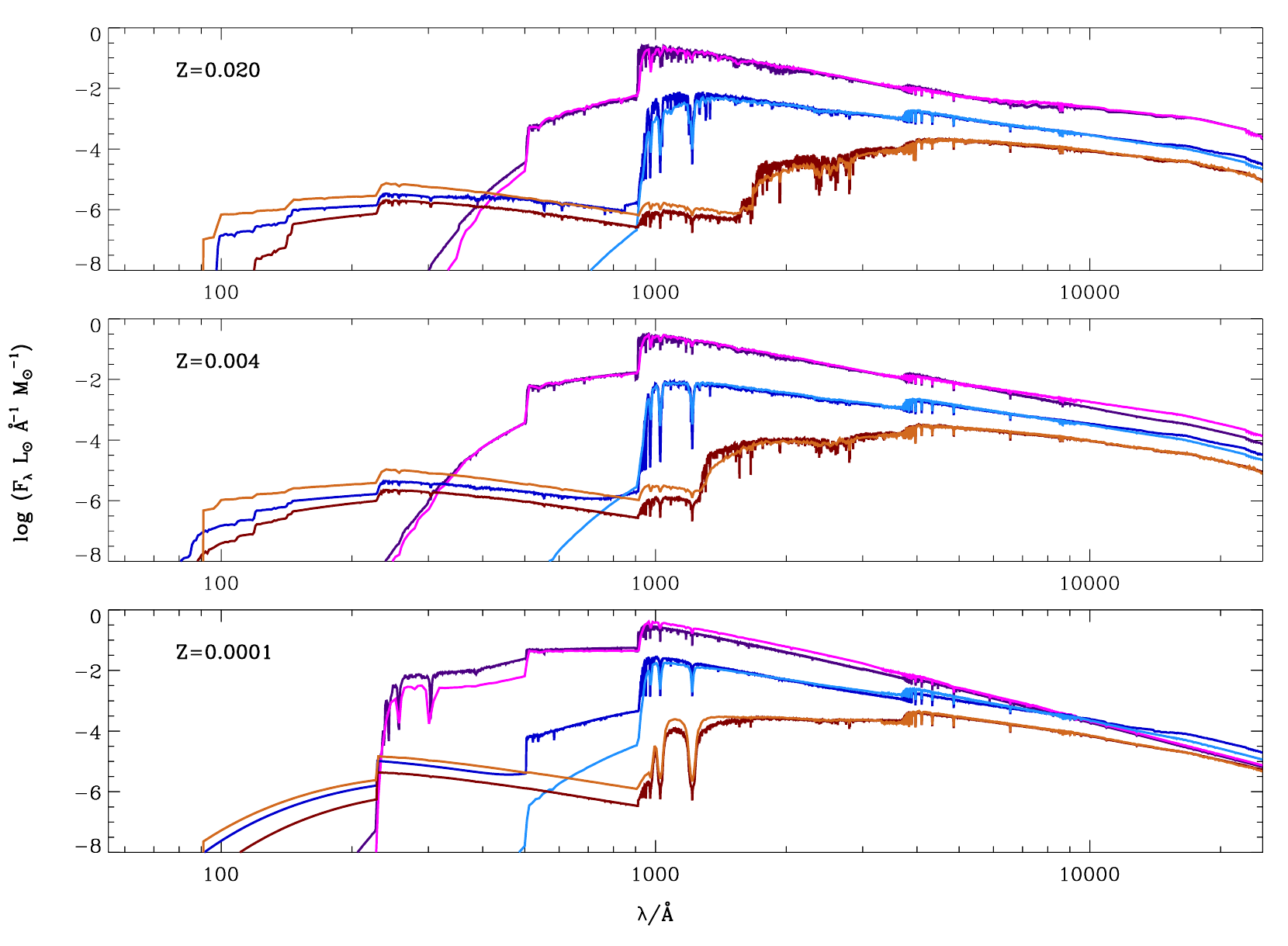}}
\end{center}
\caption{Spectral energy distributions of SSPs computed using the latest (see Section~\ref{sec:popsyn}; dark colours) and original (light colours) versions of the BC03 models, at ages  $10^7\,$yr (purple), $10^8\,$yr (blue) and $10^9\,$yr (brown), for three metallicities, $Z=0.020$ (top panel), $Z=0.004$ (middle panel) and $Z=0.0001$ (bottom panel) in the wavelength range $55\,\AA\leq\lambda\leq 2.5\,\mu$m.}
\label{fig:comparison_seds}
\end{figure*}

\begin{figure*}
\begin{center}
\resizebox{\hsize}{!}{\includegraphics{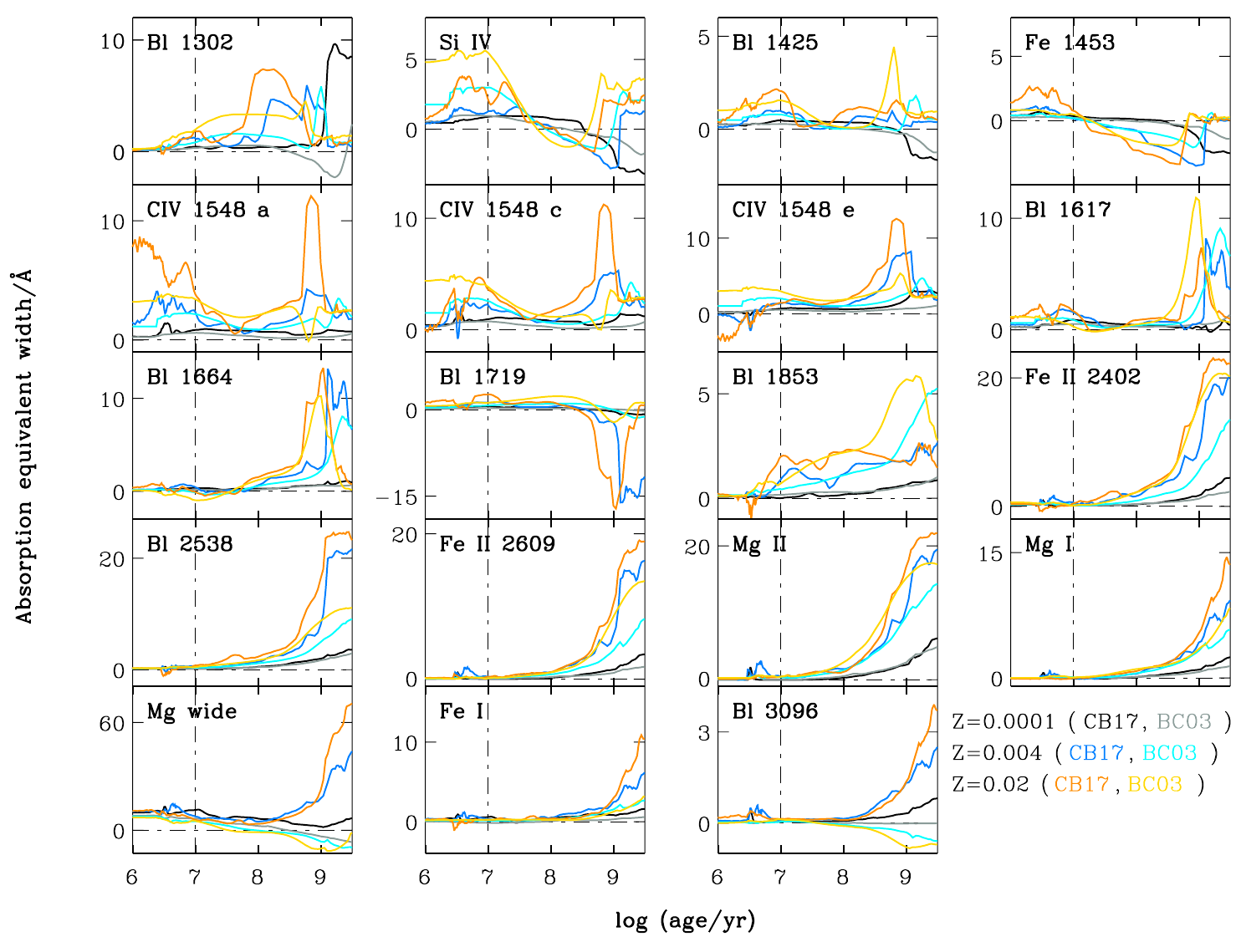}}
\end{center}
\caption{Strengths of the 19 \citet{Fanelli1992} ultraviolet spectral indices defined in Table~\ref{tab:tabindices} plotted against age, for SSPs with metallicities $Z=0.0001$, 0.004 and 0.020, computed using the latest (see Section~\ref{sec:popsyn}; dark colours, with same colour coding as in Fig.~\ref{fig:indmet}) and original (light colours) versions of the BC03 models.}
\label{fig:comparison_ind}
\end{figure*}

The version of the  \citet[][BC03]{Bruzual2003} stellar population synthesis code used in this paper (Charlot \& Bruzual, in preparation; see Section~\ref{sec:popsyn}) has not  yet been published. We show here how spectral energy distributions and \citet{Fanelli1992} ultraviolet-index strengths computed using this updated version compare with those computed using the original BC03 code in the age and metallicity ranges relevant to the present study. To perform this comparison, we consider SSP models with a smoothly sampled \citet{Chabrier2003} IMF truncated at 0.1 and 100\,\Msun.

Fig.~\ref{fig:comparison_seds} shows the spectral energy distributions at wavelengths $55\,\AA\leq\lambda\leq 2.5\,\mu$m of SSPs computed using the latest (dark colours) and original (light colours) versions of the BC03 models, at ages $10^7\,$yr (purple), $10^8\,$yr (blue) and $10^9\,$yr (brown), for three metallicities, $Z=0.020$ (top panel), $Z=0.004$ (middle panel) and $Z=0.0001$ (bottom panel). As expected, at any metallicity, the ultraviolet flux drops sharply as a function of age as short-lived, hot massive stars leave the main sequence. The two model versions exhibit some differences in the predicted ionizing spectra, such as a larger flux of \lhei-ionizing photons ($\lambda<504\,\AA$) at an age of $10^7\,$yr in the latest (dark purple) relative to the original (light purple) models. After some time, the appearance of PAGB stars makes the ionizing radiation rise again, an effect already visible at $10^8\,$yr in the most recent version of the models (based on the stellar evolutionary tracks of \citealt{Bressan2012}; dark blue spectra) and at $10^9\,$yr in both model versions (brown spectra). This delay is caused by a change in the upper-mass limit for degenerate carbon ignition, and hence PAGB evolution, from 5\,\Msun\ (turnoff age of $1.0\times10^8\,$yr) in the stellar evolution prescription used in BC03, to 6\,\Msun\ (turnoff age of $6.7\times10^7\,$yr) in that used in the latest models. At $10^9\,$yr, the change in stellar evolution prescription also makes the far-ultraviolet luminosity fainter in the latest models relative to the original ones. The ultraviolet spectral resolution of the revised models, corresponding to a resolving power ranging from $R\approx3000$ to 1000 at wavelengths from $\lambda=1000$ to 3000\,\AA\ (section~\ref{sec:ismmodelsex}), is much higher than that of the BC03 models, which have $R\approx300$ in this wavelength range (table~2 of BC03). This allows better modelling of stellar absorption features, whose strengths increase with age and metallicity as Fig.~\ref{fig:comparison_seds} shows (note in particular the series of prominent absorption lines between 1800 and 2200\,\AA\ usually attributed to \lfeiii; see \citealt{Rix2004,Leitherer2011}). The difference in the predictions of the two code versions at optical and near-infrared wavelengths arise primarily from differences in the prescriptions for post-main sequence stellar evolution \citep{Bressan2012,Marigo2013}. These are less relevant to the present study.

As a complement to Fig.~\ref{fig:comparison_seds}, we show in Fig.~\ref{fig:comparison_ind} the time evolution of the strengths of the 19 \citet{Fanelli1992} ultraviolet indices measured in the spectra of SSP with metallicities $Z=0.0001$, 0.004 and 0.020, computed using the latest (dark colours) and original (light colours) versions of the models. For simplicity, in this figure we have adopted the same colour coding as in Fig.~\ref{fig:indmet} for models computed using the most recent version of the code. In general, the two  versions predict roughly the same qualitative trends of index strengths with age. However, the much higher spectral resolution of the latest models relative to the original ones allows better sampling of deep, narrow features, hence leading to typically stronger indices.

\section{Correction of ultraviolet index strengths for Milky-Way absorption}\label{app:mwcorr}

\begin{table*}
\caption{\label{tab:ews}Equivalent widths of the 19 \citet{Fanelli1992} indices defined in Table~\ref{tab:tabindices} in the spectra of the 10 LMC clusters observed by \citet{Cassatella1987}, along with the associated measurement errors (both quantities in \AA), after correction for contamination by strong Milky-Way absorption lines.}
\begin{threeparttable}
\centering
\footnotesize\setlength{\tabcolsep}{1.9pt}
  \begin{tabular}{l r r r r r r r r r r}
  \toprule
         Index         & NGC~1711          & NGC~1805          & NGC~1818          & NGC~1847          & NGC~1850           & NGC~1866         & NGC~1984         & NGC~2004          & NGC~2011         & NGC~2100          \\
  \midrule
       \bl1302       & $3.23 \pm  0.49$  & $4.23 \pm 0.66$   & $2.93 \pm 0.94$   & $3.63 \pm  0.66$  & $3.53 \pm  0.57 $  & $3.23 \pm  0.76$ & $3.03 \pm  1.04$ & $3.33 \pm  0.41$  & $3.03 \pm 0.85$  & $3.03 \pm  0.49$  \\
       \siliv        & $1.33 \pm  0.54$  & $2.03 \pm  0.73$  & $1.13 \pm  0.82$  & $2.33 \pm  0.63$  & $0.33 \pm  0.63$   & $-0.16 \pm 0.73$ & $3.73 \pm  0.92$ & $3.23 \pm  0.44$  & $3.03 \pm  0.73$ & $2.93 \pm  0.44$  \\
       \bl1425       & $-0.30 \pm  0.50$ & $1.20 \pm  1.70$  & $0.40 \pm 0.80$   & $0.90 \pm  0.70$  & $-0.30 \pm  0.60$  & $-0.10 \pm 0.70$ & $1.60 \pm  1.00$ & $0.60 \pm  0.40$  & $0.90 \pm  0.70$ & $0.30 \pm  0.50$  \\
       \fe\          & $-0.90 \pm  0.70$ & $-0.50 \pm  1.10$ & $-1.70 \pm  1.10$ & $-0.70 \pm 1.00$  & $-2.10 \pm  1.00$  & $-2.30 \pm 1.20$ & $1.00 \pm  1.50$ & $-1.00 \pm  0.60$ & $-0.60 \pm 1.10$ & $-0.90 \pm 0.70$  \\
       \civa\        & $1.09 \pm  0.52$  & $1.39 \pm  0.81$  & $0.79 \pm  0.91$  & $0.99 \pm  0.71$  & $1.29 \pm  0.61$   & $-0.01 \pm 0.61$ & $1.69 \pm  1.11$ & $2.49 \pm  0.42$  & $3.69 \pm  0.61$ & $1.39 \pm  0.52$  \\
       \civc\        & $0.95 \pm  0.43$  & $1.05 \pm  0.72$  & $0.75 \pm  0.72$  & $1.85 \pm  0.62$  & $0.75 \pm  0.62$   & $-0.15 \pm 0.53$ & $1.75 \pm  1.01$ & $2.15 \pm  1.51$  & $3.45 \pm  0.62$ & $1.45 \pm  0.53$  \\
       \cive\        & $0.67 \pm  0.51$  & $0.90 \pm  0.80$  & $1.37 \pm  0.71$  & $1.37 \pm  0.71$  & $1.67 \pm  0.61$   & $0.57 \pm 0.61$  & $0.97 \pm  1.10$ & $0.17 \pm  0.41$  & $2.47 \pm  0.61$ & $1.47 \pm  0.41$  \\
       \bl1617       & $0.80 \pm  0.60$  & $-0.70 \pm  1.10$ & $0.90 \pm  1.10$  & $0.90 \pm  0.90$  & $1.20 \pm  0.70$   & $1.20 \pm 0.70$  & $1.80 \pm  1.50$ & $0.90 \pm  0.50$  & $1.80 \pm  0.80$ & $1.70 \pm  0.50$  \\
       \bl1664       & $0.96 \pm  0.52$  & $-0.94 \pm  0.82$ & $0.26 \pm  1.21$  & $0.16 \pm  0.82$  & $0.46 \pm  0.62$   & $1.16 \pm 0.62$  & $-0.14 \pm 1.51$ & $-0.54 \pm  0.62$ & $0.46 \pm  0.82$ & $1.16 \pm  0.53$  \\
       \bl1719       & $1.46 \pm  0.40$  & $1.26 \pm  0.60$  & $1.56 \pm  1.00$  & $1.06 \pm  0.60$  & $1.56 \pm  0.40$   & $1.86 \pm 0.40$  & $1.56 \pm  1.30$ & $1.96 \pm  0.40$  & $1.86 \pm  0.60$ & $1.86 \pm  0.50$  \\
       \bl1853       & $0.89 \pm  0.44$  & $0.89 \pm  0.53$  & $1.09 \pm  1.02$  & $1.89 \pm  0.63$  & $1.69 \pm  0.35$   & $1.79 \pm 0.63$  & $1.39 \pm  1.61$ & $1.29 \pm  0.63$  & $0.99 \pm  0.92$ & $1.29 \pm  0.63$  \\
       \feii2402     & $1.33 \pm  1.52$  & $3.03 \pm  2.01$  & $0.63 \pm  1.32$  & $0.73 \pm  1.72$  & $-0.17 \pm  1.82$  & $1.53 \pm 1.62$  & $2.83 \pm  1.52$ & $-0.07 \pm  0.83$ & $2.03 \pm  0.74$ & $1.23 \pm  1.91$  \\
       \bl2538       & $-0.65 \pm  1.41$ & $0.85 \pm  1.71$  & $-0.15 \pm  1.11$ & $-0.75 \pm 1.11$  & $-2.15 \pm  1.51$  & $0.45 \pm 1.41$  & $-0.25 \pm 1.11$ & $-0.75 \pm  0.82$ & $1.45 \pm  0.53$ & $-1.95 \pm  1.51$ \\
       \feii2609     & $1.04 \pm  0.94$  & $1.44 \pm  1.23$  & $0.64 \pm  0.75$  & $0.84 \pm  0.75$  & $0.64 \pm  0.94$   & $1.44 \pm 0.94$  & $1.04 \pm  0.66$ & $0.74 \pm  0.66$  & $0.84 \pm  0.41$ & $0.34 \pm  0.94$  \\
       \mgii\        & $3.86 \pm  0.89$  & $3.26 \pm  1.17$  & $2.66 \pm  0.72$  & $2.46 \pm  0.72$  & $3.86 \pm  0.99$   & $4.86 \pm 0.64$  & $1.36 \pm  0.64$ & $3.96 \pm  0.72$  & $3.76 \pm  0.72$ & $4.66 \pm  0.99$  \\
       \mgi\         & $1.32 \pm  0.91$  & $0.52 \pm 1.21$   & $1.12 \pm  0.72$  & $0.92 \pm  0.62$  & $2.32 \pm  0.91$   & $1.92 \pm 0.52$  & $1.02 \pm  0.42$ & $1.52 \pm  0.62$  & $0.02 \pm  0.71$ & $1.92 \pm  1.11$  \\
       \mgw\         & $1.38 \pm  6.81$  & $4.98 \pm  9.31$  & $3.48 \pm  5.52$  & $0.88 \pm  5.11$  & $5.88 \pm  7.11$   & $9.28 \pm 5.22$  & $-4.82 \pm 4.82$ & $9.98 \pm  4.22$  & $1.68 \pm  2.63$ & $-2.42 \pm  8.11$ \\
       \fei\         & $2.70 \pm  2.20$  & $3.20 \pm  3.20$  & $0.00 \pm  1.50$  & $-2.10 \pm 1.90$  & $0.10 \pm  2.50$   & $1.40 \pm 1.40$  & $1.00 \pm  1.30$ & $0.30 \pm  1.30$  & $-1.20 \pm 1.20$ & $1.10 \pm  3.00$  \\
       \bl3096       & $-1.10 \pm  1.30$ & $-2.10 \pm  2.20$ & $0.10 \pm  1.00$  & $-1.30 \pm 0.90$  & $-1.40 \pm  1.50$  & $-0.80 \pm 1.00$ & $-0.30 \pm 0.60$ & $ -0.70 \pm 0.60$ & $-0.40 \pm 0.30$ & $0.00 \pm  1.40$ \\
\midrule  
       $V$ mag\tnote{a} & $10.11 \pm 0.03$\tnote{b} & $10.63 \pm 0.03$\tnote{b} & $9.85 \pm 0.02$\tnote{b} & $11.06 \pm 0.02$\tnote{c} & $9.36 \pm 0.06$\tnote{b} & $9.89 \pm 0.01$\tnote{b} & $9.99 \pm 0.04$\tnote{d} & $9.60 \pm 0.02$\tnote{c} & $10.58 \pm 0.02$\tnote{c} & $9.60 \pm 0.04$\tnote{b} \\
       $A_{V}$ mag\tnote{e}  & $0.56 \pm 0.01$& 0$.32 \pm 0.02$ & $0.39\pm 0.02$ & $0.49 \pm 0.02$ & $0.33 \pm 0.01$ & $0.28 \pm 0.06$ & $0.36 \pm 0.02$ & $0.33 \pm 0.02$ & $0.47 \pm 0.02$ & $0.80 \pm  0.02$ \\
\bottomrule
 \end{tabular}
 \begin{tablenotes}
\item [a] Apparent $V$-band magnitude not corrected for reddening. To transform to absolute magnitude, we adopt the extinction $A_{V}$ listed in this table and a distance modulus to the LMC of 18.49$\pm$0.05\,mag \citep[i.e., a distance of 49.97$\pm$1.11\,kpc;][]{Pietrzynski2013}.\\
\item [b] Apparent $V$-band magnitude and error from \citet{vdBerg1968}. \\
\item [c] Apparent $V$-band magnitude from \citet{Bernard1974} and error from \citet{Pessev2008}.\\
\item [d] Apparent $V$-band magnitude and error from \citet{Pessev2008}.\\
\item [e] $V$-band extinction from \citet[][and references therein]{Pessev2008}.

\end{tablenotes} 
 \end{threeparttable}
 \label{tab:idxcorr}
\end{table*}

Resonant transitions of different species in the neutral and ionized ISM of the Milky Way produce discrete absorption features in the ultraviolet spectra of the LMC clusters observed by \citet{Cassatella1987}. We must correct for these features before fitting the spectra with pure stellar population synthesis models. \citet{Savage2000} measure the absorption equivalent widths of ionized species from the Milky-Way ISM in the ultraviolet spectra of background quasars along $83$ different lines of sight observed with the Faint Object Spectrograph on board \textit{HST}. They find that weakly ionized species, such as \lsilii\ or \lcii, are in general stronger (with equivalent widths of up to $\sim0.8$\AA) than highly ionized species, such as \lsiliv\ or \lciv\  (with equivalent widths of up to $\sim0.4$\AA). \citet[][see their table~3]{Leitherer2011} compute the median equivalent widths of the $24$ strongest absorption lines in this sample (the values listed by these authors in units of km\,s$^{-1}$ can be converted into \AA\ by dividing by the line frequency). Here, we use these data to correct, a posteriori, the  equivalent widths of the 19 \citet{Fanelli1992} indices defined in Table~\ref{tab:tabindices} measured by \citet{Maraston2009} in the {\it IUE} spectra of \citet{Cassatella1987}. We find that 15 of the 24 strong ISM lines can affect the central and continuum bandpasses of 14 indices. For two indices, \bl2538\ and \feii2609, Milky-way lines contaminate both the central and a continuum bandpasses. If a Milky-Way line falls in a central index bandpass, we correct the index equivalent width as
\begin{equation}
{\rm EW_{cluster}^{corr}=EW_{cluster}^{M09}-EW_{MW}}\,,
\end{equation}
where ${\rm EW_{MW}}$ is the equivalent width of the Milky-Way line(s) affecting the central bandpass and ${\rm EW_{cluster}^{M09}}$ is the index equivalent width measured by \citet{Maraston2009}. The term $\Delta\mathrm{(EW/\AA)}$ listed in Table~\ref{tab:tabindices} is in this case simply ${\rm -EW_{MW}}$. We compute the error in ${\rm EW_{cluster}^{corr}}$ by propagating those in ${\rm EW_{cluster}^{M09}}$ \citep{Maraston2009} and ${\rm EW_{MW}}$ \citep{Leitherer2011}. The indices to which we have applied this correction are \bl1302, \siliv, \civa, \civc, \civa, \bl1664, \bl1853, \feii2402, \feii2609, \mgii, \mgi\ and \mgw.

When a Milky-Way absorption line falls in a pseudo-continuum bandpass, we estimate a mean correction to that index strength for all clusters in the sample as follows. We consider a unit spectrum, in which we carve, in the central index bandpass, an absorption line (with Gaussian profile and unit standard deviation) of equivalent width the median value measured by \citet{Maraston2009} in the \citet{Cassatella1987} cluster spectra. Then, we recompute the index strength after carving the Milky-Way absorption line(s) (with Gaussian profile and unit standard deviation)  in the continuum bandpass. The $\Delta\mathrm{(EW/\AA)}$ listed in Table~\ref{tab:tabindices} in this case is the difference between the first and second estimates. We compute the error in ${\rm EW_{cluster}^{corr}}$ by repeating the process for Milky Way-line equivalent widths differing by $\pm\sigma_{i}$, where $\sigma_{i}$ is the line-measurement error from \citet{Leitherer2011}. The indices to which we have applied this correction are \bl1719, \bl1853, \bl2538\ and \feii2609. For the last two indices,  \bl2538\ and \feii2609, the total correction is the sum of that affecting the central bandpass and that affecting the pseudo-continuum bandpass. The final corrected index strengths are presented in Table~\ref{tab:ews} and the corrections terms listed in Table~\ref{tab:tabindices}.

\end{document}